%% file: curl2.tex
\documentclass[prd, superscriptaddress, tightenlines, longbibliography, nofootinbib, eqsecnum, amsfonts, amsmath, floatfix, twocolumn, notitlepage]{revtex4-2}
\pdfoutput=1

\usepackage[utf8]{inputenc}
\usepackage{mathrsfs}
\usepackage{mathtools}
\usepackage{euscript}
\usepackage{graphics}
\usepackage{graphicx}
\usepackage{amsmath}
\usepackage{amssymb}
\usepackage{gensymb}
\usepackage{bm}
\usepackage[usenames,dvipsnames,svgnames,table]{xcolor}
\definecolor{linkcolor}{rgb}{0.6,0,0}
\definecolor{citecolor}{rgb}{0,0,0.75}
\definecolor{urlcolor}{rgb}{0.12,0.46,0.7}
\usepackage{xspace}
\usepackage{wasysym}
\usepackage{times}
\usepackage{appendix}
\usepackage{comment}
\usepackage{lipsum}
\usepackage[nolist,nohyperlinks]{acronym}
\usepackage{float}
\usepackage{simplewick}
\usepackage{natbib, ifthen}
\usepackage[breaklinks, colorlinks, urlcolor=urlcolor, linkcolor=linkcolor,citecolor=citecolor,pdfencoding=auto]{hyperref}
\usepackage{longtable}
\usepackage{listings}
\usepackage{bbm}
\usepackage{hyperref}

\include{texbase/macros}

\defcitealias{Aghanim:2018oex}{PL2018}

\defcitealias{Pratten:2016dsm}{PB16}

\defcitealias{Robertson:2023xkg}{RL23}
\newcommand{\RL}{\citetalias{Robertson:2023xkg}\xspace}

\newcommand{\nhat}{\boldsymbol{\hat{n}}}
\newcommand{\ehat}{\boldsymbol{\hat{e}}}

\newcommand{\isdraft}[1]{}

\newcommand{\plancklens}{\texttt{plancklens}}

\begin{document}
\title{Detectable signals of post-Born lensing curl B-modes}

\newcommand{\Sussex}{Department of Physics \& Astronomy, University of Sussex, Brighton BN1 9QH, UK}

\author{Mathew Robertson}
\affiliation{\Sussex}

\author{Giulio Fabbian}
\affiliation{Institute of Astronomy, Madingley Road, Cambridge CB3 0HA, UK}
\affiliation{Kavli Institute for Cosmology Cambridge, Madingley Road, Cambridge CB3 0HA, UK}
\affiliation{School of Physics and Astronomy, Cardiff University, The Parade, Cardiff, CF24 3AA, UK}
\affiliation{Center for Computational Astrophysics, Flatiron Institute, New York, New York 10010, USA}

\author{Julien Carron}
\affiliation{Universit\'e de Gen\`eve, D\'epartement de Physique Th\'eorique et CAP, 24 Quai Ansermet, CH-1211 Gen\`eve 4, Switzerland}

\author{Antony Lewis}
\affiliation{\Sussex}

  \begin{abstract}

Curl lensing, also known as lensing field-rotation or shear B-modes, is a distinct post-Born observable caused by two lensing deflections at different redshifts (lens-lens coupling). For the Cosmic Microwave Background (CMB), the field-rotation is approximately four orders of magnitude smaller than the CMB lensing convergence. Direct detection is therefore challenging for near-future CMB experiments such as the Simons Observatory (SO) or CMB `Stage-4' (CMB-S4). Instead, the curl can be probed in cross-correlation between a direct reconstruction and a template formed using pairs of large-scale structure (LSS) tracers to emulate the lens-lens coupling. In this paper, we derive a new estimator for the optimal curl template specifically adapted for curved-sky applications, and test it against non-Gaussian complications using N-body cosmology simulations. We find non-foreground biases to the curl cross-spectrum are purely Gaussian at the sensitivity of SO. However, higher-order curl contractions induce non-Gaussian bias at the order of $1\sigma$ for CMB-S4 using quadratic estimators (QE). Maximum a-Posteriori (MAP) lensing estimators significantly reduce biases for both SO and CMB-S4, in agreement with our analytic predictions. We also show that extragalactic foregrounds in the CMB can bias curl measurements at order of the signal, and evaluate a variety of mitigation strategies to control these biases for SO-like experiments. Near-future observations will be able to measure post-Born lensing curl B-modes.
  \end{abstract}


   \maketitle

\section{Introduction}
Measurements of the Cosmic Microwave Background (CMB) radiation --- the relic light emitted at recombination in the early universe --- offer a direct window into the physics of the late universe (redshifts $z\lesssim20$) through gravitational lensing \cite{Lewis:2006fu}. This is the deflection of photons from their background geodesics due to the gravitational potential of matter along the line-of-sight. Therefore, the mass distribution of the large scale structure (LSS) imprints on CMB observations as a secondary signature, characterized by the lensing deflection field. 

The deflection vector field can be Helmholtz-decomposed into gradient and curl components.  The gradient component, reconstructed through measurements of the lensing convergence (or shear-E mode), has been detected with high significance by various CMB experiments \cite{BICEP2:2016rpt,POLARBEAR:2019snn,Carron:2022eyg,ACT:2023dou,SPT:2023jql}. In contrast, the curl component, described by the lensing field rotation (or shear-B mode), has yet to be detected\footnote{Negative curl power was detected at $\sim 2.9\sigma$ by Planck \cite{PL2018}. However, since the amplitude of its spectrum was negative it could not have been a physical signal, and was likely due to unknown systematics or an unusual fluctuation. The signal was not seen by subsequent CMB lensing measurements from ACT \cite{ACT:2023dou} and SPT \cite{SPT:2023jql}.}. This is expected because, at linear order in matter fluctuations (for scalar perturbations), the lensing deflection field is identically curl-free: a single lensing event cannot produce rotation because there is no handedness to a scalar field. While vector and tensor perturbations can source curl lensing at first order, the signal is negligible in $\Lambda$CDM \cite{Book:2011dz,Yamauchi:2012bc,Rotti:2011aa,Saga:2015apa}.

However, a small but non-negligible curl signal does exist at second order in scalar perturbations, as described by the lens-lens coupling term in post-Born corrections to the lensing deflection field \cite{Cooray:2002mj, Krause:2009yr, Pratten:2016dsm}. The origin of post-Born rotation is conceptually simple: consider a photon beam deflected sequentially at two different redshifts along the line-of-sight. Although the beam cannot be rotated instantaneously at either lensing event, if it is sheared twice in different directions then the resulting image has a net rotation compared to the original unlensed field, with handedness given by the relative orientation of the two shears. This signal is small enough that prospects for direct measurement will remain difficult even for optimistic future surveys \cite{Pratten:2016dsm}. However, Ref.~\cite{Robertson:2023xkg} (\RL hereafter) demonstrated that post-Born field rotation can instead be measured as a cross-correlation signal. The method involves mimicking the lens-lens coupling by using pairs of LSS matter tracers at different redshifts to create an optimal template for the lensing field rotation. The cross-spectrum of the template with a rotation measurement therefore probes the rotation-tracer-tracer bispectrum signal, which has the advantage of leveraging the high $S/N$ of LSS observables. 
Assuming appropriately chosen tracers, under idealized assumptions the rotation-template cross-spectrum is then detectable with the Simons Observatory (SO,~\cite{Ade:2018sbj}) at up to $6\sigma$, or CMB-S4 \cite{Abazajian:2019eic} at $15\sigma$ (or $21\sigma$ assuming deep South Pole observations). 

Although the \RL template is optimal and computationally tractable, it was derived in the flat-sky approximation. In this paper, we derive a new expression for the template that remains optimal and tractable, but can also be applied to curved-sky analyses. We test application of this method in the presence of non-Gaussian and foreground signals.

One significant complication of the cross-template method is the presence of reconstruction bias. While the signal has the advantage of being a cross-spectrum, effectively nulling the large reconstruction noise in CMB lensing auto-measurements, even small cross-spectrum biases could be worrisome for the curl which itself is a small signal. \RL showed that disconnected contractions of lensing-tracer pairings do bias the cross-spectrum at the order of the signal, even when no rotation is present. At this level the bias can safely be subtracted using simulations, since estimation of Gaussian terms is robust assuming the corresponding spectra are well measured. However, reliability of non-Gaussian statistics in simulations is less certain, so any presence of non-Gaussian bias could be a limiting factor in the detection of curl lensing. Another problem lies in the foreground contamination of CMB observations. Extragalactic processes, themselves tracers of the underlying matter distribution, are also secondary sources of anisotropies in the CMB. Even if these foregrounds are curl-free, they could contribute additional contractions with lensing and/or tracers, leading to more bias terms. In this paper we make use of the non-linear information contained in full-sky N-body simulations, together with the new curved-sky rotation template, to investigate the importance of non-Gaussian and foreground contributions to the curl cross-spectrum bias. 

This paper is organized as follows. Derivation of a curved-sky field rotation template, which we then use throughout the rest of the paper, is presented in Section~\ref{sec:tem}. Description of the DEMNUni N-body simulation \cite{Carbone:2016nzj, Castorina:2015bma} and its derived observables is provided in Section~\ref{sec:sim}. The simulation is then used to validate the curved-sky template in Section~\ref{sec:results}, before the investigation proceeds to non-Gaussian cross-spectrum biases. In Section~\ref{sec:fg} we switch to the AGORA N-body simulation \cite{Omori:2022uox} to examine extragalactic foreground biases. Finally, the main conclusions are summarized in Section~\ref{sec:conclusion}. Some supplementary analyses are provided in the appendices: Appendix \ref{app:magbias} investigates the effect of magnification bias on the curl template and Appendix \ref{sec:fisher} gives forecasts for the cosmological constraining power of post-Born rotation. Throughout we assume a fiducial flat $\Lambda$CDM cosmology with standard General Relativity.

\section{Full-sky rotation template}\label{sec:tem}
The weak gravitational lensing deflection field, $\boldsymbol{\alpha}$, is identically curl free to linear perturbative order in density fluctuations\footnote{We do not consider any first order contributions to the curl that could arise from enhanced vector or tensor perturbations (e.g. \cite{Namikawa:2013wda}).}
\begin{flalign}
\phantom{\text{[1st order]}}&&\alpha_a=\grad_a\phi,
&&\text{[1st order]}
\end{flalign}
and hence only depends on the gradient of a scalar field -- the lensing potential, $\phi$. Post-Born corrections to $\boldsymbol{\alpha}$ do produce non-zero curl at second order \cite{Pratten:2016dsm}
\begin{flalign}\label{eq:alpha}
\phantom{\text{[2nd order]}}&&\alpha_a=\grad_a\phi+\epsilon_{ab}\grad^b\Omega,
&&\text{[2nd order]}
\end{flalign}
so now another field is required to fully describe the lensing deflections -- the curl potential, $\Omega$. The anti-symmetric tensor $\boldsymbol{\epsilon}$, allows the 2-dimensional cross-product of scalar $\Omega$ to be written in the Einstein summation convention. We probe this through the clock-wise field rotation observable\footnote{Equivalently, one could instead consider the degenerate shear B-mode signal \cite{Krause:2009yr}.}, $\omega$, described here following a post-Born expansion in $\boldsymbol{\alpha}$
\begin{multline}\label{eq:alpha_pert}
\omega(\hat{\boldsymbol{n}})\equiv-\frac{1}{2}\boldsymbol{\grad}^2\Omega(\boldsymbol{\hat{n}})=-\frac{1}{2}\epsilon^{ab}\nabla_{b}\alpha_{a}(\hat{\boldsymbol{n}}) \\= -2\int^{\chi_s}_0d\chi W(\chi,\chi_s)
\int^{\chi}_0d\chi'W(\chi',\chi)\qquad\\\times\epsilon^{ab}\nabla_a\nabla_c\Psi(\hat{\boldsymbol{n}},\chi)\nabla_b\nabla^c\Psi(\hat{\boldsymbol{n}},\chi')
+\mathcal{O}(\Psi^3),
\end{multline}
at some point on the sky, $\hat{\boldsymbol{n}}$. The window function, $W(\chi,\chi_s)$, returns the fractional contribution to the total signal of a lens at radial distance $\chi$, given a source image at $\chi_s$.

The shear-shear post-Born term that gives rise to $\omega$ is evident in Eq.~\eqref{eq:alpha_pert} from the radially-coupled dependence between spin-2 fields involving two derivatives of the gravitational Weyl potential, $\nabla_a\nabla_c\Psi(\chi)$ and $\nabla_b\nabla^c\Psi(\chi')$.
We simplify the above equation by defining two tensor fields roughly analogous to the high- and low-$z$ shear contributions. The high-$z$ tensor
\begin{equation}
\mathcal{A}_{ab}(\boldsymbol{\hat{n}},\chi) \equiv \nabla_a\nabla_b\Psi(\boldsymbol{\hat{n}},\chi),
\end{equation}
is coupled to a low-$z$ Born lens
\begin{equation}
\mathcal{B}_{ab}(\boldsymbol{\hat{n}},\chi) \equiv \int^{\chi}_0d\chi'W(\chi',\chi)\mathcal{A}_{ab}(\boldsymbol{\hat{n}},\chi'),
\end{equation}
which allows $\omega$ to be written in a compact form that encapsulates the shear-shear coupling
\begin{equation}\label{eq:omega_ab}
\omega(\boldsymbol{\hat{n}})=-2\int^{\chi_s}_0d\chi W(\chi,\chi_s)\epsilon^{ab}\mathcal{A}_{ac}(\boldsymbol{\hat{n}},\chi)\mathcal{B}_{b}^c(\boldsymbol{\hat{n}},\chi).
\end{equation}

By introducing a general spin-2 field $P^{X} \equiv e^a_+e^b_+X_{ab}$, for complex vectors $\boldsymbol{e}_{\pm}=\boldsymbol{e}_1\pm i\boldsymbol{e}_2$ defined in terms of orthogonal basis vectors, $\omega$ can be re-expressed as 
\begin{equation}\label{eq:omega_P}
\omega(\boldsymbol{\hat{n}})=(-1)^h\int^{\chi_s}_0d\chi W(\chi,\chi_s)\operatorname{\mathbb{I}m}\{(P^{\mathcal{A}})^*P^{\mathcal{B}}\}(\boldsymbol{\hat{n}},\chi).
\end{equation}
This new form allows covariant conversion from a flat-sky Fourier space $\omega(\boldsymbol{L})$ to an approximate full-sky spherical harmonic version, $\omega_{LM}$. We demonstrate how this works in practice by using the new form of Eq.~\eqref{eq:omega_P} to create a curl template on the curved sky.

Note that sign of Eq.~\eqref{eq:omega_P} is controlled by integer index $h$ which represents the handedness of the coordinate system. We stick to the convention consistent with Ref.~\cite{Pratten:2016dsm} that positive $\omega$ defines clock-wise rotation, thus $h=1$ for left-handed systems and $h=0$ for right-handed. See Appendix \ref{app:hand} for more details.

\subsection{Flat-sky template}
In \RL, an optimal flat-sky template, $\hat{\omega}^{\textrm{tem}}$, was proposed using quadratic combinations of LSS tracers, $a^i$, to mimic the lens-lens interaction. Forecasts for the estimator showed it was a high-fidelity template, and when cross-correlated with a CMB lensing rotation reconstruction, $\hat{\omega}$, should produce detectable signals. The predictions were less promising for rotation measurements in upcoming galaxy surveys, therefore we only consider CMB lensing $\omega$ throughout the paper. This simplifies the source distance $\chi_s=\chi_*$, approximated as a single source plane at the surface of last scattering. 

Our aim is to write down a full-sky version of $\hat{\omega}^{\textrm{tem}}$. Before we get there, let's recap the flat-sky template
\begin{multline}\label{eq:omega_qe_split_ft}
\hat{\omega}^{\textrm{tem}}(\boldsymbol{L})=-2F_L^{-1}\int^{\chi_*}_0d\chi W_{\kappa}(\chi, \chi_*)\epsilon_{rs}\\
\mathcal{F}\left\{\left[ \Gamma^{pr}\Lambda_p^{s}\right](\boldsymbol{\hat{n}},\chi)\right\}.
\end{multline}
We now specify the kernel, $W_{\kappa}(\chi, \chi_*)$, as the CMB lensing window function
\begin{equation}\label{eq:window_cmb}
	W_{\kappa}(\chi,\chi_s)=\frac{3}{2}\Omega_m H_0^2\left(1+z\right)\chi\frac{\chi_s-\chi}{\chi_s}\Theta(\chi_s-\chi),
\end{equation}
for Heaviside function, $\Theta$, redshift $z\equiv z(\chi)$, fractional matter density, $\Omega_m$, and Hubble constant, $H_0$. The template normalization is simply the per-mode Fisher information for the amplitude of a perfect curl-template cross-spectrum 
\begin{equation}\label{eq:template_norm} F_L\!=\!\!\frac{1}{2C^{\omega\omega}_{L}}\!\!\int\frac{d^2\boldsymbol{L}_1}{(2\pi)^2}b^{\omega ij}_{(-\boldsymbol{L})\boldsymbol{L}_1\boldsymbol{L}_2}(\boldsymbol{C}_{\textrm{LSS}}^{-1})^{ip}_{L_1}(\boldsymbol{C}^{-1}_{\textrm{LSS}})^{jq}_{L_2}b^{\omega pq}_{(-\boldsymbol{L})\boldsymbol{L}_1\boldsymbol{L}_2}.
\end{equation}
Here, theory rotation-tracer-tracer bispectra, $b^{\omega ij}$, contract with bispectra weighted by the inverse of their LSS covariance $\boldsymbol{C}_{\textrm{LSS}}$. The presence of the fiducial post-Born rotation auto-spectrum, $C^{\omega\omega}_{L}$, ensures the template is normalized such that $\langle C^{\hat{\omega}^{\textrm{tem}}\omega}_{L}\rangle= C^{\omega\omega}_{L}$. The forward and backward Fourier transforms are respectively denoted $\mathcal{F}$ and $\mathcal{F}^{-1}$, and the functions $\Gamma = \mathcal{F}^{-1}(\gamma)$, and $\Lambda = \mathcal{F}^{-1}(\lambda)$ are defined as
\begin{equation}\label{eq:E_gamma}
\begin{aligned}
\gamma^{pr}(\boldsymbol{L},\chi)&=\frac{L^pL^r}{L^2}\frac{W_i(\chi)}{\chi^2}P_{\delta\delta}\left(k\approx\frac{L}{\chi},z(\chi)\right)(\boldsymbol{C}^{-1}_{\textrm{LSS}})^{ij}_La_j({\boldsymbol{L}})\\
&\equiv\frac{L^pL^r}{L^2}E^{\gamma}(\boldsymbol{L}, \chi),
\end{aligned}
\end{equation}
and
\begin{equation}\label{eq:E_lambda}
\begin{aligned}
\lambda^{pr}(\boldsymbol{L},\chi)&=\frac{L^pL^r}{L^2}C^{i\kappa}_L(\chi)(\boldsymbol{C}^{-1}_{\textrm{LSS}})^{ij}_La_j({\boldsymbol{L}})\\
&\equiv\frac{L^pL^r}{L^2}E^{\lambda}(\boldsymbol{L}, \chi).
\end{aligned}
\end{equation}
We have now introduced the power spectrum, $P_{\delta\delta}$, of the comoving-gauge matter density perturbations, $\delta$. Equally, one could instead use the gravitational Weyl potential spectrum, $P_{\Psi\Psi}$, which would require redefinition of Eq.~\eqref{eq:window_cmb}.   

\subsection{Curved-sky template}\label{sec:fullsky_tem}
We see that Eq.~\eqref{eq:omega_ab} and Eq.~\eqref{eq:omega_qe_split_ft} share the same form, so we can re-express the flat-sky template in the covariant from of
\begin{multline}\label{eq:tem_flat_P}
\hat{\omega}^{\textrm{tem}}(L)=(-1)^hF_L^{-1}\int^{\chi_*}_0d\chi W(\chi,\chi_*)\\
\times\mathcal{F}\left[\operatorname{\mathbb{I}m}\left\{(P^{\Gamma})^*P^{\Lambda}\right\}(\boldsymbol{\hat{n}}, \chi)\right], 
\end{multline}
which involves the product of spin-2 fields
\begin{equation}\label{eq:flat_P}
P^X(\boldsymbol{\hat{n}}, \chi)=-\mathcal{\prescript{}{+2}F}^{-1}\left[E^x(\boldsymbol{L}, \chi)+ iB^x(\boldsymbol{L}, \chi)\right].
\end{equation}
The B-mode components of $P^{\Gamma}$ and $P^{\Lambda}$ are identically zero, while the E-modes are defined in Eq.~\eqref{eq:E_gamma} and Eq.~\eqref{eq:E_lambda}. Notice that a spin-2 version of the Fourier transform $\mathcal{F}\xrightarrow{}{\prescript{}{\pm2}F}$ is now required. We verify that Eq.~\eqref{eq:tem_flat_P} and Eq.~\eqref{eq:omega_qe_split_ft} are exactly equivalent to numerical precision.

This new flat-sky form of $\omega^{\textrm{tem}}(L)$ has only implicit dependence on derivatives, since there are no explicit $L$-factors in Eq.~\eqref{eq:tem_flat_P} or \eqref{eq:flat_P}. These have been packed away and are now handled exclusively through spin-2 transformations. Therefore, the jump to a curved-sky template consistent with the covariant form of Eq.~\eqref{eq:omega_P} is achieved by simply replacing the Fourier transform with its spherical harmonic counterpart
\begin{multline}\label{eq:omega_tem_full}
\hat{\omega}^{\textrm{tem}}_{LM}=(-1)^hF_L^{-1}\int^{\chi_s}_0d\chi W(\chi,\chi_s)\\
\times\int d\boldsymbol{\hat{n}}Y_{LM}^*(\boldsymbol{\hat{n}})\operatorname{\mathbb{I}m}\left\{(P^{\Gamma})^*P^{\Lambda}\right\}(\boldsymbol{\hat{n}},\chi).
\end{multline}
Hence, the spin-2 fields take on the spherical harmonic definition
\begin{equation}
P^X(\boldsymbol{\hat{n}},\chi)=-\sum_{LM}\left[E^X_{LM}(\chi)+ iB^X_{LM}(\chi)\right]\prescript{}{+2}Y_{LM}(\boldsymbol{\hat{n}}),
\end{equation}
with corresponding harmonic-space template E-modes
\begin{equation}\label{eq:E_gamma_full}
E^{\gamma}_{LM}(\chi)\equiv \frac{W_i(\chi)}{\chi^2}P_{\delta\delta}\left(\frac{L}{\chi},z(\chi)\right)(\boldsymbol{C}^{-1}_{\textrm{LSS}})^{ij}_L (a_{j})_{LM},
\end{equation}
and
\begin{equation}\label{eq:E_lambda_full}
E^{\lambda}_{LM}(\chi)\equiv C^{i\kappa}_L(\chi)(\boldsymbol{C}^{-1}_{\textrm{LSS}})^{ij}_L (a_{j})_{LM}.
\end{equation}
Again the $B_{LM}$ are zero because the underlying tensors are pure gradients. We now have a covariant expression for the template on the curved sky, and will later show with simulations that this approximated result is sufficient to robustly measure rotation on the full sky. The convention throughout the rest of this paper is to work in the right-handed curved-sky polar basis $\boldsymbol{\hat{n}}=(\boldsymbol{\hat{e}}_{\theta}, \boldsymbol{\hat{e}}_{\phi})$, consistent with the HEALPIX\footnote{\href{https://healpix.jpl.nasa.gov/}{https://healpix.jpl.nasa.gov/}} convention, which sets $h=0$. This is different from \RL, in which their curl formalism was derived in flat-sky Cartesian coordinates (left-handed, $h=1$). 

\subsubsection{Fiducial modelling}
We keep the curved-sky template normalization $F_L$ unchanged from its flat-sky version. While use of flat-sky normalization will affect the accuracy of curved-sky results on the largest scales, we only present results at $L>30$ where curved-sky effects are limited. The coordinate handedness is also not important because rotation bispectra enter Eq.~\eqref{eq:template_norm} twice, thus rendering their sign irrelevant. Therefore, fiducial power spectra and bispectra also keep their flat-sky form. 

The cross spectra for the LSS density tracers are modelled to lowest order in the Limber approximation \cite{Kaiser92}
\begin{equation}\label{eq:density_tracer_ps}
	C^{ij}_{\ell}=\int d\chi \frac{W_{i}(\chi)W_{j}(\chi)}{(\chi)^2}P_{\delta\delta}\left(k\approx\frac{\ell}{\chi},z(\chi)\right).
\end{equation}
This ensures the spectra are modelled to consistent order with the bispectra, which follow the analytic post-Born formalism from \RL
\begin{multline}\label{eq:rot_bi}
    \begin{split}
    b^{\omega ij}_{(-\boldsymbol{L})\boldsymbol{L}_1\boldsymbol{L}_2}=(-1)^h2\frac{(\boldsymbol{L}_1\cdot \boldsymbol{L}_2)[\boldsymbol{L}_1
    \times \boldsymbol{L}_2]}{(L_1)^2(L_2)^2}\big[\mathcal{M}^{ij}(L_1,L_2)\\-\mathcal{M}^{ji}(L_2,L_1)\big],
        \end{split}
\end{multline}
with $\boldsymbol{L}_2=\boldsymbol{L}-\boldsymbol{L}_1$, and mode couplings defined by
\begin{multline}\label{eq:mode_coupling}
    \begin{split}
    \mathcal{M}^{ij}(L,L')=\int^{\chi_*}_{0}d\chi \frac{W_{\kappa}(\chi)W_{i}(\chi)}{\chi^2}P_{\delta\delta}\left(\frac{L}{\chi},z(\chi)\right)\\
    \times\int^{\chi}_{0} d\chi' \frac{W_{j}(\chi')W_{\kappa}(\chi',\chi)}{(\chi')^2}P_{\delta\delta}\left(\frac{L'}{\chi'},z(\chi')\right).
        \end{split}
\end{multline}
Here, $W_{\kappa}(\chi)\equiv W_{\kappa}(\chi,\chi_*)$. 

Finally, the analytic post-Born rotation spectrum is also modelled to lowest order in the Limber approximation \cite{Cooray:2002mj, Krause:2009yr}
\begin{equation}\label{eq:omega_ps}
    C^{\omega\omega}_{L}=4\int \frac{d^2\boldsymbol{l}}{(2\pi)^2}\frac{
    \left(\boldsymbol{l}\cdot\boldsymbol{l}'\right)^2\left[\boldsymbol{l}\times\boldsymbol{L}\right]^2}{l^4\left(l'\right)^4}
    \mathcal{M}^{\omega\omega}\left(l,l'\right),
\end{equation}
where the rotation mode-coupling matrix has a slightly different definition
\begin{multline}
    \begin{split}\label{eq:omega_mc}
    \mathcal{M}^{\omega\omega}(l,l')&=\int^{\chi_*}_{0}d\chi \frac{W_{\kappa}(\chi,\chi_*)^2}{\chi^2}
    P_{\delta\delta}\left(\frac{l}{\chi},z(\chi)\right)\\
    &\times\int^{\chi}_0d\chi'\frac{W_{\kappa}(\chi',\chi)^2}{\chi'^2}
    P_{\delta\delta}\left(\frac{l'}{\chi'},z(\chi')\right),
        \end{split}
\end{multline}
and here $\boldsymbol{l}'\equiv\boldsymbol{L}-\boldsymbol{l}$.

\section{Non-Gaussian simulation}\label{sec:sim}
Reference \RL used Gaussian realizations of matter tracers to test the curl cross-spectrum method in the null hypothesis of no rotation-tracer-tracer bispectra. The use of Gaussian simulations also restricted the analysis to focus only on Gaussian biases to the curl cross-spectrum. For actual recovery of the field rotation signal, we require simulations of tracers with the correct higher-order correlations in place. This is achieved using N-body simulations and ray-tracing  techniques, to include both the effect of non-linear clustering of matter and post-Born lensing. We used the DEMNUni N-body simulation suite \cite{Carbone:2016nzj, Castorina:2015bma} and the full-sky ray tracing approach presented in Refs.~\cite{Fabbian:2017wfp, Fabbian:2019tik}. 

The DEMNUni simulation suite was designed to study the impact of a massive neutrino cosmology, but for our work we used the $\Lambda$CDM reference simulation based on the Planck 2013 cosmology assuming massless neutrinos\footnote{The fiducial cosmological parameters are: the fractional cold dark matter density $\Omega_c=0.27$, baryon density $\Omega_b=0.05$, the spectral index $n_s=0.96$, reduced Hubble parameter $h=0.67$, amplitude of density fluctuations on 8 $h^{-1}$Mpc scales, $\sigma_8=0.83$, and finally the optical depth to reionization $\tau=0.0925$.}. The simulation consists of 62 snapshots of $(2\,\rm{Gpc}/h)^3$ volume with $2048^3$ dark matter particles. A full-sky light cone is constructed from this series of outputs for an observer at the centre of the box. This light cone is then divided into 62 spherical shells of surface mass density $\Sigma_i$ within the redshift range $0 \leq z \leq 99$ \cite{Calabrese:2014gla}, using Healpix pixelization. From each shell, the projected surface mass overdensity from the $i$th matter density sheet is computed as
\begin{equation}
    \Delta_i = \frac{\Sigma_i -\bar{\Sigma_i}}{\bar{\Sigma}_i}.
\end{equation}
The $\Delta_i$ maps are then used in the curved-sky multiple-lens ray-tracing algorithm \cite{Hilbert:2008kb,Das:2007eu}, which propagates all components of the distortion tensor of the CMB \cite{Becker:2012qe} beyond the Born approximation, as implemented in the LenS$^2$HAT code \cite{Fabbian:2013owa}. The code outputs post-Born CMB lensing convergence $\kappa^{\textrm{sim}}$ and rotation $\omega^{\textrm{sim}}$, as well as the lensing convergence computed in the Born approximation, $\kappa^{\textrm{born}}$, using the same matter distribution as in the full post-Born case. The ray tracing was carried out at  $\sim 3$ arcsec resolution on an Equidistant Cylindrical Projection (ECP \cite{Muciaccia:1997pi}) grid following the approach of Ref.~\cite{Fabbian:2017wfp}. Because the components of the ray-traced distortion tensor are not strictly bandlimited, their full resolution version was  transformed in harmonic space and filtered to remove all power at $\ell\geq 8000$ prior to being transformed back in real space on an Healpix grid with $\texttt{nside}=4096$ to avoid aliasing.

\begin{figure}[t]
 	\includegraphics[width=\linewidth]{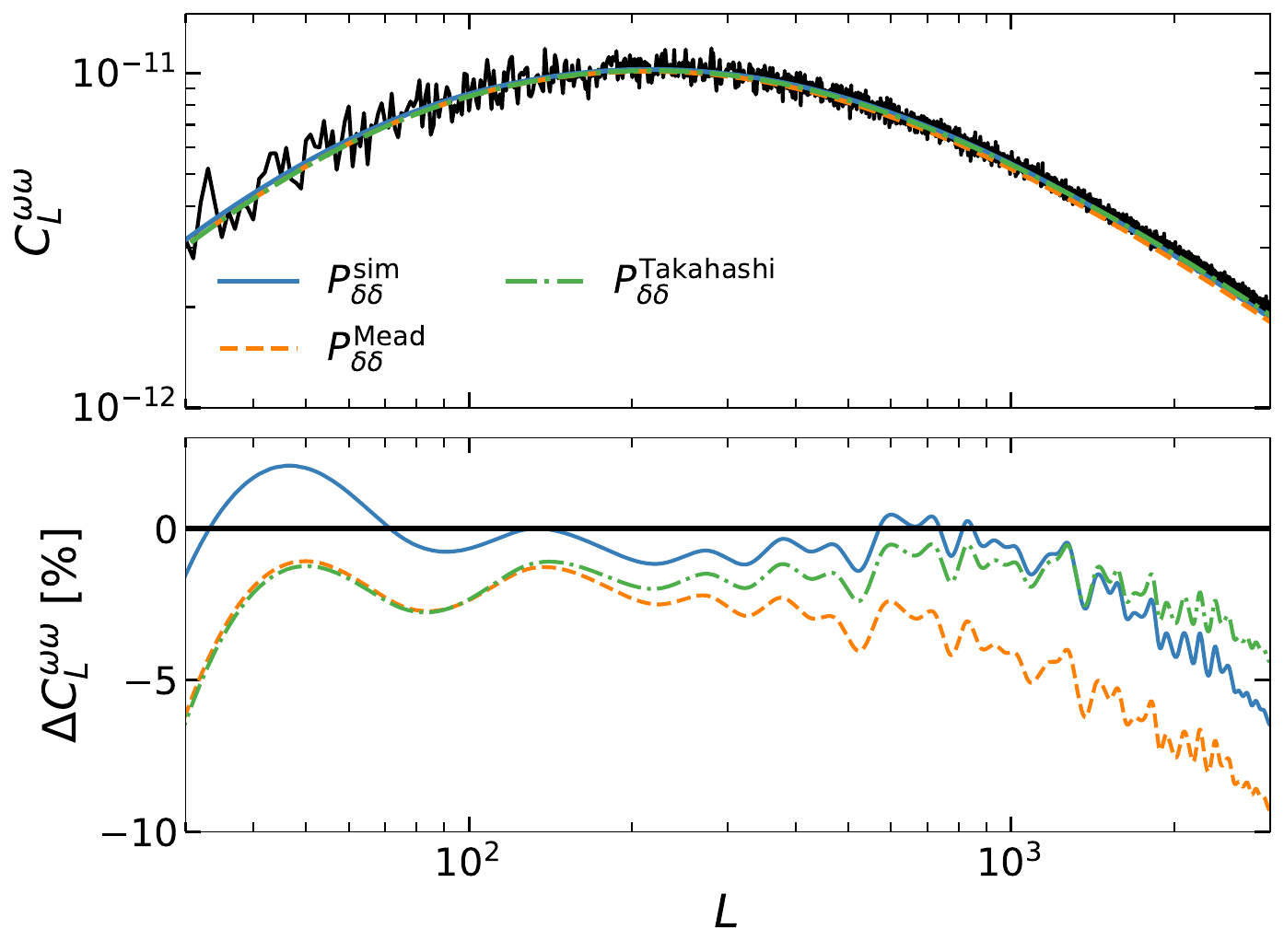}	\caption{\small{The total field rotation auto-spectrum as measured from ray tracing through one realization of the DEMNUni N-body simulation (solid black), compared to the fiducial post-Born rotation spectrum (solid blue) calculated to leading order in Eq.~\eqref{eq:omega_ps} with the empirical matter distribution of the simulation, $P^{\textrm{sim}}_{\delta\delta}$. Also shown for comparison are theory curves using non-linear prescriptions of $P_{\delta\delta}$ such as HMcode (Mead) \cite{Mead:2016zqy} (dashed orange), and Takahashi \cite{Takahashi:2012em} (dash-dot green). The bottom panel displays the percentage deviation of each theory curve to the (smoothed) ray-traced rotation. }}
	\label{fig:delta_omega}
\end{figure}

The autospectrum of $\omega^{\textrm{sim}}$ is plotted in Fig.~\ref{fig:delta_omega} and compared to the analytic flat-sky post-Born prediction using Eq.~\eqref{eq:omega_ps}.
To model the rotation mode-coupling matrix of Eq.~\eqref{eq:omega_mc}, we used the empirical matter power spectrum, $P^{\textrm{sim}}_{\delta\delta}$, which was calculated directly from the DEMNUni simulation without correcting for simulation shot noise\footnote{Where the empirical power spectrum was not resolved, we instead used the non-linear prescription from HMcode (Mead) \cite{Mead:2016zqy}. This came into effect for the smallest scales, $k>1$, only for the higher redshift shells, $z>2.5$, or very small scales $k>10$ at lower redhsifts $z<2.5$. The choice of prescription here did not affect the results in Fig.~\ref{fig:delta_omega}.}. The rotation field is consistent with theory on scales $L\lesssim 1000$, but is biased on smaller scales. The discrepancy could be coming from low-$z$ contributions to $\omega$ where the magnitude of the rotation angles are comparable to the resolution of the ray tracing and shot noise itself. This was tested in figure 2 of Ref.~\cite{Fabbian:2019tik}, in which the DEMNUni ray-traced rotation exhibits better agreement with theory at higher redshift and worse at low redshift. We also compare using non-linear prescriptions HMcode (Mead) and Takahashi \cite{Takahashi:2012em} to model $P_{\delta\delta}$, and likewise find they exhibit a similar bias trend on small scales.

\subsection{CMB lensing}\label{sec:cmb_lens}

One leg of the cross-spectrum signal is the CMB lensing reconstructed rotation, $\hat{\omega}$. The reconstructed CMB convergence, $\hat{\kappa}$, is also used as one of the input density tracers to the rotation template. Here we describe the process used to simulate these observables for different experimental configurations.

A deflection field can be decomposed into gradient and curl components, and calculated from harmonic convergence and rotation fields 
\begin{equation}\label{eq:alpha_comps}
\alpha^{\phi}_{LM} = \frac{2}{\sqrt{L(L+1)}}\kappa_{LM} \quad\text{and}\quad \alpha^{\Omega}_{LM} = \frac{2}{\sqrt{L(L+1)}}\omega_{LM}
\end{equation}
We derive three different $\boldsymbol{\alpha}$ using combinations of DEMNUni lensing maps as input to Eq.~\eqref{eq:alpha_comps}
\begin{equation}
    \boldsymbol{\alpha}^{(\textrm{sim})}\equiv \boldsymbol{\alpha}\left[\kappa^{\textrm{sim}},\omega^{\textrm{sim}}\right],
\end{equation}
\begin{equation}
    \boldsymbol{\alpha}^{(\phi)}\equiv \boldsymbol{\alpha}\left[\kappa^{\textrm{sim}},0\right],
\end{equation}
\begin{equation}
    \boldsymbol{\alpha}^{(\Omega)}\equiv \boldsymbol{\alpha}\left[0,\omega^{\textrm{sim}}\right].
\end{equation}
We then generate 100 different Gaussian realizations of unlensed CMB temperature and polarization fields, $T$, $E$, and $B$, from their auto- and cross-spectra, $C^{TT}_{\ell}$, $C^{EE}_{\ell}$, and $C^{TE}_{\ell}$. Primordial B-modes are not considered here, so the initial $B$ maps have $C^{BB}_{\ell}=0$. Each map is lensed in real space\footnote{The {\sc lenspyx} \cite{Reinecke:2023gtp} software package performed the lensing deflections on the unlensed fields.} by $\boldsymbol{\alpha}^{(\textrm{sim})}$, creating a set of 100 lensed CMBs. We repeat for $\boldsymbol{\alpha}^{(\phi)}$ and $\boldsymbol{\alpha}^{(\Omega)}$ using the same initial unlensed CMB fields, and thus procure 3 sets of lensed CMB simulations. A further set of CMB maps were generated using Gaussian lensing deflections
\begin{equation}
    \boldsymbol{\alpha}^{(G)}_i\equiv \boldsymbol{\alpha}\left[\kappa^{G}_i,0\right],
\end{equation}
in which each $\kappa^{G}_{LM}$ is a different realization with variance $C^{\kappa^{\textrm{sim}}\kappa^{\textrm{sim}}}_L$. For this set only, each unlensed CMB realization was lensed by a unique lensing realization.  

Simulated map-level noise is also required to emulate the data quality of near-future CMB experiments. We primarily consider noise contamination at 2 levels; for a Simons Observatory (SO) \cite{Ade:2018sbj} like setup at ``goal sensitivity'', and also a CMB-S4 \cite{Abazajian:2019eic} like experiment. The noise spectra, $N^{TT}_{\ell}$, $N^{EE}_{\ell}$, and $N^{BB}_{\ell}$, are the same curves as described in \RL, and include detector noise, beam deconvolution, and contributions from foreground mitigation using the Harmonic-space Internal Linear Combination (HILC) model \cite{Ade:2018sbj}. Gaussian isotropic noise realizations were generated from these spectra and added to the lensed CMB maps. The same noise is given to each of the 4 lensed CMB sets. We occasionally also consider a foreground free ``CMB-S4 deep'' experimental configuration from \RL, which has much lower noise compared to CMB-S4 but covers a significantly smaller patch of the sky (assuming a South Pole observation within the CMB-S4 setup). 

Finally, the lensing observables, $\hat{\kappa}$ and $\hat{\omega}$, were reconstructed\footnote{Lensing reconstruction performed using \plancklens ~\cite{PL2018} software.} from the simulations using the full global minimum variance (GMV) quadratic estimator (QE) \cite{Maniyar:2021msb}. To account for potential uncertainties in the noise modelling of the ground-based experiments considered, multipole cuts were applied to the maps at $30\leq L\leq 3000$ for $T$, and $30\leq L\leq 5000$ for $E$ and $B$.

\subsubsection{MAP reconstruction}\label{sec:map}
Although QEs are near optimal for current CMB experiments, other methods will be required for effective lensing reconstruction with improved polarization data. The maximum a posteriori (MAP) lensing estimator is the most optimal estimator of a single lensing reconstruction map~\cite{Hirata:2003ka, Carron:2017mqf}, thus we also consider MAP-derived observables, $\hat{\kappa}^{\textrm{MAP}}$ and $\hat{\omega}^{\textrm{MAP}}$, in our analysis. While the details of the MAP procedure are complex, it can be summarized as follows: the method iteratively delenses the CMB maps using a newly estimated quadratic lensing template from the previous iteration. Each subsequent set of delensed maps has reduced variance due to the removal of the lensing signal, leading to progressively better lensing measurements. This technique is particularly effective for experiments which are limited by the variance of lensed B modes. 

For MAP lensing we use the \texttt{delensalot}\footnote{\href{https://github.com/NextGenCMB/delensalot}{https://github.com/NextGenCMB/delensalot}} package to perform simultaneous joint gradient and curl reconstruction \cite{Belkner:2023duz}. The same scale cuts were applied as for the QE case: $30\leq L\leq 3000$ for $T$, and $30\leq L\leq 5000$ for $E$ and $B$. The reconstruction is stopped at 15 iterations. The normalization of the output lensing maps is approximated as an inverse Wiener-filter \cite{Legrand:2023jne}
\begin{equation}
A^{\hat{\omega}^{\textrm{MAP}}}_L=\left[\frac{C^{\omega\omega}_L}{C^{\omega\omega}_L+N^{\omega,\textrm{MAP}}_0(L)}\right]^{-1}\!\!,
\end{equation}
and similarly for $A^{\hat{\kappa}^{\textrm{MAP}}}_L$. The MAP reconstruction biases, $N^{\textrm{MAP}}_0$, are estimated from the usual analytic GMV $N_0$\footnote{In this paper, we stick to the convention of describing CMB lensing reconstruction auto-spectrum biases by $N_0$, $N_1$, etc. in which the order in $C^{\phi\phi}$ is placed in the subscript. Later, for curl cross-spectrum biases we instead place the order in the superscript, i.e. $N^{(0)}$, $N^{(1)}$, etc.} by replacing the lensed gradient CMB spectra with their delensed counterparts in an iterative procedure \cite{Legrand:2021qdu}. We used \texttt{plancklens} to compute the analytic $N^{\textrm{MAP}}_0$.

Iterative reconstruction unfortunately produces anisotropies in the lensing response, even for full-sky maps, which can bias the normalization at $\sim5\%$ \cite{Legrand:2021qdu,Legrand:2023jne}. We account for this by computing an additional MC correction using 20 noiseless CMB simulations lensed by new Gaussian realizations $\kappa^{\textrm{inp}}$ and $\omega^{\textrm{inp}}$ (different from the $\kappa^{\textrm{G}}$ realizations defined previously)
\begin{equation}
    A^{\hat{\omega}^{\textrm{MAP}},\textrm{MC}}_L=\left[\frac{\langle C^{\hat{\omega}^{\textrm{MAP}}\omega^{\textrm{inp}}}_L\rangle_{\textrm{CMB}}}{C^{\omega^{\textrm{inp}}\omega^{\textrm{inp}}}_L}\right]^{-1}\!\!,
\end{equation}
and similarly for $A^{\hat{\kappa}^{\textrm{MAP}},\textrm{MC}}_L$. 

\subsection{Tracers}

The rotation template in Eq.~\eqref{eq:omega_tem_full} is designed to predict $\omega$ using observations of density tracers, $\hat{a}^i$. We continue to follow \RL and consider three LSS observables in our analysis, $\hat{a}\in\{\hat{\kappa}, \hat{g},\hat{I}\}$. Simulations of $\hat{\kappa}$ were described in Section \ref{sec:cmb_lens}. For galaxy density, $g$, and the Cosmic Infrared Background (CIB), $I$, maps were derived directly from the DEMNUni overdensity shells 
\begin{equation}\label{eq:make_tracer}
a = \left(W_a\right)^i \Delta_i,
\end{equation}
where the $i$th kernel is the integrated window function spanning the redshift range of the shell 
\begin{equation}
(W_a)^i\equiv\int^{z^i_{\textrm{max}}}_{z^i_{\textrm{min}}} \frac{dz}{H(z)} W_a\left(\chi\right).
\end{equation}
It is implicit that $\chi\equiv\chi(z)$. We use the same window functions as described in Appendix B of \RL. Hence, $W_g$ produces a one-bin LSST-like galaxy density \cite{LSSTScience:2009jmu}, while $W_I$ models the Planck 2018 CIB map at $353$ GHz \cite{Aghanim:2016pcc}. Modelling of the galaxy bias is included within the window functions. For maps derived directly from $\Delta_i$, such as $g$ and $I$, we also apply a pixel window function correction so their auto- and cross-spectra are unbiased.

\begin{figure}[t]
 	\includegraphics[width=\linewidth]{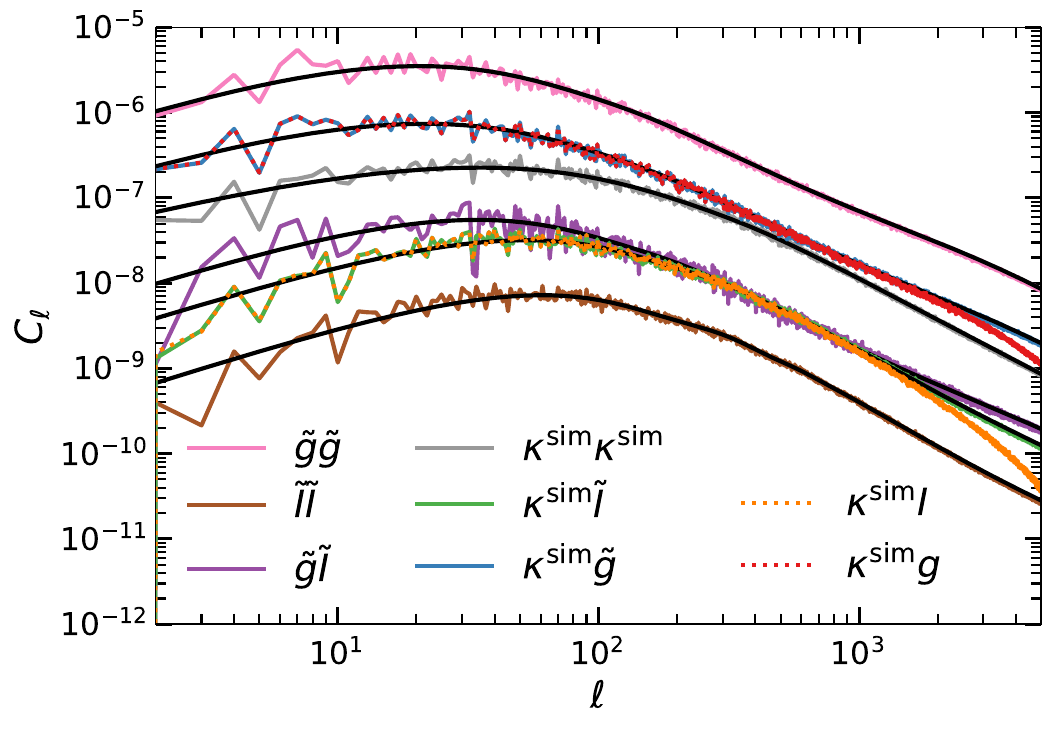}
	\caption{\small{The auto- and cross-spectra of matter density tracer fields derived from the DEMNUni simulation compared to the respective fiducial curves (solid black) at lowest order in the Limber approximation. The cross-spectra between the unlensed galaxy density, $g$, or CIB, $I$, with the ray-traced CMB lensing convergence, $\kappa^{\textrm{sim}}$, (dashed orange or dashed red) experience loss of power on small scales due to mismatched perturbative orders. This is corrected by replacing the unlensed fields with lensed galaxy density, $\tilde{g}$, and lensed CIB, $\tilde{I}$, as demonstrated by the solid-line spectra. The CIB maps have units of MJy sr$^{-1}$.}}
	\label{fig:lss_cls}
\end{figure}

As the $g$ and $I$ fields were constructed from unlensed density perturbations, they are by definition modelled to 1st perturbative order; adequate for correlating with other 1st order observables. However, $\kappa^{\textrm{sim}}$ includes post-Born corrections at 2nd perturbative order. Cross-correlations between observables at different orders experience unphysical loss of correlation towards smaller scales \cite{Fabbian:2019tik,Bohm:2019bek}. As an example, in Fig.~\ref{fig:lss_cls} we show the impact of the lack of correct post-Born corrections in $C_\ell^{\kappa^{\textrm{sim}}I}$ and $C_\ell^{\kappa^{\textrm{sim}}g}$. Therefore, we also build a set of {\it lensed} tracers, $\tilde{g}$ and $\tilde{I}$, such that they can be crossed with $\kappa^{\textrm{sim}}$ to consistent perturbative order. To do this, we require the lensing convergence, $\kappa_i$, of each density shell in the simulation. This is calculated by summing the projected density fluctuations of all shells at $z<z_i$

\begin{figure*}[t]
 	\includegraphics[width=\linewidth]{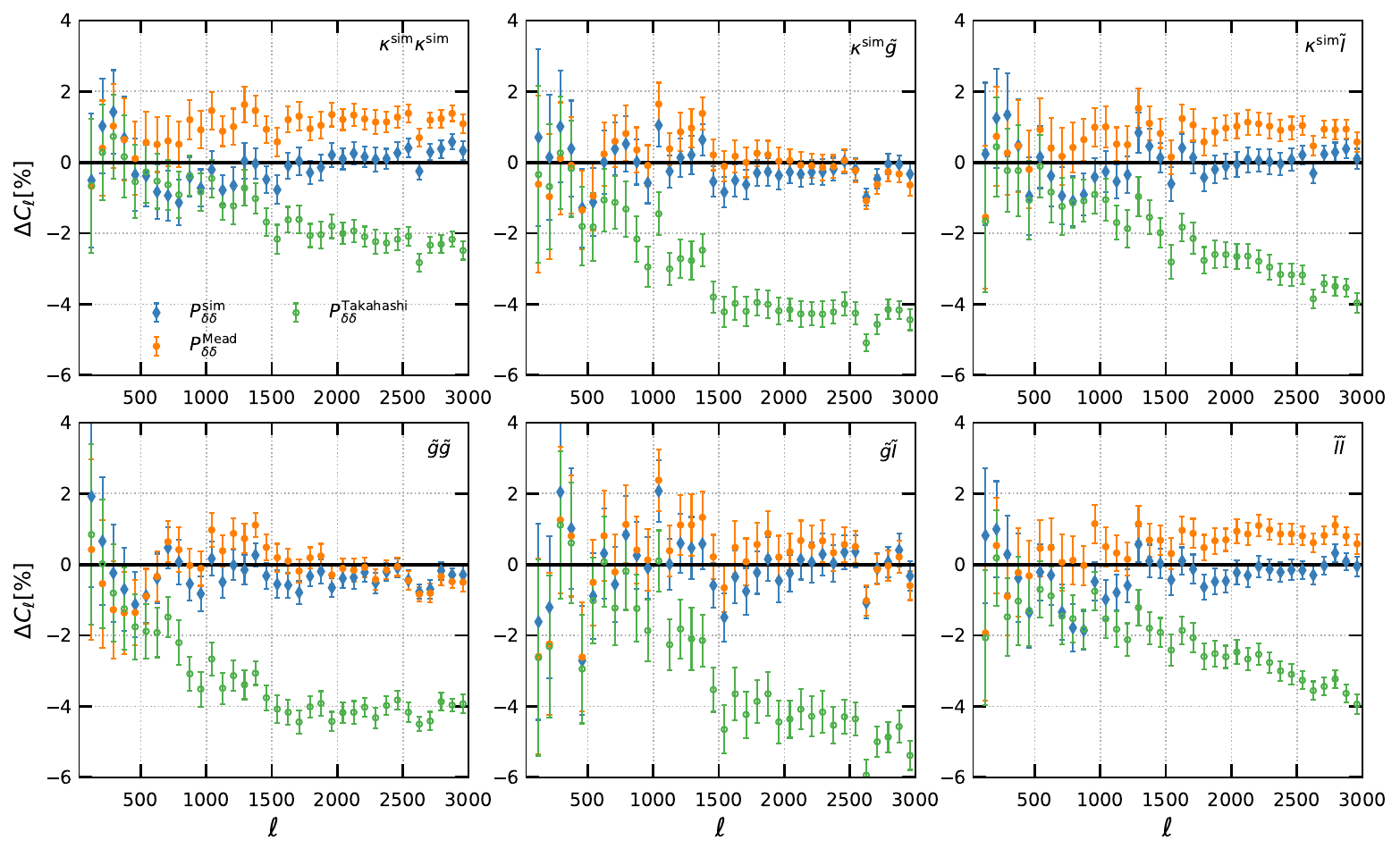}
	\caption{\small{Each panel plots the percentage difference between auto- and cross-spectra of the lensed DEMNUni tracer fields compared to the fiducial lowest-order Limber models. The blue diamonds show that the spectra agree with theory to sub-percent levels on scales $30\leq\ell\leq3000$ when accounting for the empirical matter distribution of the N-body simulation, $P^{\textrm{sim}}_{\delta\delta}$. For comparison, we also show the difference of the measured spectra to model predictions using alternative theory prescriptions for $P_{\delta\delta}$, HMcode (Mead) \cite{Mead:2016zqy} (orange dots), and Takahashi \cite{Takahashi:2012em} (green circles).}}
	\label{fig:lss_diff}
\end{figure*}

\begin{equation}
\kappa_i=\sum^{j}_{j<i}\left[W_{\kappa}\left(\chi_i\right)\right]^j\Delta_j,
\end{equation}
assuming that $\Delta_i$ can be treated as a single source plane. The value of $W_{\kappa}\left(\chi_i\right)$ at the $j$th shell is taken to be the radially averaged value
\begin{equation}
\left[W_{\kappa}\left(\chi_i\right)\right]^j=\int^{z^j_{\textrm{max}}}_{z^j_{\textrm{min}}} \frac{dz}{H(z)} W_{\kappa}\left(\chi, \chi_i\right).
\end{equation}
Each density sheet $\Sigma_i$ was subsequently lensed by the corresponding deflection field, $\boldsymbol{\alpha}\left[\kappa_i,0\right]$, to create a set of {\it lensed} densities $\tilde{\Sigma}_i$. The new lensed observables were constructed as before
\begin{equation}
\tilde{a} = \left(W_a\right)^i \tilde{\Delta}_i.
\end{equation}
We note that this procedure does not account for post-Born corrections at the fully correct order given the $\kappa_i$s were computed in the Born approximation. However, as discussed in Ref.~\cite{Fabbian:2019tik}, this is enough to account for the relevant post-Born corrections at leading order in cross-correlation. 

The cross-spectra of the lensed tracers with $\kappa^{\mathrm{sim}}$ now agree with the fiducial predictions, as shown in Fig.~\ref{fig:lss_cls} and Fig.~\ref{fig:lss_diff}. We only use maps at $L\geq 30$ where the Limber approximation is adequate \cite{LoVerde:2008re}. There are significant deviations on the largest scales $L<10$ (not shown in Fig.~\ref{fig:lss_diff}), however these are primarily due to the finite box size of the simulation and cosmic variance. On the same figure we also compare with $C_{\ell}$ predictions using different models for $P_{\delta\delta}$ that are less consistent with the measured N-body simulation power.

\section{Non-Gaussian results}\label{sec:results}
We now make use of the mock CMB lensing reconstructions, $\hat{\kappa}$ and $\hat{\omega}$, and matter tracer observables, $\tilde{g}$ and $\tilde{I}$, derived from the DEMNUni N-body simulation in the previous section. Unless explicitly specified otherwise, $\hat{\omega}^{\textrm{tem}}$ represents a full-sky field rotation template constructed using Eq.~\eqref{eq:omega_tem_full} for all 3 density tracers $a\in\{\hat{\kappa},\tilde{g},\tilde{I}\}$. 

Reference \RL showed that CMB map-level noise significantly biases the rotation cross-spectrum. Therefore, $\hat{\omega}$ and $\hat{\kappa}$ were reconstructed from CMB maps that included noise to ensure all bias terms were captured. On the other hand, noise in observations of $\hat{\tilde{g}}$, or $\hat{\tilde{I}}$, do not bias the signal assuming noise is independent of the cosmological fields, and independent between tracers. So, to reduce simulation scatter noise was not added to the simulated galaxy and CIB maps, instead $\tilde{g}$ and $\tilde{I}$ were applied directly to the template (thus the hat notation is dropped). Modelling of noise for {\it all} tracers was however included in the template weights, $(\boldsymbol{C}_{\textrm{LSS}})^{ij}_{\ell}=C^{ij}_{\ell}+\delta_{ij}N^{ij}_{\ell}$, used for optimal filtering of tracers in Eq.~\eqref{eq:E_gamma_full} and Eq.~\eqref{eq:E_lambda_full}. The $C_{\ell}$s were modelled using the fiducial lowest order limber spectra (Eq.~\ref{eq:density_tracer_ps}) and empirical $P^{\textrm{sim}}_{\delta\delta}$, assuming diagonal covariance. The noise spectra, $N_{\ell}$, were taken from Appendix B of \RL. For galaxy clustering, $N^{gg}_{\ell}$ was modelled as the shot noise forecast for the LSST gold sample~\cite{LSSTScience:2009jmu}, while $N^{II}_{\ell}$ contains shot noise and dust contamination from fits to the Planck 2018 CIB observations \cite{Aghanim:2016pcc}. CMB noise was added as appropriate for SO or CMB-S4 experimental configurations as described in Sec.~\ref{sec:cmb_lens}.

Before moving on to interpreting the results, let's first establish a threshold at which reconstruction biases are relevant. For the foreseeable future, achieving sub-percent precision on $C^{\omega\hat{\omega}^{\textrm{tem}}}$ is unrealistic, so effects at this level are negligible. In the context of SO or CMB-S4, optimistic Fisher forecasts predict detections at $\sim 6\sigma$ and $\sim 15 \sigma$ respectively (\RL). Consequently, only biases exceeding approximately $15\%$ (for SO) or $7\%$ (for CMB-S4) of the signal amplitude would meaningfully impact detection prospects at the $\agt1\sigma$ level. We should use these thresholds to evaluate the importance of any observed bias.

\subsection{Recovering rotation power}\label{sec:curl_only_test}
The rotation template was derived such that it returns the rotation auto-spectrum when crossed with fiducial $\omega$,  i.e. $\la C^{\omega\hat{\omega}^{\textrm{tem}}}\ra=C^{\omega\omega}$ [\RL]. To test this, we simulate a full-sky $\hat{\omega}^{\textrm{tem}}$ and contract it directly with the ray-traced realization of the DEMNUni field rotation, $\omega^{\textrm{sim}}$. 

Here we constructed the template for tracers $a\in\{\kappa^{\textrm{sim}},\tilde{g},\tilde{I}\}$. We take $\kappa^{\textrm{sim}}$ directly from the simulation to lower the variance. As the left leg of the cross-spectrum was also taken directly from the simulation, $\omega^{\textrm{sim}}$, there should therefore be no biases present on the $C^{\omega^{\textrm{sim}}\hat{\omega}^{\textrm{tem}}}$ measurement. While a reconstructed $\hat{\kappa}$ could be used without biasing the measurement, it would add unnecessary scatter to the results.

\begin{figure}[t]
 	\includegraphics[width=\linewidth]{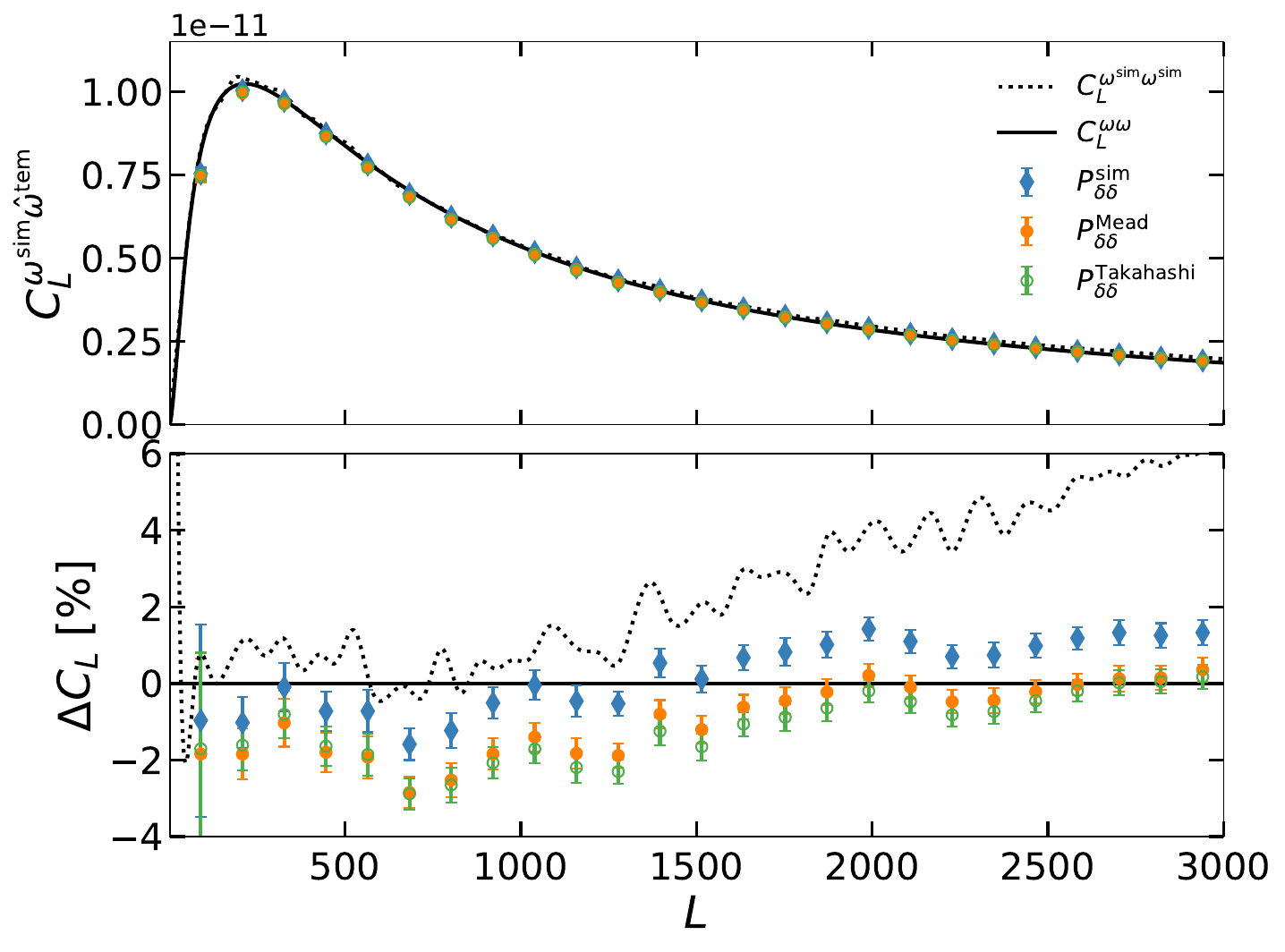}
	\caption{\small{Confirmation from the DEMNUni N-body simulation that a rotation template, $\hat{\omega}^{\textrm{tem}}$, constructed using Eq.~\eqref{eq:omega_tem_full}, can probe post-Born lensing rotation as a cross-spectrum signal. The blue diamonds show that $C^{\omega^{\textrm{sim}}\hat{\omega}^{\textrm{tem}}}$ is in agreement with theory (solid black) at the $\lesssim2\%$ level, as shown by the percentage error in the lower panel. The results do not follow the ray-traced auto-spectrum $C^{\omega^{\textrm{sim}}\omega^{\textrm{sim}}}$ (black dots), implying that the ray-tracing resolution may have been insufficient to resolve $\omega$ on small scales. Also displayed are cross-spectra with templates built using different non-linear prescriptions of $P_{\delta\delta}$ such as HMcode (Mead) \cite{Mead:2016zqy} (orange dots), and Takahashi \cite{Takahashi:2012em} (green circles). This illustrates that the cross-spectrum is not overly sensitive to mismodelling in the fiducial template weights.}}
	\label{fig:omega_only}
\end{figure}

Fig.~\ref{fig:omega_only} demonstrates that $C^{\omega^{\textrm{sim}}\hat{\omega}^{\textrm{tem}}}$ agrees with the fiducial $C^{\omega\omega}$ to $\lesssim2\%$ accuracy. This confirms that the cross-template method can probe post-Born field rotation with percent-level precision, sufficient for unbiased measurements at SO and CMB-S4 sensitivities. It also validates the full-sky template derivation in Sec.~\ref{sec:fullsky_tem}.
The cross-spectrum exhibits similar small-scale discrepancies with $C^{\omega^{\textrm{sim}}\omega^{\textrm{sim}}}$ as observed between $C^{\omega\omega}$ and $C^{\omega^{\textrm{sim}}\omega^{\textrm{sim}}}$ in Fig.~\ref{fig:delta_omega}. This supports the hypothesis that $\omega^{\textrm{sim}}$ may be affected by resolution and shot noise limitations at small scales. Consequently, the cross-spectrum result provides the best current validation of the analytic post-Born curl prediction (Eqs.~\ref{eq:omega_ps} and \ref{eq:omega_mc}).

We also tested the template's sensitivity to inaccurate theory modelling by varying the non-linear $P_{\delta\delta}$ prescriptions used in the template weights/normalization. The combined mismodelling of $C^{\omega\omega}$ (Fig.~\ref{fig:delta_omega}) and LSS $C_{\ell}$s (Fig.~\ref{fig:lss_diff}) results in a $\sim1\%$ offset when using either $P^{\textrm{Mead}}_{\delta\delta}$ or $P^{\textrm{Takahashi}}_{\delta\delta}$ compared to the true $P^{\textrm{sim}}_{\delta\delta}$ of the simulation (Fig.~\ref{fig:omega_only}). This indicates that at near-future sensitivities, the rotation cross-spectrum is not sensitive to small inaccuracies in the fiducial model used for the template. Further investigation of template mismodelling, with particular focus on magnification bias in $\hat{g}$, is explored in Appendix \ref{app:magbias}.

\subsection{Non-Gaussian biases}\label{sec:bias}
Gaussian contributions to the curl cross-spectrum biases were extensively investigated in \RL. The terms $N^{(0)}_{\hat{\kappa}\hat{\kappa}}$, $N^{(1)}_{\hat{\kappa}}$, $N^{(2)}_{\textrm{A}1}$, $N^{(2)}_{\textrm{C}1}$, and $N^{(2)}_{\hat{\kappa}}$ were found to adequately describe the bias in Gaussian null-hypothesis simulations. Analytic derivations were provided for each of these terms except $N^{(2)}_{\hat{\kappa}}$ which was shown to be of minimal importance at SO noise levels. We extend upon the analysis by exploiting the non-linear matter distribution of the DEMNUni simulation, and the multi-lens plane ray tracing, to probe additional biases arising from non-Gaussian sources. 

\subsubsection{Curl-free bias}\label{sec:phi_bias}
In this section, we work within the null hypothesis of zero $\omega$, but now specifically target non-Gaussian contributions to the bias. The non-Gaussian biases, $N^{\textrm{nG}}_L$, were extracted from the total measured bias (in the null hypothesis), $N^{\textrm{tem}}_L$, by subtracting off the measured Gaussian component, $N^{\textrm{G}}_L$, 
\begin{equation}
N^{\textrm{nG}}_L=N^{\textrm{tem}}_L-N^{\textrm{G}}_L.
\end{equation}
The $N^{\textrm{tem}}_L$ bias is simply the rotation-template cross-spectrum in the absence of $\omega$. Therefore, lensing observables $\hat{\omega}^{(\phi)}$ and $\hat{\kappa }^{(\phi)}$ were reconstructed from CMB maps that were lensed by the gradient-only deflection field, $\boldsymbol{\alpha}^{(\phi)}$. The template $\omega^{\textrm{tem},(\phi)}$ was constructed for $a\in\{\hat{\kappa}^{(\phi)},\tilde{g},\tilde{I}\}$, and the bias was then evaluated by averaging over different CMB realizations to reduce the variance
\begin{equation}\label{eq:N_L_bias}
N^{\textrm{tem}}_L=\langle C^{\hat{\omega}^{(\phi)}\hat{\omega}^{\textrm{tem},(\phi)}}_L\rangle_{\textrm{CMB}}.
\end{equation}
Similarly, Gaussian lensing observables $\hat{\omega}^{(\textrm{G})}$ and $\hat{\kappa}^{(\textrm{G})}$ were reconstructed from the CMB maps lensed by $\boldsymbol{\alpha}^{(\textrm{G})}$. We used Cholesky decomposition to generate Gaussian realizations $\tilde{g}^{\textrm{G}}$ and $\tilde{I}^{\textrm{G}}$ such that they correlate with $\kappa^{\textrm{G}}$, and each other, as expected from their empirical DEMNUni cross-spectra (Fig.~\ref{fig:lss_cls} and Fig.~\ref{fig:lss_diff}). The tracers $a\in\{\hat{\kappa}^{(\textrm{G})},\tilde{g}^{\textrm{G}},\tilde{I}^{\textrm{G}}\}$ were then used to construct the template $\hat{\omega}^{\textrm{tem,(G)}}$, and the Gaussian bias is then 
\begin{equation}
N^{\textrm{G}}_L=\langle C^{\hat{\omega}^{(\textrm{G})}\hat{\omega}^{\textrm{tem},(\textrm{G})}}_L\rangle_{\textrm{CMB}}.
\end{equation}

\begin{figure}[t]
 	\includegraphics[width=\linewidth]{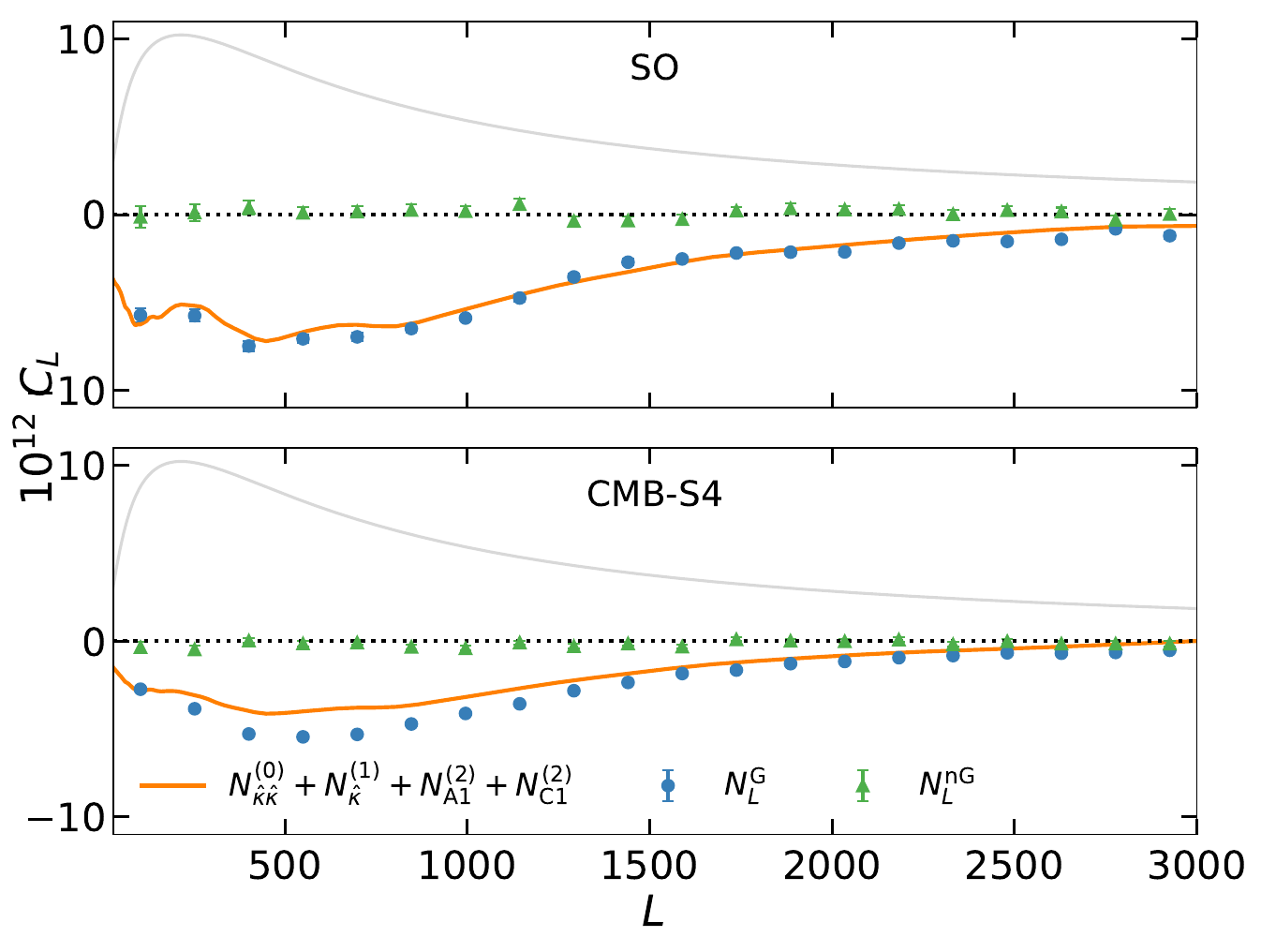}
	\caption{\small{The total Gaussian bias, $N^{\textrm{G}}$ (blue dots), to $C^{\hat{\omega}\hat{\omega}^{\textrm{tem}}}$, computed using 100 Gaussian realizations of DEMNUni tracer fields in the null hypothesis of zero $\omega$. The analytic predictions for the Gaussian bias terms $N^{(0)}_{\hat{\kappa}\hat{\kappa}}$, $N^{(1)}_{\hat{\kappa}}$, $N^{(2)}_{\textrm{A}1}$, and  $N^{(2)}_{\textrm{C}1}$ (solid orange) show that $N^{\textrm{G}}$ is consistent with the results of \RL (the large scale discrepancy for CMB-S4 is due to an $N^{(2)}_{\hat{\kappa}}$ bias term not included in plotted theory curve). The non-Gaussian bias, $N^{\textrm{nG}}$, (green triangles) defined as the residual of total and Gaussian bias, is consistent with a null signal for SO (top panel) and CMB-S4 (bottom panel) in the null hypothesis.}
}
	\label{fig:phi_bias}
\end{figure}

The Gaussian and non-Gaussian biases are presented in Fig.~\ref{fig:phi_bias} for 100 CMB realizations at SO and CMB-S4 noise levels. On the same figure, we plot the analytic Gaussian predictions for $N^{(0)}_{\hat{\kappa}\hat{\kappa}}$, $N^{(1)}_{\hat{\kappa}}$, $N^{(2)}_{\textrm{A}1}$, and  $N^{(2)}_{\textrm{C}1}$, computed such that they are compatible with the DEMNUni cosmology. The simulated Gaussian biases agree with the flat-sky results of \RL; the same measured discrepancy between simulated and predicted Gaussian bias for CMB-S4 is due to (as yet unaccounted for analytically) $N^{(2)}_{\hat{\kappa}}$ bias. The results show no significant non-Gaussian contributions to the bias at any scale. To within the accuracy of 100 simulations, the non-Gaussian bias is consistently near zero for SO and CMB-S4. Thus, consideration of $N^{\textrm{nG}}_L$ can be neglected for upcoming CMB experiments.

\subsubsection{Other biases}
\begin{figure}[t]
 	\includegraphics[width=\linewidth]{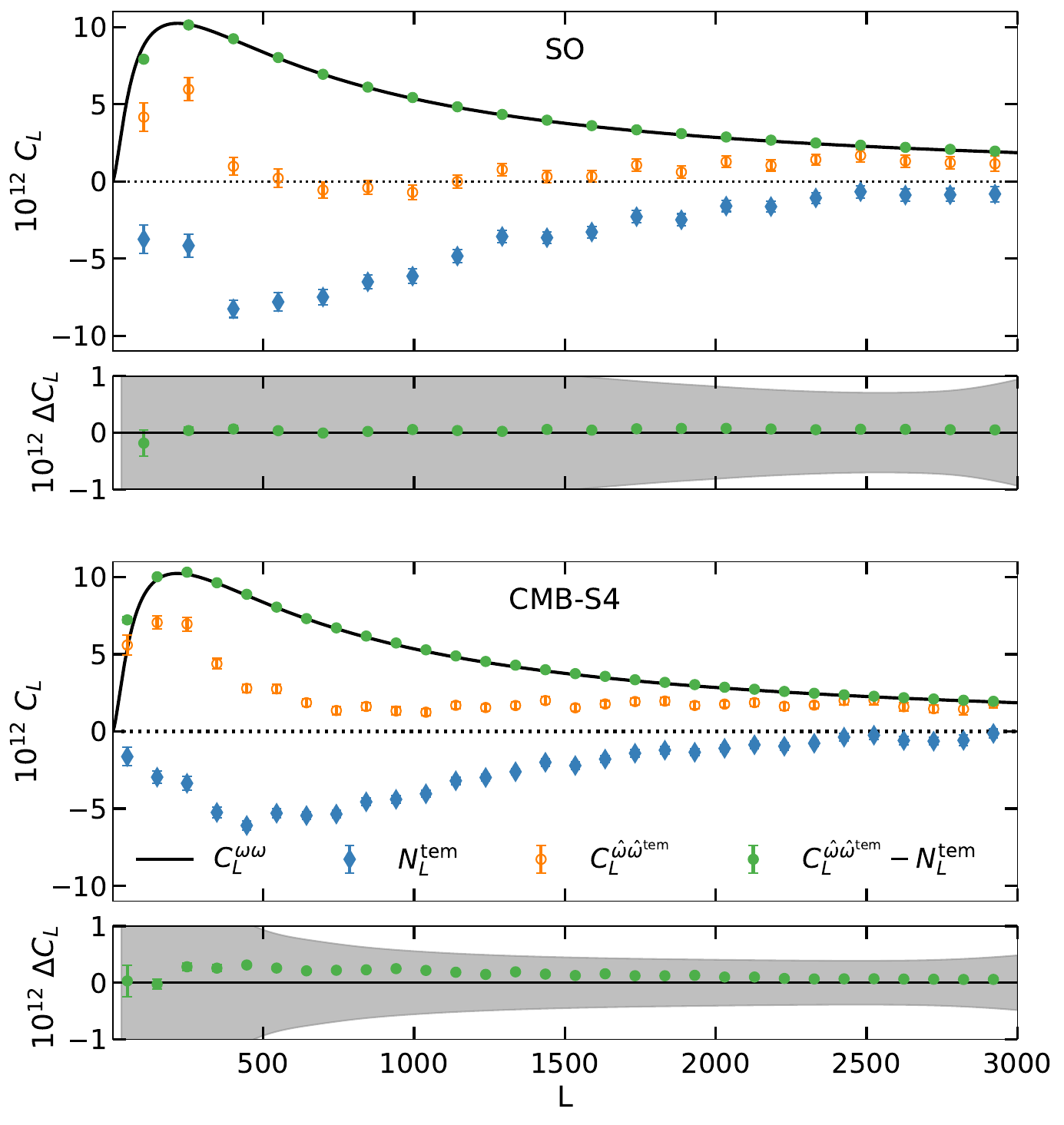}
	\caption{\small{The green dots represent the residual signal when subtracting the total null-hypothesis bias, $N^{\textrm{tem}}$, (blue diamonds) from the total rotation-template cross-spectrum, $C^{\hat{\omega}\hat{\omega}^{\textrm{tem}}}$ (orange circles). It agrees with the fiducial $C^{\omega\omega}$ (solid black) at $\lesssim3$\% for SO (top 2 panels) and $\lesssim5$\% for CMB-S4 (bottom 2 panels). The small deviations imply the existence of unaccounted curl biases. Nonetheless, the observed differences are small with respect to near future survey sensitivities. This is represented by the grey bands showing the predicted error assuming mode-accumulated variance (minimum possible error on a measurement at $L$ in which all modes $\leq L$ contribute to the measurement assuming a constant reconstruction noise defined at L (Eq.~\ref{eq:error})).}}
	\label{fig:sig_minus_bias}
\end{figure}

We now check for the existence of additional biases not picked up in the curl-free bias, $N^{\textrm{tem}}_L$, by relaxing the null-hypothesis requirement. The lensing observables $\hat{\omega}$ and $\hat{\kappa}$ were reconstructed from maps lensed by the full DEMNUni deflection field, $\boldsymbol{\alpha}^{\textrm{sim}}$, and the template was constructed from tracers $a\in\{\hat{\kappa},\tilde{g},\tilde{I}\}$. The total cross-spectrum, $\langle C^{\hat{\omega}\hat{\omega}^{\textrm{tem}}}\rangle_{\textrm{CMB}}$, is shown in Fig.~\ref{fig:sig_minus_bias} along with curl-free $N^{\textrm{tem}}_L$ (Eq.~\ref{eq:N_L_bias}) for 20 CMB realizations. We find that subtracting $N^{\textrm{tem}}_L$ from the total cross-spectrum does recover $C^{\omega\omega}$ well, suggesting that $N^{\textrm{tem}}_L$ describes most of the bias. However, variance cancellation in the subtraction significantly reduced the error-bars such that offsets are observed at the positive $\lesssim3$\% and $\lesssim5$\% level for SO and CMB-S4 (larger than the offsets in Fig.~\ref{fig:omega_only}). Such deviations are expected if the curl-free $N^{\textrm{tem}}_L$ does not contain all bias contributions. This implies additional biases are present from non-Gaussian contractions involving $\Omega$. Although the size of the discrepancies are insignificant at SO noise levels, for CMB-S4 this extra bias could lead to a spurious near-1$\sigma$ enhancement of rotation signals. To illustrate this, the projected error on the cross-spectrum is shown in the lower panels of Fig.~\ref{fig:sig_minus_bias} compared to the mode-accumulated variance\footnote{The variance of the cross-spectrum at $L$ on the full-sky is approximately $C^{\omega\omega}_LN^{\omega}_0(L)F^{-1}_L/(2L+1)$ assuming $C^{\omega\omega}\ll N^{\omega}_0(L)$. We define the mode-accumulated variance as the error on a binned measurement at $L$, in which all modes $\leq L$ are included in the bin. We also approximate the lensing reconstruction noise, and template noise, for all points in the bin to be constant and defined at $L$. Hence, Eq.~\eqref{eq:error} has an extra factor of $L$ in the denominator, representing the extreme example of best possible measurement at a given $L$.}

\begin{equation}\label{eq:error}
\sigma^2_L=\frac{C^{\omega\omega}_LN^{\omega}_0(L)F^{-1}_L}{L(2L+1)f_{\textrm{sky}}}.
\end{equation}
The bias does not exist in the null hypothesis, so this does not actually inhibit detection (rather the opposite, as it is an extra rotation signal that increases the amplitude of the rotation cross-spectrum). It would however be problematic for cosmological inference. One way to reduce the bias, and weak lensing biases generally, is to use the more optimal MAP lensing estimators.

\subsection{MAP bias}\label{sec:it_bias}
The delensing aspect of the iterative process in MAP estimators is particularly effective for reducing lensing biases. There are two related effects that lead to bias suppression. The first is simply that delensing removes the lensing signal from the maps, therefore lensing-induced biases (e.g. standard $N_{1}$, $N_{3/2}$, etc.) now come from just the residual lensing 
\begin{equation}\label{eq:delen_kappa}
\kappa^{\textrm{delen}}(\boldsymbol{L})\sim\sqrt{\left(1-\rho^2_L\right)} \kappa(\boldsymbol{L}),
\end{equation}
which is approximated using the delensing efficiency
\begin{equation}\label{eq:delen_eff}
    \rho^2_L=\frac{C^{\kappa\kappa}_L}{C^{\kappa\kappa}_L+N^{\kappa,\textrm{MAP}}_0(L)}.
\end{equation}
The delensed curl signal is not considered here because possible curl-induced biases are small (e.g. see Fig.~\ref{fig:sig_minus_bias}), and curl reconstruction is poor for near-future sensitivities.

The second effect is that the lensing reconstruction is computed using a partially-delensed sky, so contractions between CMB fields produce partially-delensed spectra, $C^{\textrm{delen}}_{\mathcal{\ell}}$, which have reduced variance compared to their fully lensed counterparts, $\tilde{C}_{\mathcal{\ell}}$. This leads to further modification of the biases, including suppression of $N_{0}$. We compute the $C^{\textrm{delen}}_{\mathcal{\ell}}$s starting from their unlensed spectra, $C_{\mathcal{\ell}}$, and assuming lensing by the residual lensing signal, Eq.~\eqref{eq:delen_kappa}. We then use {\texttt{CAMB}}\footnote{\href{https://camb.readthedocs.io/}{https://camb.readthedocs.io/}} to output the partially-delensed versions (using the full-sky non-perturbative lensing correlation functions~\cite{Lewis:2006fu}).

\subsubsection{Bias predictions}

Here we are interested in how MAP lensing measurements, $\hat{\omega}^{\textrm{MAP}}$ and $\hat{\kappa}^{\textrm{MAP}}$, affect curl cross-spectrum biases. To approximate the leading bias contributions analytically, we modify the QE results for $N^{(0)}_{\hat{\kappa}\hat{\kappa}}$, $N^{(1)}_{\hat{\kappa}}$, $N^{(2)}_{A1}$, and $N^{(2)}_{C1}$. We do not review the QE results here, instead we direct the reader to \RL where they were first derived. The terms were altered as follows
\begin{itemize}
\item{The analytic QE normalization ($A_L$), template normalization ($F_L$), and LSS covariance matrix ($\boldsymbol{C}_{\textrm{LSS}}$), were recomputed to account for the lowered lensing reconstruction noise. This is a simple replacement of $N_0^{\textrm{QE}}\rightarrow N_0^{\textrm{MAP}}$.}
\item{Both lensed and grad-lensed spectra\footnote{Lensed gradient CMB spectra \cite{Lewis:2011fk} enter the response in QEs. We assume the difference between partially delensed gradient spectra and $C^{\textrm{delen}}_{\mathcal{\ell}}$ is inconsequential to the bias predictions.} were replaced by $C^{\textrm{delen}}_{\mathcal{\ell}}$. This includes all spectra in the QE weights and response functions.}
\item{For $N^{(1),\textrm{MAP}}_{\hat{\kappa}}$, $N^{(2),\textrm{MAP}}_{A1}$, and $N^{(2),\textrm{MAP}}_{C1}$, we also account for the residual lensing signal using Eqs.~\eqref{eq:delen_kappa} and \eqref{eq:delen_eff}. We therefore modify the lensing cross-spectra within the LSS mode-coupling functions (equations 6.6 and 6.10 in \RL) via 
\begin{equation}
\qquad C^{\kappa a}_L\rightarrow\sqrt{\left(1-\rho^2_{L}\right)}C^{\kappa a}_L.
\end{equation}
This does not affect spectra modelled within the LSS covariance, $\boldsymbol{C}_{\textrm{LSS}}$, or fiducial bispectra, $b^{\omega ij}$. 
}
\end{itemize}
We note that a similar procedure for estimating $N^{\kappa,\textrm{MAP}}_0$ and $N^{\kappa,\textrm{MAP}}_1$ worked remarkably well when validated against polarization-only simulations in Ref.~\cite{Legrand:2021qdu}.

\begin{figure*}[t]
 	\includegraphics[width=\linewidth]{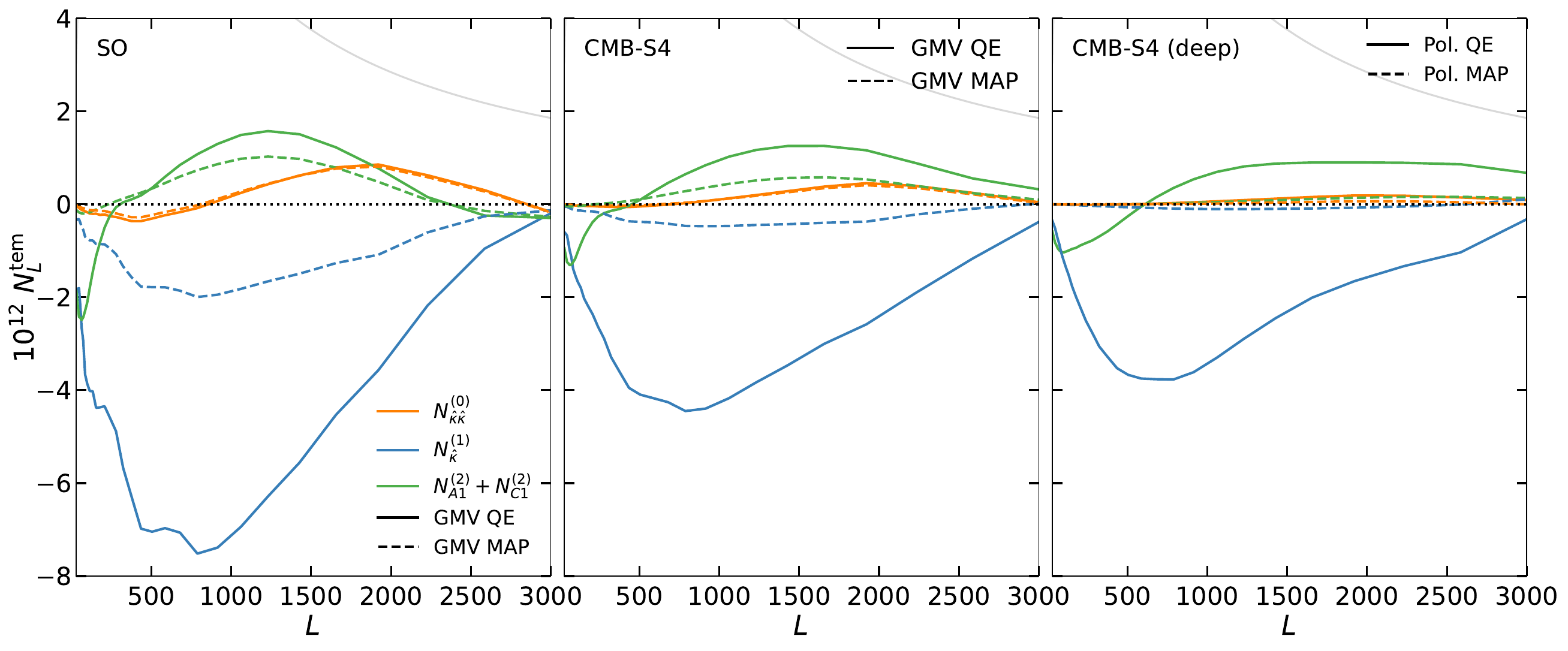}
	\caption{\small{The analytic approximation of individual MAP bias terms are shown by the dashed lines in the three panels for SO, CMB-S4, and CMB-S4 deep experimental configurations. The equivalent quadratic estimator biases, from which the MAP results are derived, are also shown in the solid lines. For context, the fiducial $C^{\omega\omega}_L$ power spectrum (solid grey) and null line (dotted black) are included in the plots.}}
	\label{fig:map_bias_analytic}
\end{figure*}

The resulting predictions for the leading curl cross-spectrum biases are displayed in Fig.~\ref{fig:map_bias_analytic} for SO and CMB-S4 with GMV estimators. The absolute size of each MAP bias is reduced on all scales compared to the equivalent QE case. Significant suppression is observed for $N^{(1),\textrm{MAP}}_{\hat{\kappa}}$, $N^{(2),\textrm{MAP}}_{A1}$, and $N^{(2),\textrm{MAP}}_{C1}$, a direct consequence of their dependence on $\phi$ which is iteratively removed during lensing reconstruction. The extent of the expected suppression is impressive considering that MAP and QE lensing reconstruction perform similarly well at these noise levels. Iterative reconstruction becomes more optimal for lower noise configurations such as a CMB-S4 deep setup, which is also included in Fig.~\ref{fig:map_bias_analytic} for polarization-only reconstruction.  Despite the similarities between the deep and wide CMB-S4 QE biases, the MAP versions are shown to be significantly more suppressed in the deep patch case. There is even a noticeable reduction in $N^{(0),\textrm{MAP}}_{\hat{\kappa}\hat{\kappa}}$ for CMB-S4 deep. The sign differences also lead to near-zero total bias for polarization MAP biases on the curl cross-spectrum.

\subsubsection{MAP results}

\begin{figure}[t]
 	\includegraphics[width=\linewidth]{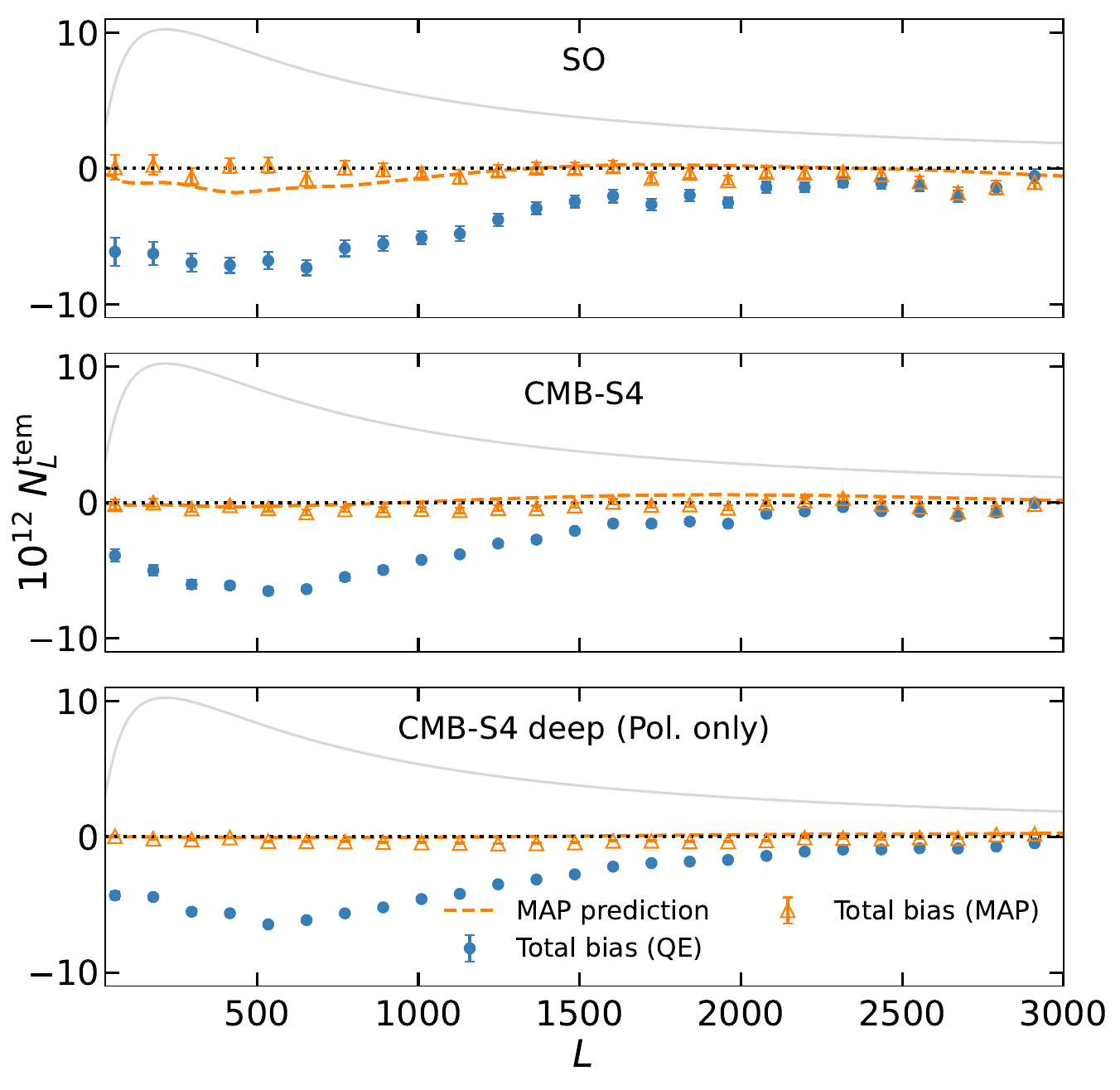}
	\caption{\small{The total bias, $N^{\textrm{tem}}$, to the curl cross-spectrum $C^{\hat{\omega}\hat{\omega}^{\textrm{tem}}}$ using MAP lensing estimators (orange triangles), compared to the usual bias with QEs (blue dots). Analytic approximations for the MAP biases (dashed orange lines) qualitatively agree with the amplitude of the bias. The fiducial rotation curve (solid grey) also shown for reference.}
}
	\label{fig:map_bias}
\end{figure}

We now test the analytic MAP predictions using the DEMNUni simulation. Curl-free CMB simulations were created through lensing with the gradient-only deflected field $\boldsymbol{\alpha}^{(\phi)}$, and the lensing observables $\hat{\omega}^{(\phi),\textrm{MAP}}$ and $\hat{\kappa}^{(\phi),\textrm{MAP}}$ were reconstructed following the procedure described in Section~\ref{sec:map}. The total bias, $N_L^{\textrm{MAP}}$, in the absence of $\omega$, is then simply

\begin{equation}
N^{\textrm{tem, MAP}}_L=\langle C^{\hat{\omega}^{(\phi),\textrm{MAP}}\hat{\omega}^{\textrm{tem},(\phi),\textrm{MAP}}}_L\rangle_{\textrm{CMB}},
\end{equation}
in which the template is computed for tracers $a\in\{\hat{\kappa}^{(\phi),\textrm{MAP}},\tilde{g},\tilde{I}\}$. For 20 CMB realizations, Fig.~\ref{fig:map_bias} confirms the bias suppression effect for SO, CMB-S4, and the CMB-S4 deep patch polarization case. The amplitude of the total rotation-template cross bias is generally reduced over all scales in each experimental configuration considered. Significant suppression is apparent on large scales, where lensing reconstruction noise is lowest. There is modest agreement between analytic predictions with the simulated results, and the observed differences are small compared to $C^{\omega\omega}$ due to the significant suppression in bias amplitude. It is unlikely that any deviations could stem from higher order bias terms, as the suppression effect scales $\propto\mathcal{O}(\phi)$. Nonetheless, the varying sign dependence of different biases could cause cancellation effects to amplify expected errors in the analytic approximations relative to the reduced bias. 

\section{Foreground biases}\label{sec:fg}

Another type of bias neglected so far is that from contamination in the observed CMB due to intermediate astrophysical processes. Foreground biases are important for CMB lensing measurements \cite{Carron:2022eyg, ACT:2023dou, SPT:2023jql}, and mitigation techniques include masking, using frequency dependencies to isolate the CMB signal  \cite{Eriksen:2004jg}, de-projecting specific foregrounds \cite{Remazeilles:2010hq}, and constructing new lensing estimators \cite{Sailer:2020lal, Schaan:2018tup}. There are two distinct types of foreground contamination: galactic and extragalactic. Galactic processes contribute to observations through local dust \cite{Planck:2014dmk} and synchrotron emission \cite{Planck:2015mvg}, which contaminate both temperature and polarization maps on large scales \cite{Kandel:2017xjx, Abril-Cabezas:2023ftf}. This can bias the CMB lensing auto-spectrum \cite{ACT:2023dou} (including curl lensing). Similarly, galactic contamination of galaxy surveys, or the CIB, can also bias cross-spectra with CMB lensing \cite{ACT:2023oei}. Nonetheless, masking of the galactic plane and/or enforcing large scale cuts can alleviate much of these biases, and testing of residual sensitivity is easily achieved by varying masking area/multipole scales used. Lensing reconstruction is also largely insensitive to the very large-scale CMB contamination, so we assume galactic foregrounds can be controlled, and leave detailed investigation of possible galactic curl cross-spectrum biases to future work. 

However, there are several extragalactic processes that leave a significant imprint on CMB observations, including millimetre emission from the CIB and radio galaxies that appear as point-like signals in temperature. Polarized emission from radio galaxies also contribute to E- and B-mode maps. Additionally, CMB photons can scatter off ionized dust surrounding galaxy clusters resulting in the well-known thermal and kinetic Sunyaev-Zel’dovich effects (tSZ and kSZ). The non-Gaussianity of each of these extragalactic processes can bias CMB lensing measurements \cite{Osborne:2013nna, vanEngelen:2013rla}. Gaussian biases could also be problematic for the curl-template cross-spectrum, potentially arising from correlations between template tracers and foregrounds in the CMB maps. We now investigate the importance of extragalactic foregrounds for near-future detection prospects of $\omega$.

\subsection{Extragalactic foreground simulation}

To study extragalactic foreground biases to the CMB curl cross-spectrum, we take advantage of the ready-made products from the AGORA simulation\footnote{\href{https://yomori.github.io/agora/index.html}{https://yomori.github.io/agora/index.html} \cite{Omori:2022uox}.}. AGORA implements a light cone on the MultiDark Planck 2 (MDPL2) N-body simulation\footnote{\href{https://www.cosmosim.org/metadata/mdpl2/}{https://www.cosmosim.org/metadata/mdpl2/}} based on a Planck cosmology\footnote{AGORA cosmology: $\Omega_m=0.307$, $\Omega_b=0.048$, $\Omega_{\Lambda}=0.693$, $n_s=0.96$, reduced Hubble parameter $h=0.6777$, and $\sigma_8=0.8288$.}, and uses multi-plane ray-tracing to extract the lensing potential (the curl potential is not given as a simulation product). This was used to create the CMB lensing convergence map, $\kappa^{\textrm{sim}}$, that includes Born and post-Born corrections. From the N-body dark matter particles and halo catalogue, maps of galaxy density, the CIB, thermal and kinetic Sunyaev-Zel’dovich effects, and radio galaxies are produced. Each map was lensed as appropriate for cross-spectra analyses. The dark matter particles were also used in conjuncture with the ray-traced output to create cosmic shear catalogues, though we do not use those here.  

Extragalactic foregrounds vary in frequency, $\nu$, so we need to be careful over choice of CMB channels for the analysis. We choose $95$, $150$, and $220$ GHz frequencies for consistency with the available AGORA products, and similarity to the expected SO channels ($93$, $145$, and $225$ \cite{Ade:2018sbj}). The CIB and radio maps were already provided at the desired frequencies, we just convert the units from Jansky per steradian to $\mu$K by applying the factor \cite{Planck:2013wmz}
\begin{equation}\label{eq:jy_to_K}
10^{-20}c^2\left(e^x-1\right),
\end{equation}
where $x=h_p\nu_c/k_{B}T_{\textrm{CMB}}$ is the Planck factor for Planck constant $h_p$, Boltzmann constant, $k_B$, speed of light, $c$, and CMB temperature $T_{\textrm{CMB}}=2.7255 {\rm K}$ \cite{Fixsen:2009ug}. We have neglected passband contributions, equivalent to approximating the spectral transmission function as a Dirac delta $\tau(\nu)=\delta(\nu-\nu_c)$ such that only the frequency of the channel, $\nu_c$, is picked up. We note that the AGORA CIB maps have had colour corrections applied \cite{Planck:2013wmz} assuming an ACT-like bandpass \cite{Omori:2022uox}. For tSZ, we take the supplied Compton-y map and convert it into its respective frequency channels (whilst also converting to $\mu$K) by multiplying the map by factor \cite{Planck:2013wmz} 
\begin{equation}
T_{\textrm{CMB}}\left[x\frac{e^x+1}{e^x-1}-4\right].
\end{equation}
Again, we assume no passband. Finally, the kSZ signal is independent of frequency and the simulated kSZ product was provided in units of $\mu$K, hence we apply no corrections to the map.

For each frequency channel, we model the map-level noise as isotropic with spectrum given by \cite{Ade:2018sbj} 

\begin{figure*}[t]
 	\includegraphics[width=0.8\linewidth]{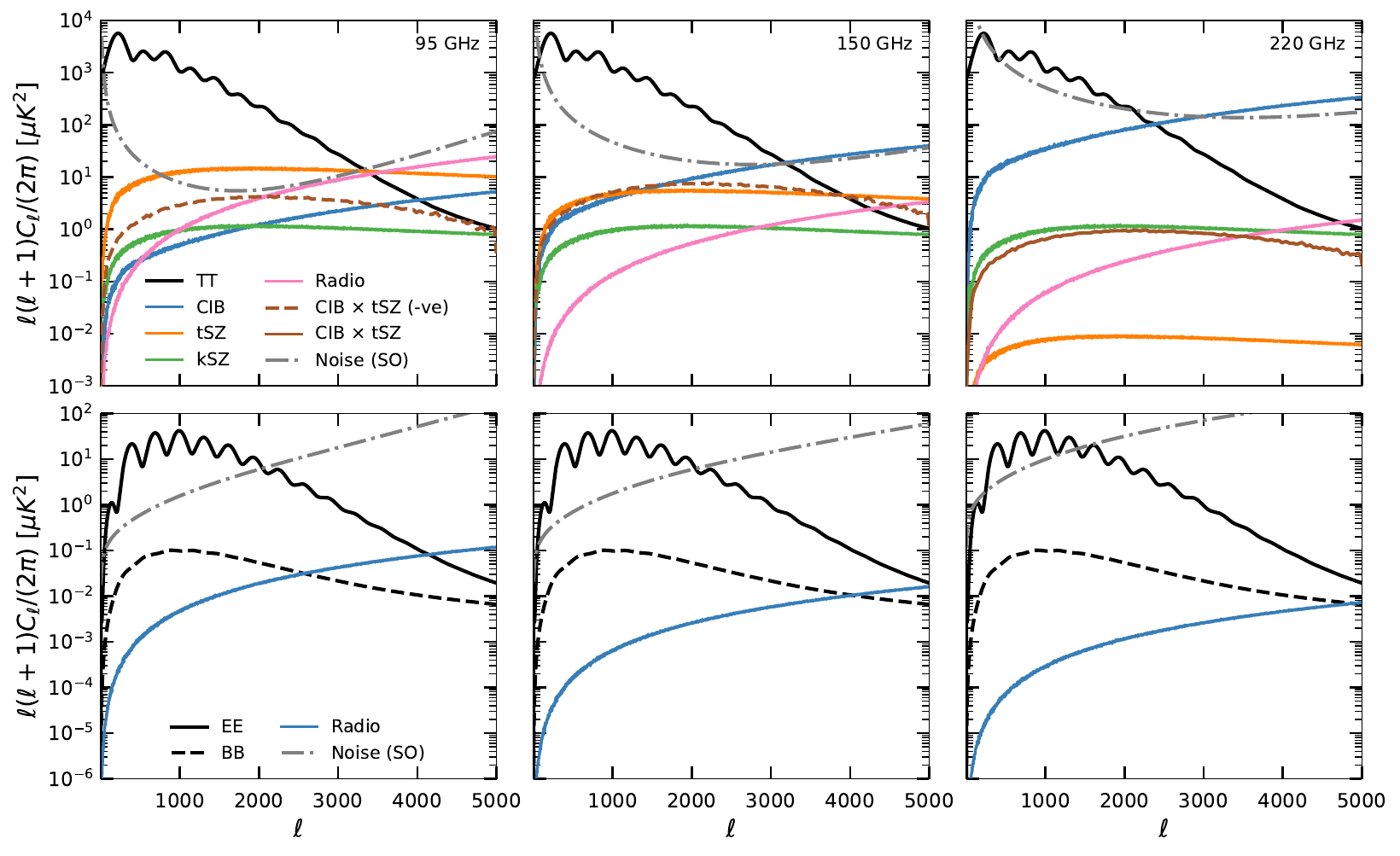}
	\caption{\small{The constituent power spectra of various signals used in the analysis of foreground bias for an SO-like experiment. Each column represents a different frequency: 95 GHz, 150 GHz, and 220 GHz (left to right). The top row shows temperature (T) components, the bottom polarization (E and B). The fiducial CMB spectra are shown in solid black for T and E, and dashed black for B. These are compared to the foreground spectra (solid colours) in each channel. The negative cross-correlation between CIB and tSZ foregrounds is also shown in the dashed brown line. For E and B polarization measurements, only radio sources were considered as contaminants. The SO map-level noise (dash-dot grey) is also illustrated for each case.}
}
	\label{fig:fg_cls}
\end{figure*}

\begin{equation}\label{eq:freq_noise}
N_{\ell}=\left[n_{\textrm{red}}^2\left(\frac{\ell_{\textrm{knee}}}{\ell}\right)^{\alpha_{\textrm{knee}}}+n_{\textrm{white}}^2\right]\exp{\left(\ell(\ell+1)\frac{\theta^2}{8\ln{2}}\right)},
\end{equation}
which accounts for $1/f$ noise due to atmospheric and electronic effects through the knee components, white noise, and $\theta$ is the full width at half maximum of the beam. We take the noise at SO goal level\footnote{ $n_{\textrm{red}}$ is 23, 58, and 194 $\mu$K-arcmin (calculated using efficiency, observing time and $f_{\rm sky}$ assumptions in Ref. \cite{Ade:2018sbj}, and values from Table 2. in Ref. \cite{Ade:2018sbj}) for 95, 150 and 220 GHz maps. $\ell_{\textrm{knee}}=1000$ for T and 700 for P, $\alpha_{\textrm{knee}}$ = 3.5 for T and 1.4 for P. $n_{\textrm{white}}$ and $\theta$ values given in Table 1. of Ref. \cite{Ade:2018sbj}.}. This is the same model that was used in Section~\ref{sec:results} for SO, except now we do not include foreground contributions within $N_{\ell}$ and use only 3 frequency bands instead of all 6. We do not consider CMB-S4-like noise levels when investigating extragalactic foreground biases. In principle, CMB-S4 can detect the curl at high fidelity using polarization-only data which is less susceptible to extragalactic contamination. SO will be more dependent on temperature data to make a detection, thus investigation of foregrounds for SO-like experimental configurations is better motivated.

The auto-spectra of the simulated foregrounds and noise are shown in Fig.~\ref{fig:fg_cls} at the considered frequencies. The tSZ spectra dominate at lower $\nu$ while the CIB becomes large at higher $\nu$. The tSZ cross CIB correlation is also plotted and has significant negative amplitude in the lower frequency channels. The radio signal would dominate over all frequencies if threshold cuts were not applied. SO is forecast to detect 15,000 radio point sources from the 95 GHz channel alone \cite{Ade:2018sbj} at goal sensitivity. We find this corresponds to a 10 mJy flux cut in AGORA. Therefore, we select pixels $>10$ mJy in the 95 GHz radio temperature map, and replace those pixels in each of the 95, 150, and 220 GHz temperature and polarization maps by the median pixel value of the respective map. This produces temperature spectra an order of magnitude larger than in Ref.~\cite{SPT:2020psp} who apply 6 mJy cuts. Similarly, the resulting polarization spectra are an order of magnitude larger than their equivalent in Ref.~\cite{Lagache:2019xto}. Our cuts are therefore very conservative, and though the amplitude of the radio signal is still important for $T$ it is negligible in polarization. No point source or cluster cuts were applied to the CIB or SZ maps, initially.

For the template input density tracers, we again use $\tilde{g}$, $\tilde{I}$, and $\hat{\kappa}$. AGORA provides a 5-binned suite of LSST-like galaxy density maps which were computed directly from the density sheets of the light-cone in a similar procedure to Eq.~\eqref{eq:make_tracer} assuming a linear galaxy bias. We convert the maps into a 1-bin version through re-normalizing them using their binned redshift distributions, $(dn/dz)^{i}$, (which were provided as an AGORA data product)

\begin{equation}
\tilde{g}(\boldsymbol{\hat{n}})=\frac{\sum_i^5\tilde{g}_i(\boldsymbol{\hat{n}})\int dz \left(dn/dz\right)^i}{\sum_i^5\int dz \left(dn/dz\right)^i}.
\end{equation}
For $\tilde{I}$, we use the provided AGORA map at $\nu=353$ GHz, which includes colour corrections assuming a Planck-like bandpass, although we still convert the map into units of $\mu$K assuming no bandpass with Eq.~\eqref{eq:jy_to_K}. For context, the redshift distributions of $\tilde{g}$ and $\tilde{I}$ are presented in figures 26 and 19 of Ref.~\cite{Omori:2022uox}. For the CMB lensing observables, reconstruction follows a similar procedure to Section~\ref{sec:cmb_lens}; the lensing deflection field $\boldsymbol{\alpha}^{(\phi)}$ is calculated from the AGORA ray-traced CMB convergence, $\kappa^{\textrm{sim}}$, only.

\subsubsection{HILC maps}
Foreground biases to the rotation cross-spectrum from individual-$\nu$ CMB maps could potentially be large and complicated. For example, the CIB has strong correlation with $\kappa$ and $g$, hence justifying its inclusion in rotation templates. However, as seen in Fig.~\ref{fig:fg_cls}, the CIB is also a dominant contaminant in the CMB at high $\nu$, and correlated with the tSZ at low $\nu$. Therefore, just from consideration of the CIB alone, a slew of Gaussian and non-Gaussian contractions involving the CIB foreground with template tracers, or other foregrounds, or $\phi$, could induce $\nu$-dependent biases on the rotation-template cross-spectrum. Similar arguments could be applied to the sensitivity of tSZ and kSZ to the underlying density distribution, or to their foreground cross-terms. Thus, to simplify the analysis we instead investigate biases of {\it residual} foregrounds from a single foreground cleaned CMB map.

\begin{figure}[t]
 	\includegraphics[width=\linewidth]{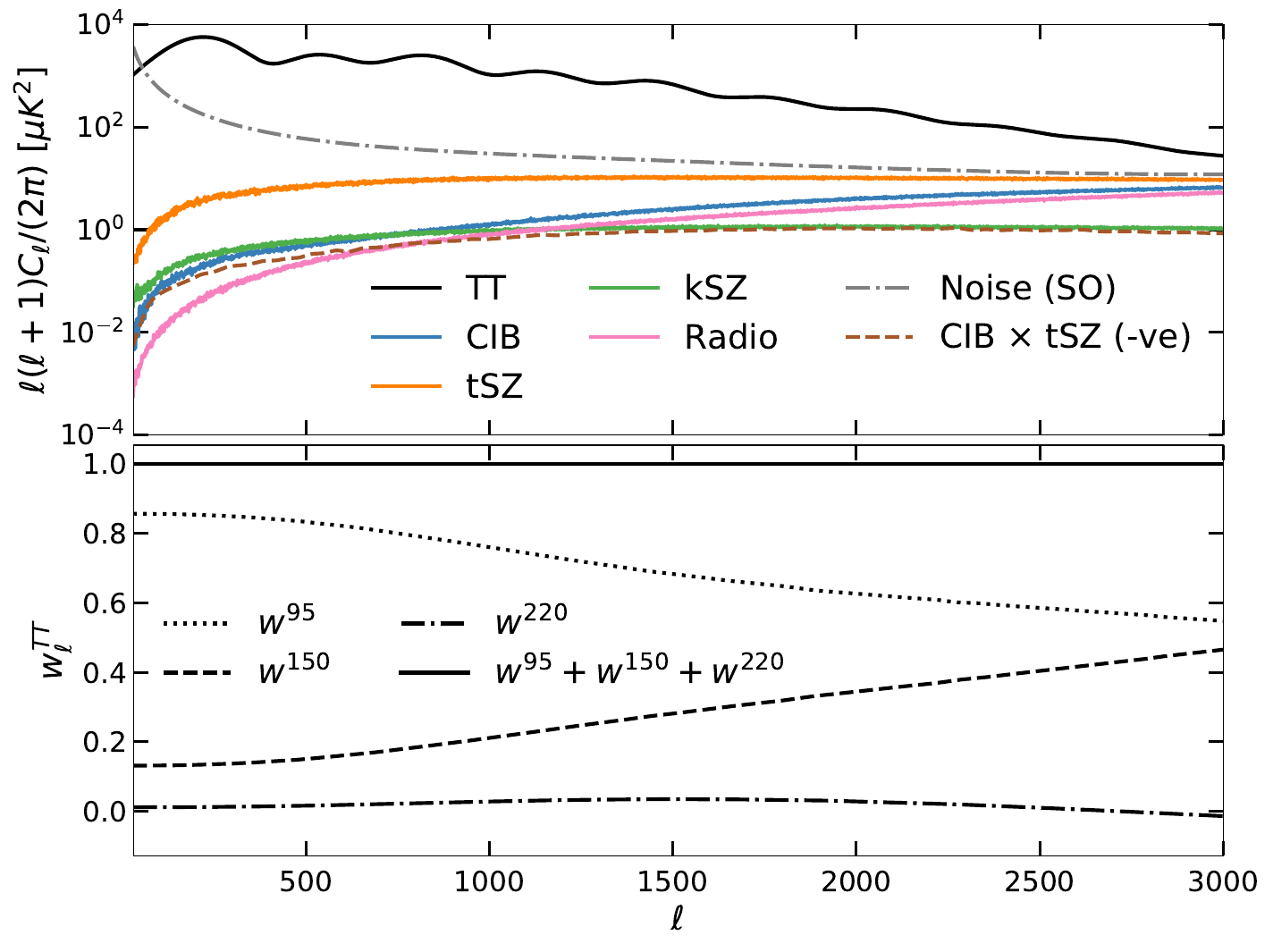}
	\caption{\small{Top panel: power spectra of the constituent components in the HILC temperature map. The CMB (solid black) is signal dominated on the scales considered, compared to both SO-like noise (dashed grey) and foregrounds (solid colours). The negative cross-correlation between CIB and tSZ foregrounds is also shown (dashed brown line). Bottom panel: the weights used to construct the HILC temperature map.}}
	\label{fig:hilc_w}
\end{figure}

We choose the simplest foreground de-projection technique: the Harmonic Internal Linear Combination (HILC) method \cite{WMAP:2003ivt,Eriksen:2004jg}. This takes individual frequency maps, $X^{\nu}=X+\textrm{fg}^{\nu}+n^{\nu}$, and outputs a single minimum variance CMB map, $X^{\textrm{HILC}}$. Frequency varying signals in the data are separated out through weighted linear combinations of maps in harmonic space

\begin{equation}
X^{\textrm{HILC}}_{\ell m}=\delta_{ij}w^{\nu_{i}}_{\ell}X^{\nu_{j}}_{\ell m}.
\end{equation}
The weights are defined through minimizing the $\nu$-dependent contributions

\begin{equation}
\boldsymbol{w}_{\ell}=\frac{\boldsymbol{C}^{-1}_{\ell}\boldsymbol{\mathbbm{1}}}{\boldsymbol{\mathbbm{1}}^{\textrm{T}}\boldsymbol{C}^{-1}_{\ell}\boldsymbol{\mathbbm{1}}},
\end{equation} 
where the covariance matrix between frequencies includes both foregrounds and noise, $C_{\ell}^{\nu_i\nu_j}=(C_{\ell}^{\rm fg})^{\nu_i\nu_j}+\delta_{ij}N_{\ell}^{\nu_i\nu_j}$, and $\boldsymbol{\mathbbm{1}}$ is a vector of ones. The empirical auto- and cross-spectra from AGORA, Fig.~\ref{fig:fg_cls}, were used for the foreground components while the analytic Eq.~\eqref{eq:freq_noise} was used for the noise curves. We generated one set of maps in which the CMB is unlensed, $X^{\textrm{HILC}}_{\ell m}$. In addition, a second set of maps were built for the lensed CMB case, $\tilde{X}^{\textrm{HILC}}_{\ell m}$, in which $\tilde{X}$ has been deflected by the gradient-only lensing field, $\boldsymbol{\alpha}^{(\phi)}$. The weights, shown for temperature in the bottom panel of Fig.~\ref{fig:hilc_w}, were used for both unlensed and lensed sets. The temperature map is dominated by contributions from the 95 and 150 GHz maps, with little information from 220 GHz due to high noise and CIB contamination. Therefore, it is the tSZ effect that is the dominant residual foreground for temperature, although it is still subdominant to noise, as illustrated in the top panel of Fig.~\ref{fig:hilc_w}. The kSZ foreground is unaffected by the HILC process due to its frequency independence. We verify that the cross-spectrum between $X^{\textrm{HILC}}_{\ell m}$ and $X_{\ell m}$ is unbiased.

\subsection{Curl-free foreground bias}\label{sec:fg_bias}
\begin{figure}[t]
 	\includegraphics[width=\linewidth]{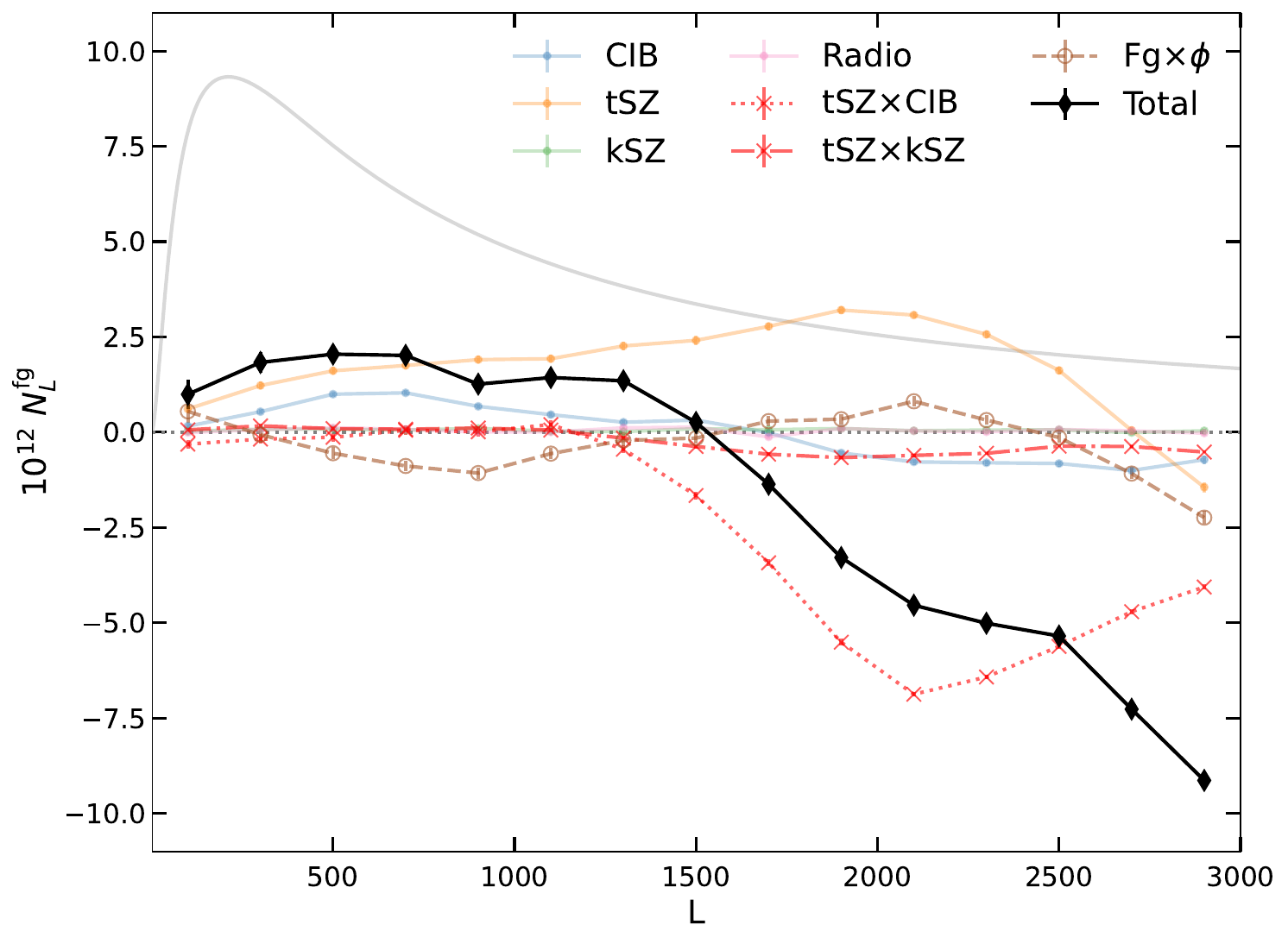}
	\caption{\small{The total foreground bias, $N^{\textrm{fg}}_L$ (black diamonds), computed on simulated SO HILC maps, compared to fiducial $C^{\omega\omega}_L$ (solid grey) and null lines (dotted grey). From individual contributions to the bias, it is tSZ (orange dots) and CIB (blue dots) that are most significant. Despite radio sources (pink dots) also contaminating polarization, the bias is negligible as for kSZ (green dots). Pure foreground cross terms (red crosses) involving tSZ  significantly affect the overall amplitude and shape of the bias. A foreground cross $\phi$ term (brown circles) is also important.}}
	\label{fig:fg}
\end{figure}

We follow a similar procedure from Section \ref{sec:bias} to investigate the effect of extragalactic foregrounds on the rotation template cross-spectrum. Working in the null hypothesis of zero curl, the  total residual bias is defined as
\begin{equation}\label{eq:fg_bias}
N^{\textrm{fg, total}}_L=\langle C^{\hat{\omega}^{(\phi, \textrm{fg})}\hat{\omega}^{\textrm{tem},(\phi, \textrm{fg})}}_L\rangle_{\textrm{CMB}}-N^{\textrm{tem}}_L,
\end{equation}
where $N^{\textrm{tem}}_L$ is the non-foreground bias investigated in Section \ref{sec:bias}. Lensing reconstructions, $\hat{\omega}^{(\phi, \textrm{fg})}$ and $\hat{\kappa}^{(\phi, \textrm{fg})}$ were measured from $\tilde{X}^{\textrm{HILC, fg}}$ which includes residual foreground components. Although noise was included in calculation of the HILC weights, it was not added to the individual CMB maps. This lowers the variance of the total cross-spectrum signal and also reduces the amplitude of $N^{\textrm{tem}}_L$ by removing noise contractions in $N^{(0)}_{\hat{\kappa}\hat{\kappa}}$, and $N^{(1)}_{\hat{\kappa}}$. The template was constructed for $a\in\{\hat{\kappa}^{(\phi)},\tilde{g},\tilde{I}\}$ in which $\tilde{g}$ and $\tilde{I}$ are noise free. To compute $N^{\textrm{tem}}_L$, the $\tilde{X}^{\textrm{HILC}}$ maps were constructed using the same CMB fields and weights, but without including the foregrounds
\begin{equation}\label{eq:nonfg_bias}
N^{\textrm{tem}}_L=\langle C^{\hat{\omega}^{(\phi)}\hat{\omega}^{\textrm{tem},(\phi)}}_L\rangle_{\textrm{CMB}}.
\end{equation}
The residual bias is shown in Fig.~\ref{fig:fg} for an SO-like experiment including tSZ, kSZ, CIB, and radio sources. Its amplitude is of comparable size to the fiducial signal but with sign changing shape. The individual contributions to the bias from respective foregrounds come primarily from tSZ and CIB sources. Although radio sources also contaminate polarization observations, we find it does not induce significant bias to the curl cross-spectrum. These individual foreground biases were approximated using modified versions of Eqs.~\eqref{eq:fg_bias} and \eqref{eq:nonfg_bias} 
\begin{equation}\label{eq:fg1_bias}
N^{\textrm{fg}^i}_L=\langle C^{\hat{\omega}^{(\textrm{unl,fg}^i)}\hat{\omega}^{\textrm{tem},( \textrm{unl,fg}^i)}}_L\rangle_{\textrm{CMB}}-N^{\textrm{tem, unl}}_L,
\end{equation}
\begin{equation}\label{eq:nonfg1_bias}
N^{\textrm{tem, unl}}_L=\langle C^{\hat{\omega}^{(\textrm{unl})}\hat{\omega}^{\textrm{tem},(\textrm{unl})}}_L\rangle_{\textrm{CMB}}.
\end{equation}
These use unlensed HILC maps, $X^{\textrm{HILC, fg}_i}$, that only include the relevant foreground (e.g. $\textrm{fg}^1=\textrm{tSZ}$). As the maps were unlensed, the lensing observables, $\hat{\omega}^{(\textrm{unl,fg})}$ and $\hat{\kappa}^{(\textrm{unl,fg})}$, only contain reconstruction noise. While this approach isolates foreground only bias contributions by neglecting possible cross terms with $\phi$, it fails to pick up any foreground term involving the $\kappa$ template tracer. Contractions between foregrounds and the lensing signal $\phi$, along with the missing $\kappa$ terms, were recovered through recomputing the total bias with unlensed HILC versions and subtracting from the total  
\begin{equation}
N^{\textrm{fg}\times\phi}_L=N^{\textrm{fg, total}}_L-N^{\textrm{fg, total (unl)}}_L.
\end{equation}
It is clear from Fig.~\ref{fig:fg} that summing over all $N^{\textrm{fg}^i}$ and $N^{\textrm{fg}\times\phi}$ terms does not account for the full size and shape of the residual bias. To capture other contributions beyond individual sources we define `cross-foreground' terms
\begin{multline}
N^{\textrm{fg}^i\times\textrm{fg}^j}_L=\langle C^{\hat{\omega}^{(\textrm{unl,fg}^i+\textrm{fg}^j)}\hat{\omega}^{\textrm{tem},(\textrm{unl,fg}^i+\textrm{fg}^j)}}_L\rangle_{\textrm{CMB}}\\-N^{\textrm{fg}^i}_L-N^{\textrm{fg}^j}_L-N^{\textrm{tem, unl}}_L,
\end{multline}
which includes excess bias due to intra-foreground contractions. Again, unlensed HILC maps were used here. The cross terms involving tSZ and CIB, or tSZ and kSZ, are found to be significant in amplitude and mostly negative. This could be driven by tSZ clusters, which imprint negatively onto the CMB at low-$\nu$. We now find good agreement between the sum of all constituent biases shown in Fig.~\ref{fig:fg} with the total foreground bias $N^{\textrm{fg}}_L$. 

The bias term involving CMB lensing $\phi$ with foregrounds, $N^{\textrm{fg}\times\phi}$, was separated out. However, the AGORA foreground maps are lensed, so a caveat is that there could still be lensing contractions within the $N^{\textrm{fg}^i}$ and $N^{\textrm{fg}^i\times\textrm{fg}^j}$ bias terms even though they were calculated with unlensed CMBs. We do not investigate the decomposition of bias terms any further, but note that foreground lensing contractions could be responsible for part of the $N^{\textrm{fg}^i}$ and $N^{\textrm{fg}^i\times\textrm{fg}^j}$ signals.

\subsection{Foreground mitigation}

Up to this point, our foreground mitigation has been limited to deprojecting frequency-dependent signals using the HILC method, combined with the baseline threshold limit applied to the radio source maps. While this approach has allowed assessment of the bias from extragalactic sources on the rotation cross-spectrum, the results do not accurately reflect the expected bias when realistic mitigation strategies are applied to actual data. We now explore various methods to further reduce the impact of foreground bias on curl measurements.

A significant source of residual extragalactic foregrounds is point sources. Our baseline setup already included a conservative flux cut on radio maps to remove their brightest sources. Emission from dusty star-forming galaxies (DSFGs) appears in the CIB as point sources and correlates with other large-scale structure observables and extragalactic foregrounds. Typically, such sources can be match-filtered and masked out. To avoid handling cut-sky effects, we adopt the same procedure used for the radio maps and apply a flux cut.

A full-sky experiment at SO goal sensitivity is forecast to detect ~21,250 DSFGs from a single band \cite{Ade:2018sbj}. We find a threshold cut of 7mJy on the CIB 220 GHz map selects 20,680 pixels in AGORA, which we then replace by the median pixel value of the respective CIB map.  This is a conservative cut, as the number of pixels does not directly represent DSFG sources, given that a single point source could affect multiple pixels on high-resolution maps. As there is strong CIB correlation with tSZ \cite{Planck:2015emq}, we perform the same source cut on tSZ maps. The analysis of Section~\ref{sec:fg_bias} is then repeated, but with HILC weights and maps recomputed.

\begin{figure*}[t]
 	\includegraphics[width=\linewidth]{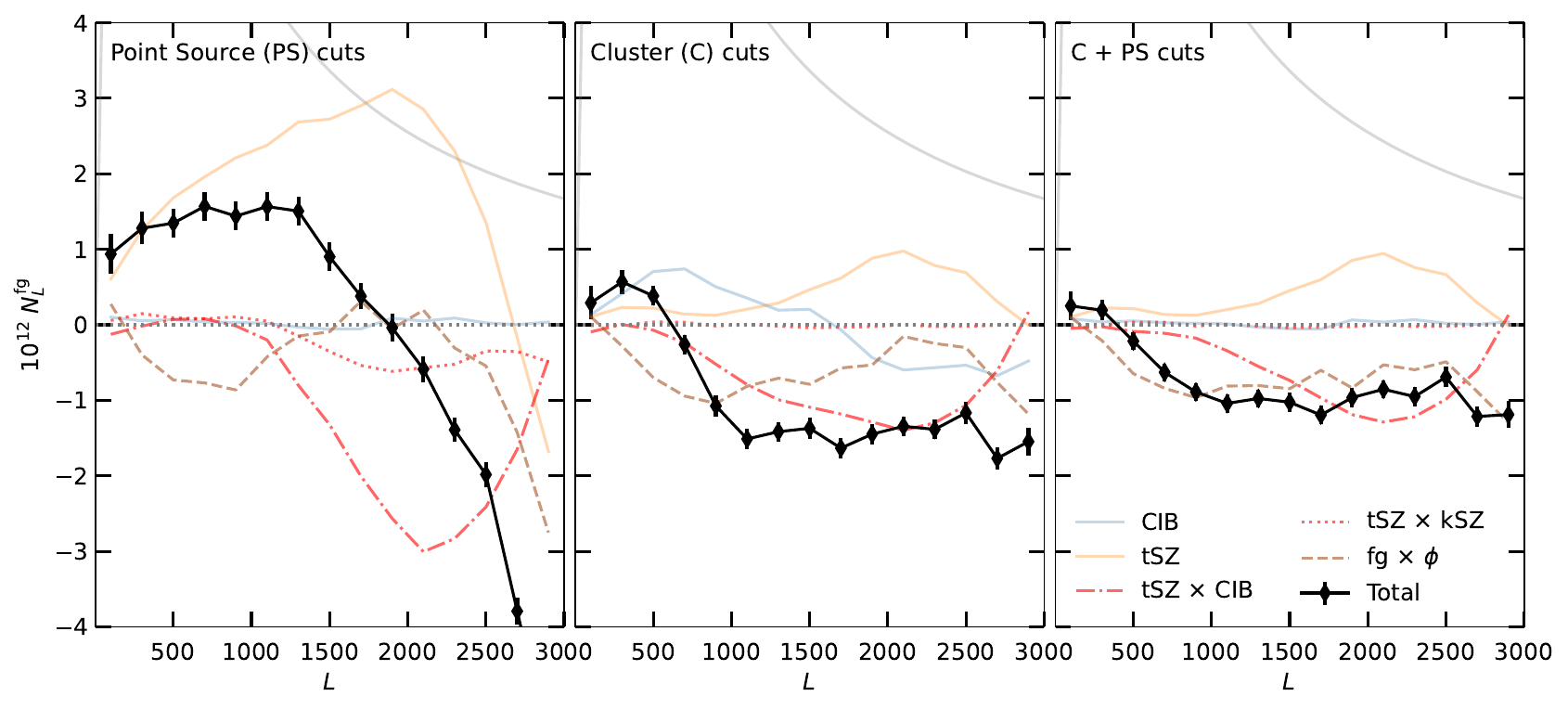}
	\caption{\small{Each panel illustrates the total and constituent foreground biases to the field rotation cross-spectrum for different mitigation strategies with an SO-like experiment. The lines are the same as those in Fig.~\ref{fig:fg} but for legibility only the total foreground bias, $N^{\textrm{fg}}$ (solid black diamonds), keep the error bars. Going left to right, the first panel shows the case where CIB point sources were cut from both tSZ and CIB maps. The second panel instead removes cluster sources directly from the tSZ map only. The third panel combines both approaches, applying both cluster and point-source cuts simultaneously.}}
	\label{fig:fg_mit}
\end{figure*}

The first panel in Fig.~\ref{fig:fg_mit} plots the updated foreground bias when the HILC has been recalculated to incorporate the point source cuts. The amplitude of constituent foreground biases are generally reduced. The individual CIB bias is now negligible, and the amplitude of the cross-term between tSZ and CIB has more than halved.

The dominant residual extragalactic foreground in the CMB is the tSZ signal (Fig.~\ref{fig:hilc_w}), which also contributes most to the foreground bias on the curl cross-spectrum (Fig.~\ref{fig:fg}), both individually and through cross terms with other foregrounds. The physical source of the largest SZ signal, individual galaxy clusters, can be detected by matched filtering of the CMB \cite{Herranz:2002kg,Melin:2006qq}. A full-sky SO-like survey is expected to detect $\sim60,000$ such clusters \cite{Ade:2018sbj} at goal sensitivity. A common approach to mitigating SZ foregrounds is to mask (or in-paint \cite{Bucher:2011nf}) the detected clusters. To avoid complications with measuring statistics on a cut sky, we do not explicitly perform any masking (or in-painting) here. Nonetheless, we do investigate the impact of removing cluster signals.

\begin{figure}[t]
 	\includegraphics[width=\linewidth]{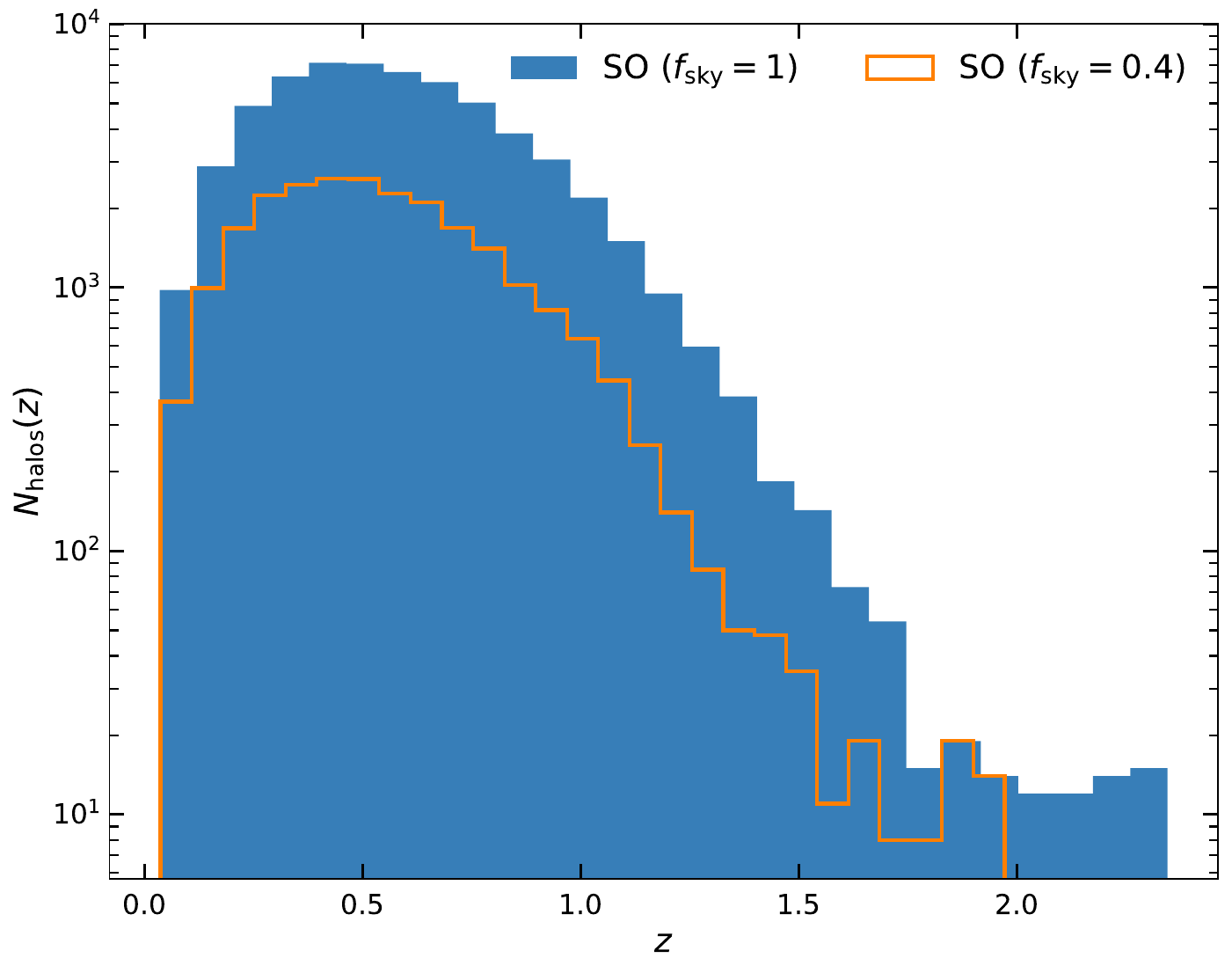}
	\caption{\small{The redshift distribution of the 60,000 halos with the largest virial mass in the AGORA simulation is shown in blue. This represents the projected number of tSZ clusters that SO is forecast to detect if it were an all-sky survey. The equivalent distribution for a survey with 40\% sky coverage is shown in orange, containing 24,000 of the most massive AGORA halos.}}
	\label{fig:halos}
\end{figure}

All tSZ pixels within 1.5 times the virial radius of the 60,000 most massive halos in the AGORA simulation were replaced by the median pixel value of the respective map. The redshift distribution of the selected halos in Fig.~\ref{fig:halos} match the predicted distribution of the tSZ clusters in Ref.~\cite{Ade:2018sbj}. We perform this process for the tSZ maps only. The second panel of Fig.~\ref{fig:fg_mit} shows that removal of SO clusters leads to an overall shape change in the foreground bias, and significant reduction in amplitude. All tSZ bias contributions have reduced amplitude, the cross-signal with kSZ is now negligible, and the cross with the CIB is significantly reduced.

The results of applying cluster and point source cuts simultaneously are displayed in the final panel of Fig.~\ref{fig:fg_mit}. Each constituent bias has significantly reduced amplitude, demonstrating the sensitivity of foregrounds to cluster and point source masking. Interestingly, the contributions from $N^{\textrm{tSZ}}$ and $N^{\textrm{tSZ}\times\textrm{CIB}}$ have similar shape but opposite sign and so cancel out. Thus, it is the $N^{\textrm{fg}\times\phi}$ term that dominates the total bias which remains non-negligible. It should be noted that our choice of cuts are conservative: all clusters were cut by an area defined by just 1.5 times the viral radius, and the number of point sources is likely underestimated. Therefore, the bias amplitudes in Fig.~\ref{fig:fg_mit} indicate realistic worst-case scenarios for SO.

\subsubsection{Extra mitigation options}

\begin{figure}[t]
 	\includegraphics[width=\linewidth]{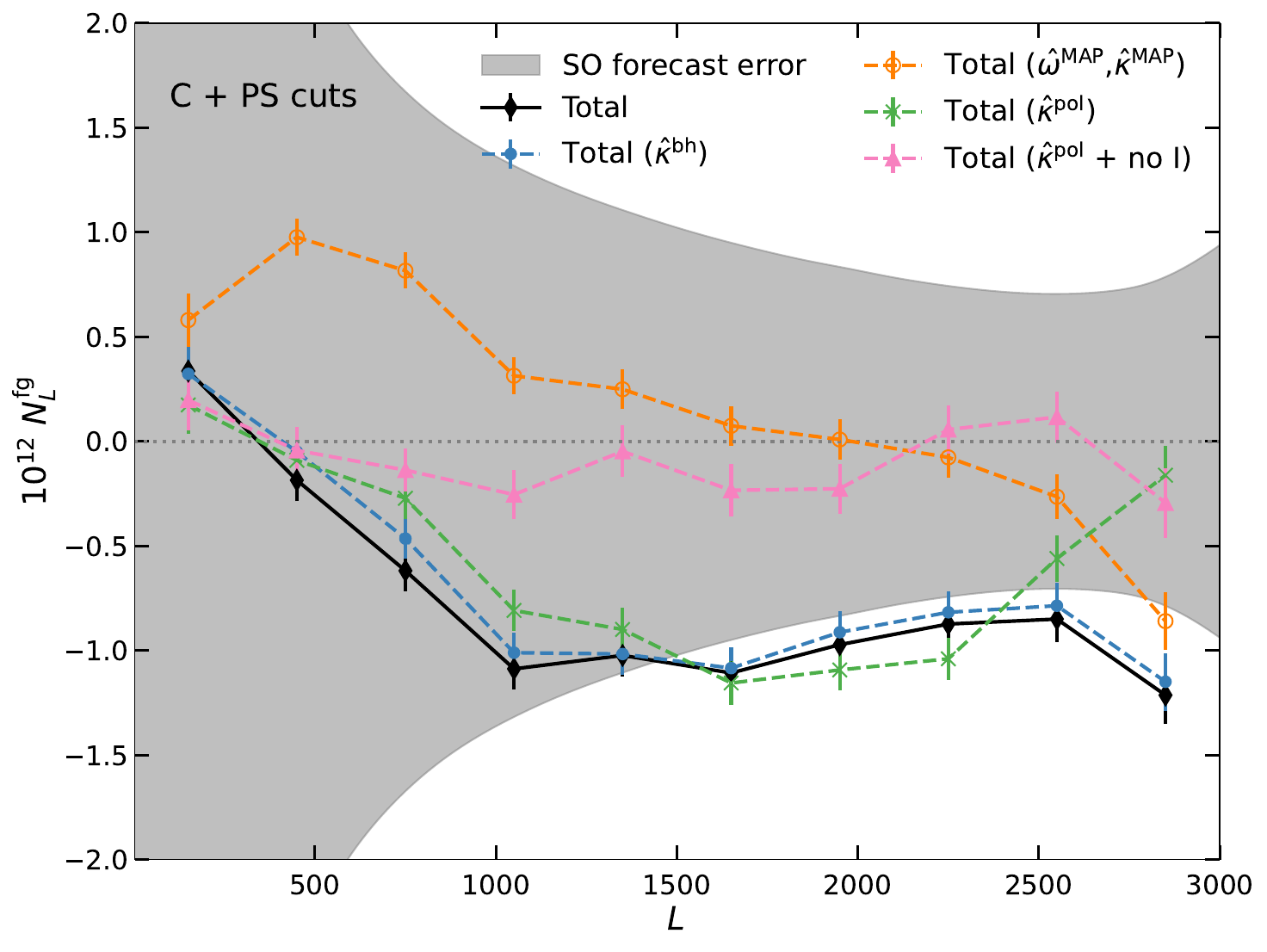}
	\caption{\small{The total foreground bias to the rotation-template cross-spectrum under different foreground mitigation strategies for an SO-like experiment. In each case, cluster and point source cuts have been applied to the tSZ maps, while only point sources were removed from CIB maps. For reference, the total foreground bias using QEs (black diamonds) is the same curve as from the third panel in Fig.~\ref{fig:fg_extra}. The blue dots represent total foreground bias with point-source hardened $\hat{\kappa}^{\textrm{bh}}$ in the template. For MAP estimators, $\hat{\omega}^{\textrm{MAP}}$ and $\hat{\kappa}^{\textrm{MAP}}$, the amplitude of the ${\textrm{fg}}\times\phi$ term decreases which reduces significance of the the total foreground bias (orange circles). Moreover, a polarization-only reconstruction, $\hat{\kappa}^{\textrm{pol}}$, removes all foreground contributions from the template leg, but has little impact on the total foreground bias (green crosses). Finally, the bias can become consistent with near-zero for a CIB-free rotation template constructed from tracers $a\in\{\hat{\kappa}^{\textrm{pol}}, \hat{g}\}$ (pink triangles). The forecast accumulated error bars (Eq.~\eqref{eq:error}) for an SO-like experiment are represented by the grey contour. This illustrates that the total extra-galactic bias, if ignored, would significantly bias rotation detection prospects without additional foreground mitigation.}}
	\label{fig:fg_extra}
\end{figure}

We continue to use the cluster and point source cuts of the previous section, but now test additional foreground mitigation strategies. Foregrounds enter through each leg in the cross-spectrum, either via quadratic estimation of $\hat{\omega}$ or the $\hat{\kappa}$ measurement used in constructing the template. Assuming there are no relevant extragalactic polarization foregrounds, using a polarization-only reconstruction of $\hat{\kappa}^{\textrm{pol}}$ ensures foregrounds only enter via the $\hat{\omega}$ leg. The cost is paid for by reduced $S/N$ of the rotation-template cross-spectrum, by about $1.4\sigma$ (estimated using the same forecast method described in \RL). However, Fig.~\ref{fig:fg_extra} shows that $\hat{\kappa}^{\textrm{pol}}$ does not significantly affect the total foreground bias amplitude. This may be misleading due to the cancellation effects observed between $N^{\textrm{tSZ}}$ and $N^{\textrm{tSZ}\times\textrm{CIB}}$ in Fig.~\ref{fig:fg_mit}. It is possible to achieve near elimination of the bias by using a combination of $\hat{\kappa}^{\textrm{pol}}$ and a CIB-free rotation template, i.e. a template constructed using $a\in\{\hat{\kappa}^{\textrm{pol}}, \hat{g}\}$, as shown in Fig.~\ref{fig:fg_extra}. Again, this significantly affects $S/N$, but could still give a detection at $\sim3\sigma$ for SO.

Another approach to foreground mitigation is with bias-hardened lensing estimators. Lensing measurements can be `foreground hardened' by computing the response to a specific type of foreground and subtracting it from the lensing reconstruction \cite{Sailer:2020lal}. This comes at the cost of a modest increase in reconstruction noise. We switch on the bias-hardened estimator in \plancklens, so that the convergence estimator $\hat{\kappa}^{\textrm{bh}}$ is hardened against Poisson sources. To avoid complications/signal loss on the already noise-dominated curl, we do not bias harden the $\hat{\omega}$ reconstruction. Fig.~\ref{fig:fg_extra} shows the bias amplitude with $\hat{\kappa}^{\textrm{bh}}$ has little effect on the total bias amplitude, which may again be impacted by cancellation effects and is expected considering that $\hat{\kappa}^{\textrm{pol}}$ (which removes all foregrounds in the $\kappa$ leg) did not significantly alter the bias.  Nonetheless, bias hardened estimators could still be useful at improving the robustness of foreground mitigation by reducing the non-Gaussianity of the bias. One could extend this further and also harden against cluster profiles, however such sources are not implementation in \plancklens\ and so beyond the scope of this paper. 

The final mitigation method explored was using MAP estimators, $\hat{\omega}^{\textrm{MAP}}$ and $\hat{\kappa}^{\textrm{MAP}}$, to reduce the lensing-foreground cross term from Fig.~\ref{fig:fg_mit}. Fig.~\ref{fig:fg_extra} shows this would lead to a significant reduction in bias amplitude on most scales. While there is an increase at $L>1000$, it is still smaller than the project SO errors. This increase may be consequence of the different response to other foreground bias terms with MAP estimators compared to the QE case. Finally, MAP estimators are the only mitigation approach considered that actually {\it increase} the $S/N$ of $C^{\hat{\omega}\hat{\omega}^{\textrm{tem}}}$, making MAP estimators an appealing option for curl detection. 

\subsection{Non-Gaussian foreground bias}
The amplitudes of the total foreground biases in Figs \ref{fig:fg_mit} and \ref{fig:fg_extra} are typically smaller than the equivalent template-lensing bias, $N^{\textrm{tem}}$ (Fig.~\ref{fig:phi_bias}). We argued in Section.~\ref{sec:results} that $N^{\textrm{tem}}$ does not limit curl detection prospects provided it can be simulated and subtracted. However, this conclusion relied on the proven Gaussianity of the $N^{\textrm{tem}}$ contractions, which can be reliably modelled with simulations. In contrast, modelling of non-Gaussian foregrounds presents greater challenges. For instance, as observed in Ref.~\cite{ACT:2023ubw}, foregrounds in the Websky \cite{Stein:2020its} simulation yield different predicted lensing biases compared to results obtained using the Seghal \cite{Sehgal:2009xv} simulation. Therefore, the extent to which $N^{\textrm{fg}}$ can be effectively removed from curl cross-spectrum measurements may be limited by any non-Gaussian contributions.

We evaluate the Gaussianity of $N^{\textrm{fg, total}}$ using a similar approach from Section~\ref{sec:bias}. Gaussian random fields $\tilde{\kappa}^{\textrm{G}}$, $\tilde{g}^{\textrm{G}}$, $\tilde{I}^{\textrm{G}}$, tSZ$^{\textrm{G}}(\nu)$, and CIB$^{\textrm{G}}(\nu)$ were generated via Choleksky decomposition of their smoothed empirical auto- and cross-spectra from AGORA. The same point source and cluster cuts of the previous sections were again applied prior to measuring the spectra used to generated the maps. Such cuts remove potential sources of non-Gaussianity from the foreground components, thus the focus here is on evaluation of the residual non-Gaussianity remaining after realistic foreground mitigation (Fig.~\ref{fig:fg_mit}). We choose to exclude kSZ and radio signals because their impact on the bias is minimal, as demonstrated in Figs \ref{fig:fg} and \ref{fig:fg_mit}. Inter-frequency correlations were also not included in the generation of the Gaussian foregrounds, so we do not construct a HILC map and instead compute the bias at each individual $\nu$. The non-Gaussian contribution to the foreground bias at $\nu$ is then inferred by subtracting the Gaussian bias from the total
\begin{equation}
N^{\textrm{fg}^{\textrm{nG}},\textrm{total}}_L(\nu)=N^{\textrm{fg},\textrm{total}}_L(\nu)-N^{\textrm{fg}^{\textrm{G}},\textrm{total}}_L(\nu),
\end{equation}
in which the total Gaussian bias is defined
\begin{equation}
N^{\textrm{fg}^{\textrm{G}},\textrm{total}}_L(\nu)=\langle C^{\hat{\omega}^{(\textrm{G}, \textrm{fg}^{\textrm{G}}(\nu))}\hat{\omega}^{\textrm{tem},(\textrm{G}, \textrm{fg}^{\textrm{G}}(\nu))}}_L\rangle_{\textrm{CMB}}-N^{\textrm{tem}}_L.
\end{equation}
The lensing observables $\hat{\omega}^{(G, \textrm{fg}^{\textrm{G}}(\nu))}$ and $\hat{\kappa}^{(G, \textrm{fg}^{\textrm{G}}(\nu))}$ were reconstructed from noise-free CMB simulations deflected by $\boldsymbol{\alpha}^{(G)}$ and include the Gaussian foregrounds tSZ$^{\textrm{G}}(\nu)$, and CIB$^{\textrm{G}}(\nu)$. The template, $\hat{\omega}^{\textrm{tem},(\textrm{G}, \textrm{fg}^{\textrm{G}}(\nu))}$, was then constructed for tracers $a\in\{\hat{\kappa}^{(G, \textrm{fg}^{\textrm{G}}(\nu))},\tilde{g}^{\textrm{G}},\tilde{I}^{\textrm{G}}\}$. 

The total bias was constructed in a similar way
\begin{equation}
N^{\textrm{fg},\textrm{total}}_L(\nu)=\langle C^{\hat{\omega}^{(\phi, \textrm{fg}(\nu))}\hat{\omega}^{\textrm{tem},(\phi, \textrm{fg}(\nu))}}_L\rangle_{\textrm{CMB}}-N^{\textrm{tem}}_L,
\end{equation}
with lensing observables $\hat{\omega}^{(\phi, \textrm{fg}(\nu))}$ and $\hat{\kappa}^{(\phi, \textrm{fg}(\nu))}$ reconstructed from noise-free CMB simulations deflected by $\boldsymbol{\alpha}^{(\phi)}$ which include the AGORA foregrounds tSZ$(\nu)$ and CIB$(\nu)$. The template, $\hat{\omega}^{\textrm{tem},(\phi, \textrm{fg}(\nu))}$, was constructed for tracers $a\in\{\hat{\kappa}^{(\phi, \textrm{fg}(\nu))},\tilde{g},\tilde{I}\}$. Finally, the lensing-template bias is the same as defined by Eq.~\eqref{eq:nonfg_bias}.

\begin{figure}[t]
 	\includegraphics[width=\linewidth]{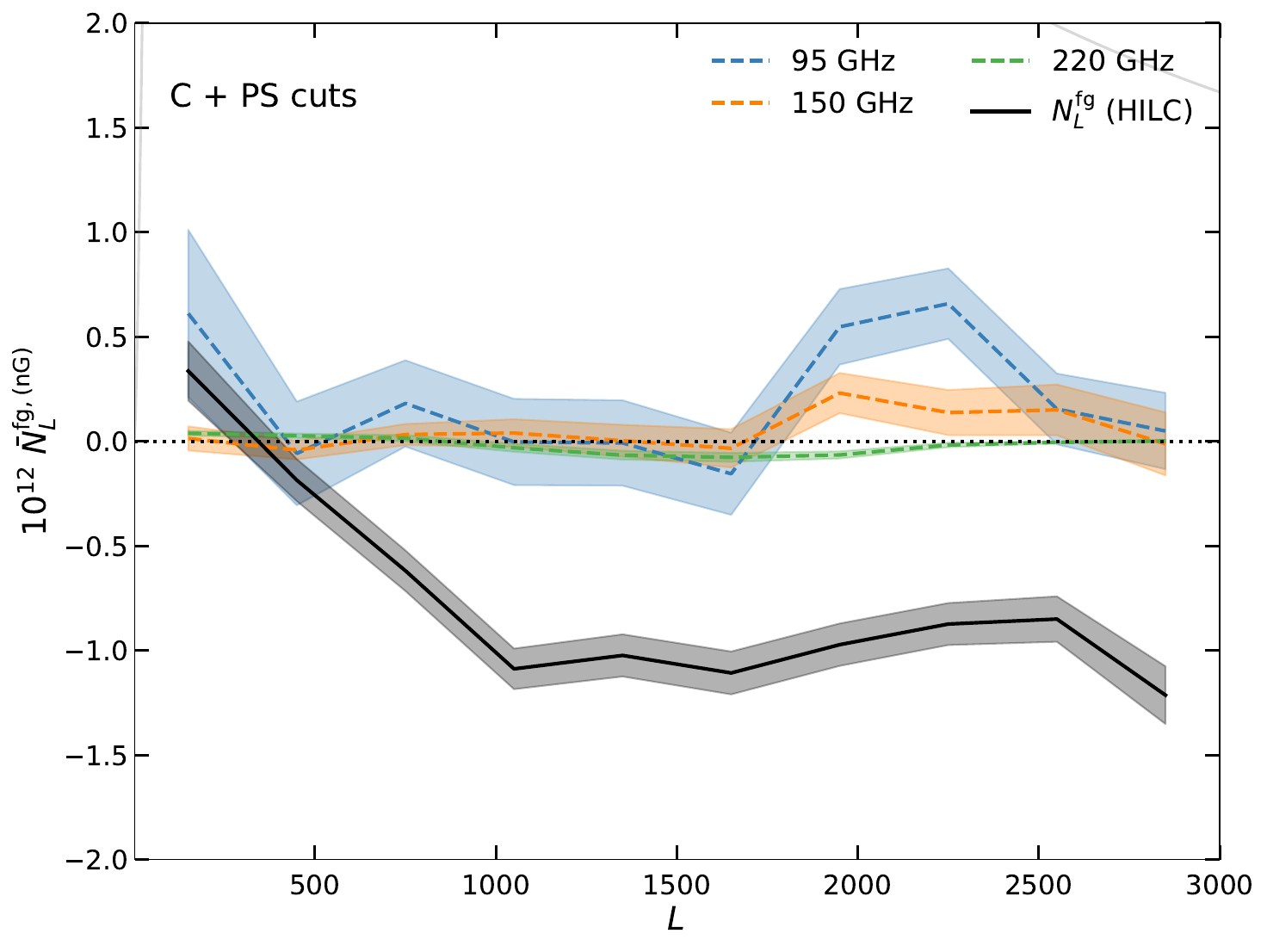}
	\caption{\small{Contributions to the HILC weighted non-Gaussian foreground bias, $\bar{N}^{\textrm{fg}^{\textrm{nG}},\textrm{total}}(\nu)$, from the 95, 150, and 220 GHz maps (dashed coloured lines), compared to the total HILC bias (solid black). In this case, cluster and point source subtraction at the SO level were applied to the tSZ and CIB maps prior to the analysis. This removes potential sources on non-Gaussianity from the foreground components, and so the results should be interpreted as the residual non-Gaussianity after realistic foreground mitigation (Figs \ref{fig:fg_mit} and \ref{fig:fg_extra}). The results are mostly consistent with near-zero non-Gaussianity, with exception of a small contribution on small scales. }}
	\label{fig:fg_gauss}
\end{figure}

We plot the resulting non-Gaussian bias normalized by HILC weight, 
\begin{equation}
\bar{N}^{\textrm{fg}^{\textrm{nG}}}_L(\nu)=w^{\nu}_LN^{\textrm{fg}^{\textrm{nG}}}_L(\nu),
\end{equation}
such that $\bar{N}^{\textrm{fg}^\textrm{nG}(\nu)}$ represents the relative contribution to the HILC bias. Fig.~\ref{fig:fg_gauss} shows the non-Gaussian foreground bias contributions are small with respect to the total HILC bias. While there is a non-zero contribution at 220 GHZ, it is down-weighted in the HILC process (Fig.~\ref{fig:hilc_w}). The 95 and 150 GHZ terms are consistent with a null signal at $L\lesssim2000$, and subdominant to the Gaussian terms at higher $L$. Therefore, after applying cluster and point source cuts, the extra-galactic foreground bias is predominantly driven by disconnected contractions from individual $\nu$. This can be predicted from simulations to sufficient precision for achieving a robust low-signal curl detection.

\subsection{Curl foreground bias}
Until now, evaluation of foreground biases have been carried out in the null-hypothesis of zero curl. Here, we extend the investigation to test the robustness of the rotation-template cross-spectrum to any additional curl-induced foreground contractions. 

\begin{figure}[t]
 	\includegraphics[width=\linewidth]{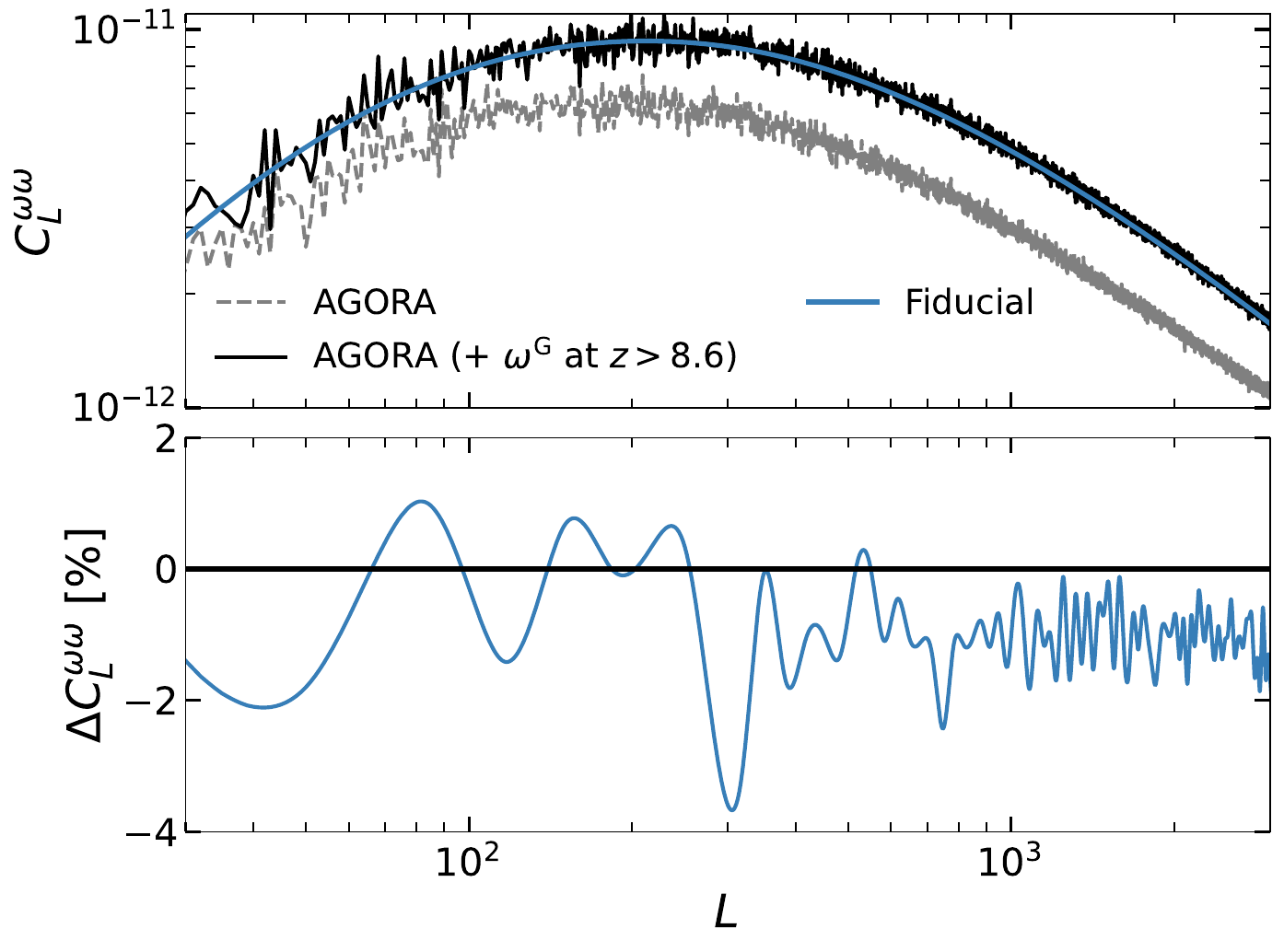}
	\caption{\small{The ray-traced field rotation power spectrum computed from the AGORA N-body simulation (dashed grey), compared to the fiducial post-Born rotation spectrum (solid blue). As the ray-tracing was only propagated through redshifts $z\leq8.62$, the spectrum is missing signal from high-redshift lensing deflections, up to $40\%$ of total power. A Gaussian realization was added to the field to account for the missing deflections (solid black), for a power spectrum computed with Eq.~\eqref{eq:omega_ps} and Eq.~\eqref{eq:omega_mc} in which the integral limits impose $z>8.62$. The bottom panel displays the percentage deviation of the theory curve to the (smoothed) corrected rotation field. }}
	\label{fig:omega_ag}
\end{figure}

The ray-traced lensing field rotation map, $\omega^{\textrm{sim}}$, was not provided as an explicit AGORA product. Instead, we were able to extract the rotation component directly from the GrayTrix \cite{Hamana:2015bwa, Shirasaki:2015dga} ray-tracing output, and processed it following a similar procedure used to produce $\kappa^{\textrm{sim}}$ in Ref.~\cite{Omori:2022uox}. The auto-spectrum of the resulting $\omega^{\textrm{sim}}$ is shown in Fig.~\ref{fig:omega_ag} to be significantly smaller in amplitude than fiducial $C^{\omega\omega}$. The ray-tracing was only propagated through the AGORA simulation for redshifts $z\leq8.62$, hence $\omega^{\textrm{sim}}$ is missing important contributions from high-$z$ lensing deflections. To combat this, we made a realization of $\omega^{\textrm{G}}$ for the missing deflection power, and added it to $\omega^{\textrm{sim}}$. The power spectrum $C^{\omega^{\textrm{G}}\omega^{\textrm{G}}}$ was computed using the post-Born Limber description of Eq.~\eqref{eq:omega_ps}, in which the $z>8.62$ cut is applied through modifying the integration limits in Eq.~\eqref{eq:omega_mc}. Although this corrects the amplitude of the auto-spectrum in Fig.~\ref{fig:omega_ag}, the field is still disconnected from the underlying AGORA matter distribution at high-$z$. The notational convention from here on, unless specified, is that $\omega^{\textrm{sim}}$ implicitly includes the corrected Gaussian realization.

\begin{figure}[t]
 	\includegraphics[width=\linewidth]{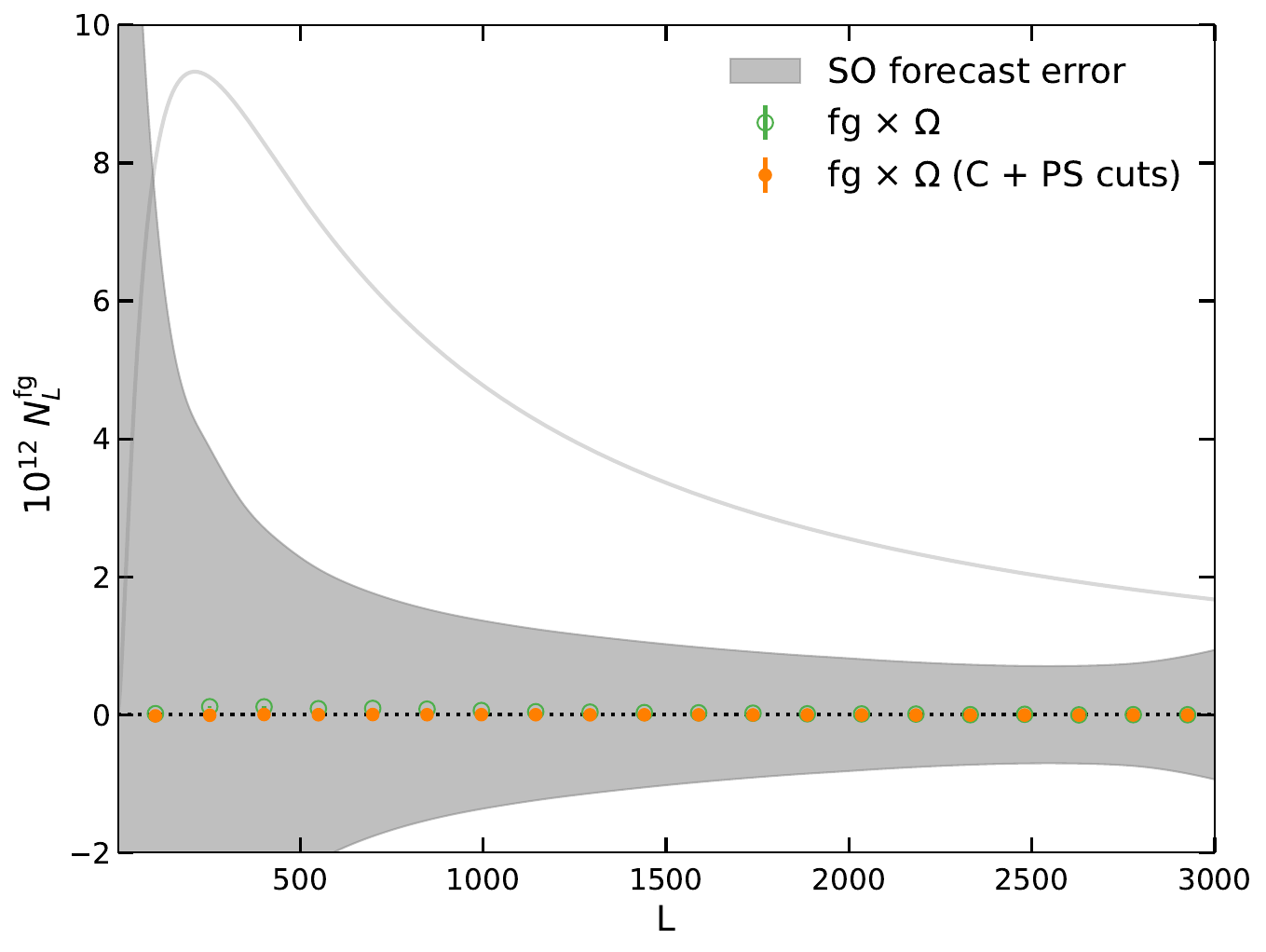}
	\caption{\small{The rotation-template cross-spectrum bias induced by higher-order contractions involving extragalactic foregrounds with the CMB lensing curl (green circles). The forecast accumulated error bars (Eq.~\ref{eq:error}) for an SO-like experiment are illustrated by the grey contour, and show that the bias is of marginal importance for SO. However, after cluster and point source cuts were applied to the foreground maps, the bias becomes negligible (orange points). For context, the fiducial $C^{\omega\omega}$ (solid grey) and the null line (dotted black) are also shown.}}
	\label{fig:fg_w}
\end{figure}

New HILC maps were built for CMB fields $\tilde{X}$ lensed by the full AGORA deflection field $\boldsymbol{\alpha}^{\textrm{sim}}$, which now includes $\omega^{\textrm{sim}}$. The foreground bias from $\Omega$-induced contractions was then isolated through 
\begin{equation}\label{eq:fg_curl_bias}
N^{\textrm{fg}\times\Omega}_L=\langle C^{\hat{\omega}^{(\textrm{sim}, \textrm{fg})}\hat{\omega}^{\textrm{tem},(\textrm{sim}, \textrm{fg})}}_L-C^{\hat{\omega}^{(\textrm{sim})}\hat{\omega}^{\textrm{tem},(\textrm{sim})}}_L\rangle_{\textrm{CMB}}-N^{\textrm{fg, total}}_L,
\end{equation}
in which $N^{\textrm{fg, total}}_L$ is the total foreground bias in the null-hypothesis, computed from Eq.~\eqref{eq:fg_bias}. Fig.~\ref{fig:fg_w} shows the $\textrm{fg}\times\Omega$ bias to be small relative to both the fiducial $C^{\omega\omega}$, and the forecast SO errors (Eq.~\ref{eq:error}). The bias amplitude could be underestimated due to missing high-$z$ contractions in $\omega^{\textrm{sim}}$. Nonetheless, if cluster and point source cuts are applied the bias becomes negligible.

\section{Conclusions}\label{sec:conclusion}

The CMB lensing curl is detectable as a cross-spectrum signal through a field-rotation reconstruction, $\hat{\omega}$, and a LSS rotation template, $\hat{\omega}^{\textrm{tem}}$. In this paper, we derived a new curved-sky estimator for $\hat{\omega}^{\textrm{tem}}$ that can be used to detect $\omega$ with upcoming wide-area CMB observations such as SO and CMB-S4. We verified with the new estimator that the cross-spectrum $C^{\hat{\omega}\hat{\omega}^{\textrm{tem}}}_{L}$ is an unbiased probe of $C^{\omega\omega}_{L}$ in idealized full-sky N-body simulations.

When the lensing fields are reconstructed with quadratic estimators, we found that $C^{\hat{\omega}\hat{\omega}^{\textrm{tem}}}_{L}$ has significant additional bias terms. Our analysis showed that the bias is Gaussian and well understood analytically in the null hypothesis, in agreement with \RL. In principle, it should therefore be straight forward to null noise contributions with data splits \cite{Madhavacheril:2020ido} and accurately subtract the remaining Gaussian bias using simulations. Relaxing the null hypothesis, we found that additional curl-induced biases will not affect detection prospects with SO, but may be marginally important for CMB-S4. Nonetheless, we demonstrated that switching to MAP lensing estimators significantly reduced biases in qualitative agreement with simplistic analytic approximations.

Further relaxation of idealized assumptions to include extragalactic foreground contamination in the CMB maps revealed significant new biases to $C^{\hat{\omega}\hat{\omega}^{\textrm{tem}}}_{L}$ arising from frequency-dependent contractions involving tSZ, kSZ, CIB, and $\phi$, in the null hypothesis. We demonstrated that the HILC method, in addition to cluster and point source cuts, effectively reduced the amplitude of the total bias to below signal level for SO. Again, we found no additional foreground bias from curl-induced contractions at SO sensitivity. 
While we found no significant non-Gaussian contributions to the extra-galactic bias, it remains uncertain how accurately simulations can model the residual foreground bias. Therefore, further foreground mitigation may be necessary for optimal curl detection strategies, of which we found MAP lensing estimators again to be the most promising approach. 

While we have devoted significant effort into developing curl detection techniques, the science prospects of measuring $C^{\hat{\omega}\hat{\omega}^{\textrm{tem}}}_{L}$ have received little attention. We rectify this in Appendix \ref{sec:fisher}, exploring the cosmological information contained within a CMB curl detection, and its implications for future surveys.

In this work we have investigated the primary complications in measuring $C^{\hat{\omega}\hat{\omega}^{\textrm{tem}}}_{L}$ using realistic N-body mock data. However, several simplifying assumptions need further exploration before switching to real data. For instance, we have not yet accounted for masking, therefore sensitivity of $C^{\hat{\omega}\hat{\omega}^{\textrm{tem}}}_{L}$ to cut-sky affects, including additional mask-induced lensing biases \cite{Lembo:2021kxc}, is still unclear. Moreover, assessment of biases from galactic foregrounds and experimental systematics is required and left to future work.

\begin{acknowledgements}
We thank Carmelita Carbone for providing the matter power spectrum of the DEMNUni simulations and Yuuki Omori for help with AGORA simulation products. MR is supported by a UK Science and Technology Facilities Council (STFC) studentship. AL was supported by the UK STFC grant ST/X001040/1. GF acknowledges the support of the European Research Council under the Marie Sk\l{}odowska Curie actions through the Individual Global Fellowship No.~892401 PiCOGAMBAS and of the European Union’s Horizon 2020 research and innovation program (Grant agreement No. 851274) during the final stages of this work.
JC acknowledges support from a SNSF Eccellenza Professorial Fellowship (No. 186879).
\end{acknowledgements}

\appendix

\section{Coordinate handedness}\label{app:hand}
The scalar lensing curl potential, $\Omega$, and scalar lensing field rotation, $\omega$, were defined with the antisymmetric tensor, $\boldsymbol{\epsilon}$, in Eq.~\eqref{eq:alpha} and Eq.~\eqref{eq:alpha_pert} respectively. We define the tensor so that in a left-handed coordinate system the components are
\begin{equation}\label{eq:ep}
\epsilon_{ij} =
\begin{cases}
+1 & \text{if } (i,j) = (1,2), \\
-1 & \text{if } (i,j) = (2,1), \\
0 & \text{if } i=j.
\end{cases}
\end{equation}
Abiding by tensorial transformation rules, in a right-handed coordinate system the components of $\boldsymbol{\epsilon}$ must change sign:
\begin{equation}
\epsilon^{\textrm{LH}}_{ij}=-\epsilon^{\textrm{RH}}_{ij},
\end{equation}
thus ensuring that the physical interpretation of $\omega$ is invariant under coordinate change. 

Note, when defining cross-products in 2-dimensions we stick to the usual definition
\begin{equation}
\boldsymbol{u}\times\boldsymbol{v}=u_1v_2-u_2v_1=|u||v|\sin(\theta_v-\theta_u),
\end{equation}
which is invariant under coordinate change.

\subsection{Flat vs curved sky}
Consider an image measured at some redshift over a small patch of sky. In the flat-sky approximation, the image can be spanned by 2D Cartesian coordinates $\nhat=(\ehat_x,\ehat_y)$, with implicit $z$-axis going into the image (page) as expected for positive redshift. Therefore, flat-sky Cartesian coordinates are a left-handed coordinate system. 

Now instead choose to work in a 2D spherical polar basis. The coordinates $\nhat=(\ehat_{\theta},\ehat_{\phi})$, defined on the surface of a sphere of radius $z$, are polar angle, $\theta$, which points North to South and azimuthal angle, $\phi$, West to East. In astronomy, we are observing from {\it inside the sphere}. Thus, $\ehat_{\phi}$ points to the left, orthogonal to $\ehat_{\theta}$, and implicit positive $z$-axis points out of the sphere (into the page). This is a right-handed system. 

Care must be taken when comparing rotations derived in the flat-sky with their curved sky version, as the handedness of the coordinate systems may differ. In this paper, the results from the DEMNUni and AGORA simulations use the right-handed curved-sky coordinates. However, the formalism in \RL was derived in the left-handed flat-sky system, so some signs may be different, but the physical interpretations are the same.

\subsection{Rotation direction}
Equations \eqref{eq:alpha}, \eqref{eq:alpha_pert}, and \eqref{eq:ep} define positive $\omega$ as a {\it clockwise} rotation from source plane to observed plane. To show this explicitly, consider the measurement of some lensed field, $\tilde{I}$, observed at position $\nhat_{\textrm{ob}}$. This lensed observation is an approximate remapping of the original \underline{s}ource image
\begin{equation}\label{eq:image}
\tilde{I}(\nhat_{\textrm{ob}})=I(\nhat_{\textrm{s}})=I(\nhat_{\textrm{ob}}+\boldsymbol{\alpha}(\nhat_{\textrm{ob}})),
\end{equation} 
in which $\boldsymbol{\alpha}$ describes the total lensing deflection. A full description of the lensing distortions is contained within the magnification matrix \cite{Bartelmann:1999yn}
\begin{equation}\label{eq:mag_mat}
A_{ij}\equiv\frac{\partial(\hat{n}_{\textrm{s}})_i}{\partial n_j}=\delta_{ij}+\frac{\partial\alpha_i}{\partial n_j}
\end{equation}
which has an antisymmetric component, $\omega$, that depends on coordinate handiness\footnote{The physical interpretation of shear is also sensitive to coordinate handedness, but we do not discuss this further. The convergence is unaffected.} (due to Eq.~\eqref{eq:ep})
\begin{align}
\boldsymbol{A}^{(\textrm{LH})}&=
\begin{pmatrix}
1-\kappa -\gamma_1 & -\gamma_2 -\omega\\ 
-\gamma_2+\omega & 1-\kappa +\gamma_1
\end{pmatrix},\\
\boldsymbol{A}^{(\textrm{RH})}&=
\begin{pmatrix}
1-\kappa -\gamma_1 & -\gamma_2 +\omega\\ 
-\gamma_2-\omega & 1-\kappa +\gamma_1
\end{pmatrix}.
\end{align}
It is the inverse matrix, $\boldsymbol{A}^{-1}$, that determines the mapping from source plane to observed plane \cite{Bartelmann:2016dvf}, which can be approximately written as 
\begin{align}
\left(\boldsymbol{A}^{(\textrm{LH})}\right)^{-1}&\sim
\begin{pmatrix}\label{eq:mag_mat_rot}
1  & +\omega\\ 
-\omega & 1
\end{pmatrix}
\begin{pmatrix}
1-\kappa +\gamma_1 & \gamma_2\\ 
\gamma_2 & 1-\kappa -\gamma_1
\end{pmatrix},\\
\left(\boldsymbol{A}^{(\textrm{RH})}\right)^{-1}&\sim
\begin{pmatrix}\label{eq:mag_mat_rot2}
1  & -\omega\\ 
+\omega & 1
\end{pmatrix}
\begin{pmatrix}
1-\kappa +\gamma_1 & \gamma_2\\ 
\gamma_2 & 1-\kappa -\gamma_1
\end{pmatrix}.
\end{align}
We see that the leading matrices in Eq.~\eqref{eq:mag_mat_rot} and Eq.~\eqref{eq:mag_mat_rot2} are rotation matrices describing the source image rotating an angle $\omega$ in the {\it clockwise} direction onto the observed image. 

\section{Mismodelling of $\hat{g}$ tracer}\label{app:magbias}
In this section, we investigate the sensitivity of $\hat{\omega}^{\textrm{tem}}$ to the inclusion of magnification bias, $\mu$, and redshift-space distortions (RSD) within the $g$ tracer.

\begin{figure}[t]
 	\includegraphics[width=\linewidth]{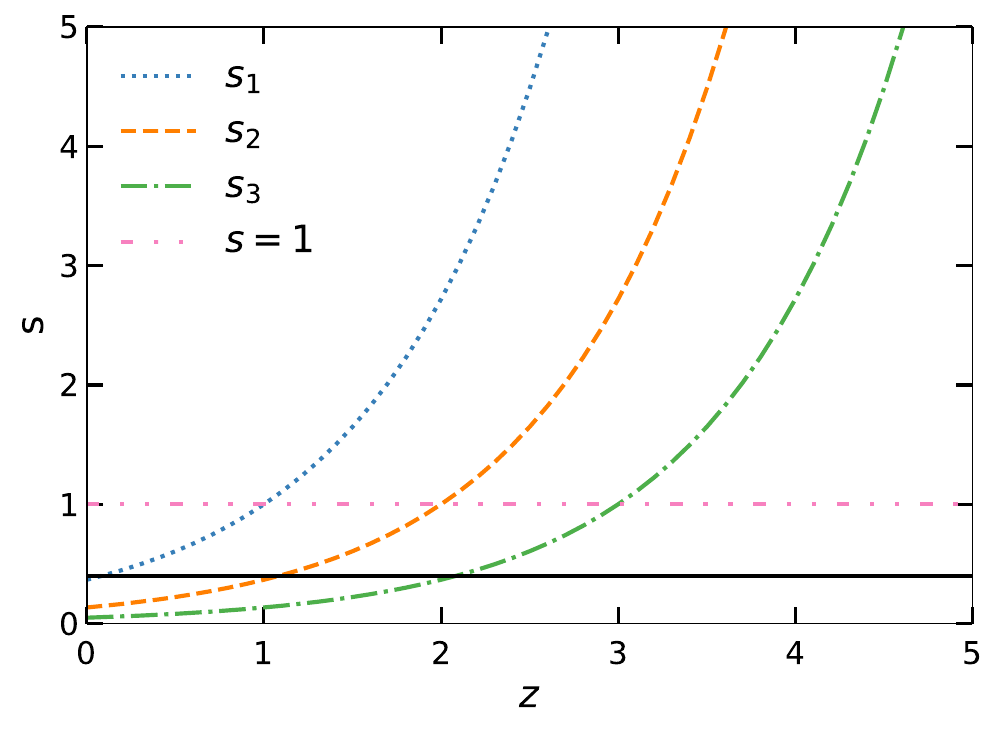}
	\caption{\small{Modelling the magnification bias, $\mu$, requires understanding the survey dependent $s$ parameter. We use 4 different toy models of $s$ (non-solid coloured lines), to test its importance for $\hat{\omega}^{\textrm{tem}}$ measurements. At $s=0.4$ (solid black) $\mu$ vanishes. As the 1-bin LSST redshift distribution peaks around $z\sim1$, it should be expected that the $\mu_{s_2}$ experiences more cancellation than other magnification biases. The $s=1$ model avoids the cancellation over all $z$ which should result in stronger (but unrealistic) $\mu$ contributions.}}
	\label{fig:magbias_s}
\end{figure}

Lensing magnification by the foreground matter distribution alters the observed number counts of background galaxies in two opposing ways. First, magnification stretches the observed galaxy distribution, reducing the local number density of galaxies. Simultaneously, magnification also increases the flux of faint galaxies, pushing more over observational flux thresholds which locally increases the number density. These competing effects modify the galaxy redshift distribution and introduce a lensing bias in the observed $\hat{g}$ field that correlates with the underlying density field \cite{Villumsen:1995ar, Moessner:1997qs, Bartelmann:1999yn}. This magnification bias is modelled as

\begin{equation}
\mu(\boldsymbol{L})=(5s-2)\int d\chi W_{\kappa^{\textrm{gal}}}(\chi)\delta(\boldsymbol{L}. \chi),
\end{equation}
where $W_{\kappa^{\textrm{gal}}}$ is the galaxy lensing window function (defined in equation 2.11 of \RL), and $s=d\log N/dm$ is the slope of the number counts with respect to the magnitude limit. We consider four different toy models for $s$: three take the form $s_i(z) = e^{z}/e^{i}$, while the fourth adopts the unphysical assumption of $s=1$. These models are illustrated in Fig.~\ref{fig:magbias_s}, and will allow exploration of $\hat{\omega}^{\textrm{tem}}$'s sensitivity to $\mu$.

The RSD complication arises from the local peculiar velocity contribution to the observed redshift of galaxies \cite{Kaiser:1987qv}, and is modelled in the (1st order) Limber approximation via \cite{Tanidis:2019teo, Euclid:2023pyq}

\begin{equation}
g^{\textrm{RSD}}(\boldsymbol{L}) = \int d\chi W_{\textrm{RSD}}(L, \chi)\delta(\boldsymbol{L}, \chi),
\end{equation}
with window function definition
\begin{equation}
W_{\textrm{RSD}}(L, \chi)=\sum^1_{i=-1}x_i(L)n(\bar{\chi}_i)f(\bar{\chi}_i),
\end{equation}
where
\begin{equation}
x_0(L) = \frac{2L^2+2L-1}{(2L-1)(2L+3)},
\end{equation} 
\begin{equation}
x_{-1}(L) = -\frac{L(L-1)}{(2L-1)\sqrt{(2L-3)(2L+1)}},
\end{equation} 
\begin{equation}
x_{+1}(L) = -\frac{2L^2+2L-1}{(2L+3)\sqrt{(2L+1)(2L+5)}},
\end{equation}
and the scaled comoving radial distance is
\begin{equation}
\bar{\chi}_i \equiv \frac{2L+1+4i}{2L+1}\chi.
\end{equation}
The growth rate, $f(z)=-(1+z)d\ln D/dz$, defined in terms of the linear growth factor, $D$, can be approximated as $f(z)\approx\Omega_m^{0.55}(z)$ \cite{PCP2018}. Note that we only model the linear RSD (Kaiser) effect here, so non-linear effects (such as the fingers of God) are not included.

\subsection{Normalization offset}

The construction of $\hat{\omega}^{\textrm{tem}}$ relies on fiducial modelling of cross post-Born bispectra, $b^{\omega ij}$, LSS tracer covariance, $\boldsymbol{C}_{\textrm{LSS}}$, and the post-Born rotation auto-spectrum, $C^{\omega\omega}$. Inaccuracies in the theory lead to incorrect normalization, $F_L$, which results in an offset to the measured $C^{\hat{\omega}\hat{\omega}^{\textrm{tem}}}$ with respect to $C^{\omega^{\textrm{true}}}$. This was demonstrated in Fig.~\ref{fig:omega_only}, different non-linear $P_{\delta\delta}$ prescriptions lead to discrepancies in the amplitude of $C^{\hat{\omega}\hat{\omega}^{\textrm{tem}}}$. 

\begin{figure}[t]
 	\includegraphics[width=\linewidth]{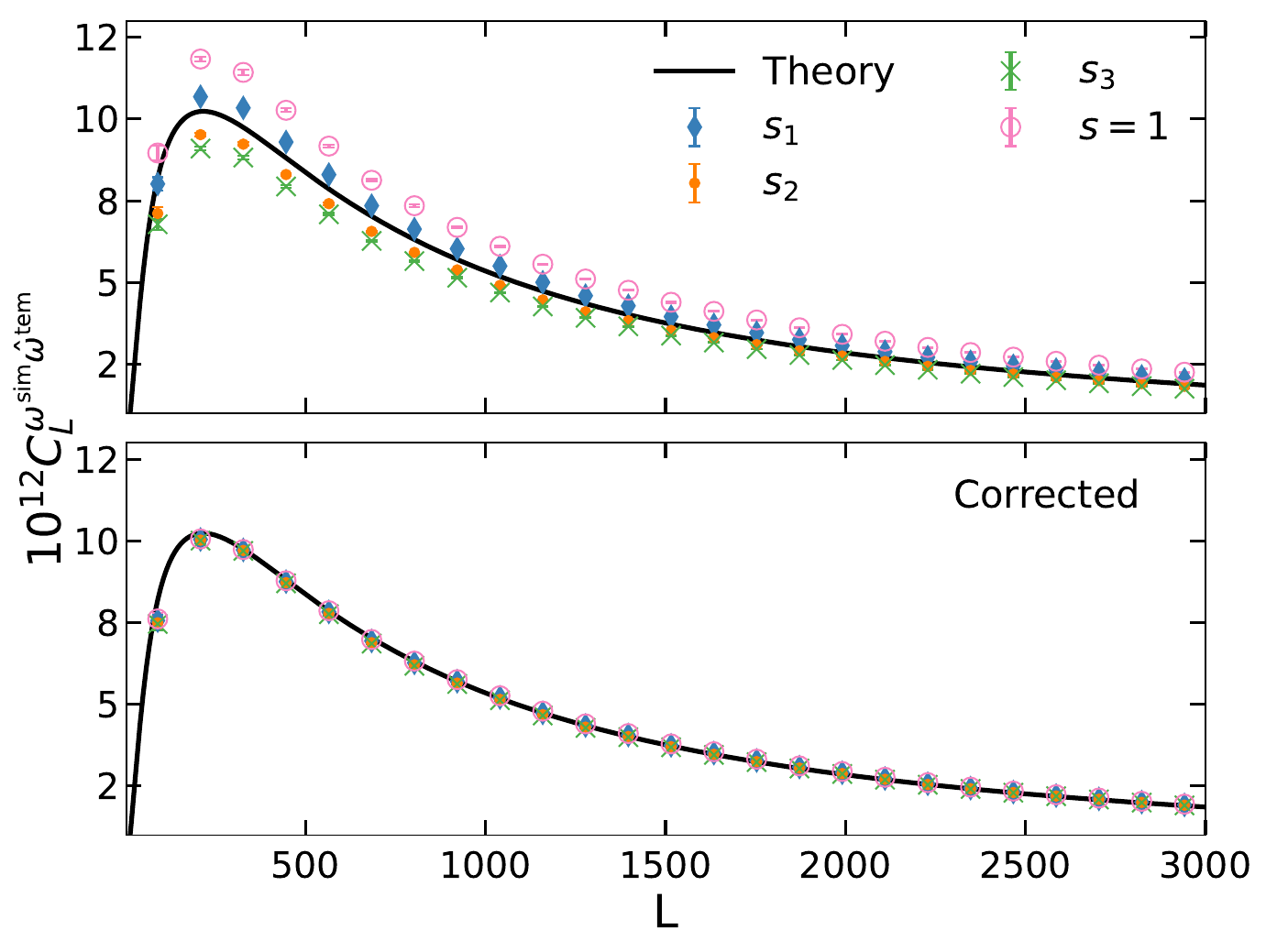}
	\caption{\small{Top panel: The offset in DEMNUni $C^{\omega^{\textrm{sim}}\hat{\omega}^{\textrm{tem}}}$ caused by including $\mu$ in the mock galaxy tracer. Without incorporating $\mu$ into the fiducial modelling of the normalization, $F_L$, the spectrum amplitude is biased with respect to $C^{\omega\omega}$ (solid black). The size and sign of the offset is dependent on modelling choice of $s$. Bottom panel: Using $\boldsymbol{C}_{\textrm{LSS}}^{\mu}$ and $F_L^{\mu}$ within $\hat{\omega}^{\textrm{tem}}$ corrects the misnormalization to within $\leq2\%$ assuming $s$ is known.}}
	\label{fig:magbias}
\end{figure}

Now consider the scenario in which $\mu$ is excluded from fiducial modelling within $\hat{\omega}^{\textrm{tem}}$ but is present in the mock observable $\hat{g}=\tilde{g}+\mu$. Failure to incorporate $\mu$ into $\boldsymbol{C}_{\textrm{LSS}}$ leads to suboptimal filtering of the tracer fields. However, as long as the same covariance is assumed within $F_L$, the cross-spectrum remains unbiased. Nevertheless, bispectra involving $\hat{g}$ now contain additional terms
\begin{equation}\label{eq:mu_bi1}
\langle\omega \hat{g}\hat{a}\rangle = \langle\omega ga\rangle + \langle\omega\mu a\rangle,
\end{equation}

\begin{equation}\label{eq:mu_bi2}
\langle\omega \hat{g}\hat{g}\rangle = \langle\omega gg\rangle + \langle\omega\mu g\rangle + \langle\omega g\mu\rangle + \langle\omega\mu\mu\rangle.
\end{equation}
If $F_L$ does not account for the new $\mu$ bispectra terms then $C^{\hat{\omega}\hat{\omega}^{\textrm{tem}}}$ will be biased. We access the significance of this normalization error by repeating Section \ref{sec:curl_only_test} but now include $\mu$ in mock $\hat{g}$. The top panel of Fig.~\ref{fig:magbias} finds that failure to model $\mu$ in the normalization induces an offset, the magnitude of which depends on $s$. In the extreme example of unphysical $s=1$, this could introduce bias of up to $\sim20\%$. 

Accounting for $\mu$ in $F_L$ is straight forward. One of the fiducial bispectra in Eq.~\eqref{eq:template_norm} must include the additional terms from Eqs.~\eqref{eq:mu_bi1} and \eqref{eq:mu_bi2}. We denote the new normalization as $F_L^{\mu}$. It is important to note that we do not change both bispectra in $F_L^{\mu}$ as we do not intend to modify the template response\footnote{In principle, the response could be optimally modified to use the magnification bias as a lensing signal and boost the $S/N$ of $\hat{\omega}^{\textrm{tem}}$. This possibility is left to future work.} as this would require non-trivial tweaks to the configuration-space estimators, Eqs.~\eqref{eq:E_gamma_full} and \eqref{eq:E_lambda_full}. To include $\mu$ in $\boldsymbol{C}_{\textrm{LSS}}^{\mu}$ the galaxy auto-spectrum is updated to
\begin{equation}\label{eq:gg_mu}
C^{\hat{g}\hat{g}}=C^{gg}+2C^{g\mu}+C^{\mu\mu},
\end{equation} 
and cross spectra similarly become
\begin{equation}\label{eq:mu_cross}
C^{\hat{g}a}=C^{ga}+C^{\mu a}.
\end{equation}
The bottom panel of Fig.~\ref{fig:magbias} demonstrates that these simple corrections to the normalization (and optimal filtering with updated $\boldsymbol{C}_{\textrm{LSS}}^{\mu}$) corrects $\mu$-related misnormalization to within $\leq2\%$.

We again repeat the method to test the effect RSD has on $\hat{\omega}^{\textrm{tem}}$ by adding $g^{\textrm{RSD}}$ to mock $\hat{g}$. We find RSD has a negligible effect, making up $\leq0.2\%$ of the rotation cross-spectrum amplitude. The bias is small because the Kaiser effect is only significant an large scales ($L\lesssim150$) while rotation bispectra are sensitive to smaller scales ($L>100$). Moreover, unlike $\mu$, the RSD effect is restricted to the $\langle\omega gg\rangle$ configuration, which is a sub-dominant contributor to the rotation cross-spectrum signal. RSD is therefore not important for near future lensing rotation measurements.

\subsection{Bias to the bias}

Magnification bias, itself a lensing signal, not only interferes with the template normalization, but also produces additional contributions to quadratic estimator biases, such as $N^{(1)}_{1\kappa}$, $N^{(2)}_{A1}$, $N^{(2)}_{C1}$, etc. To capture these extra biases we follow the approach outlined in Section.~\ref{sec:bias} and compute the total curl free bias, $N^{\textrm{tem},\mu}_L$, in which the magnification bias is included in mock $\hat{g}$. Then the difference with $N^{\textrm{tem}}_L$, Eq.~\eqref{eq:N_L_bias}, isolates the  contribution from $\mu$
\begin{equation}\label{eq:N_L_bias_magbias}
\Delta N^{\mu}_L=N^{\textrm{tem},\mu}_L-N^{\textrm{tem}}_L.
\end{equation}

\begin{figure}[t]
 	\includegraphics[width=\linewidth]{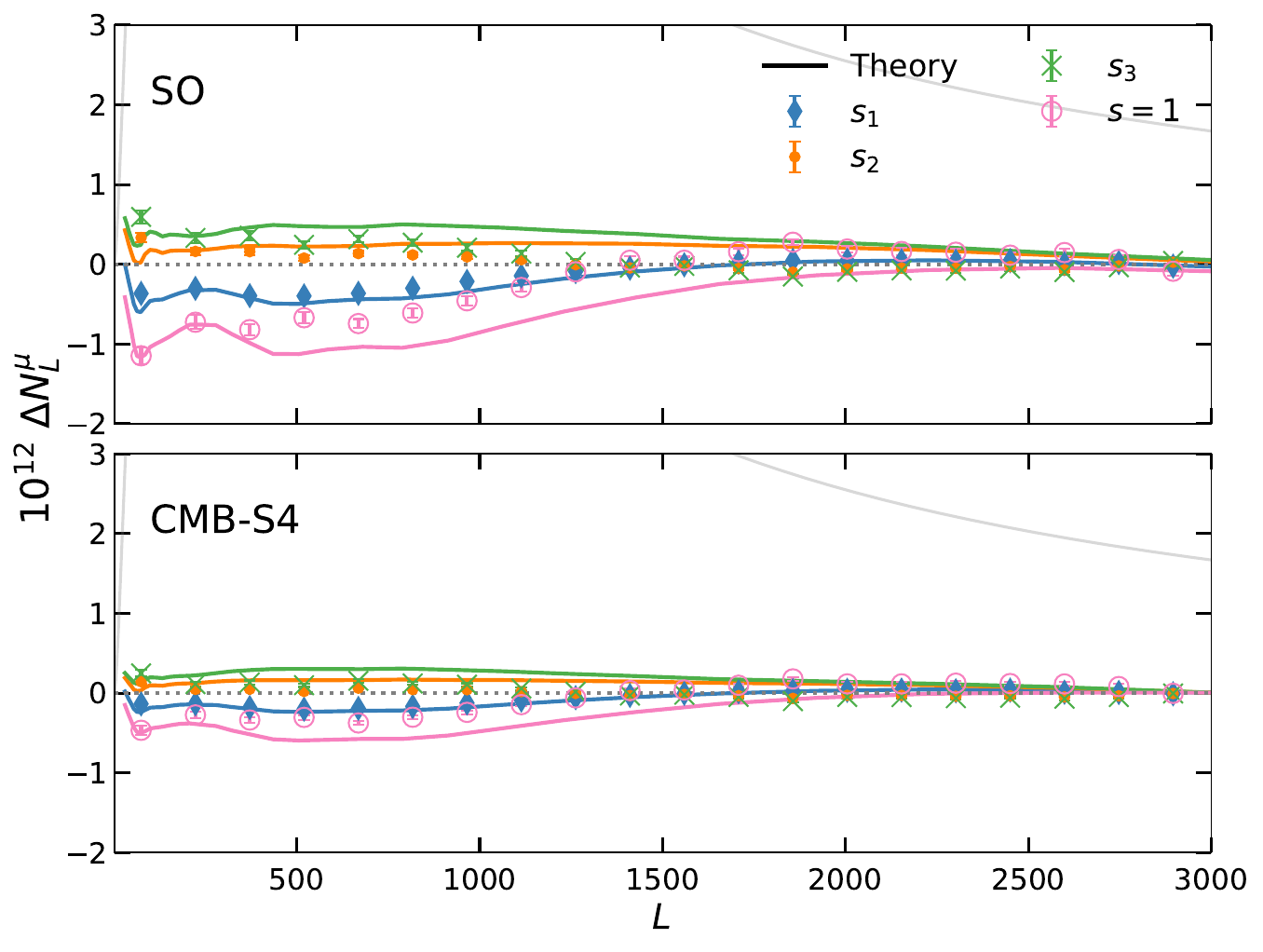}
	\caption{\small{Additional contributions to quadratic estimator biases from contractions involving $\mu$ are described by the $\Delta N^{\mu}$ bias and plotted for different $s$ using the DEMNUni simulation. Analytic predictions for the $\mu$ biases (coloured lines) qualitatively agree with shape and sign but overestimate the amplitude. The biases are small for SO (top panel), CMB-S4 (bottom panel), and all $s$ considered.}}
	\label{fig:magbias_bias}
\end{figure}

When computing $N^{\textrm{tem},\mu}_L$ we update the template modelling to incorporate $\boldsymbol{C}_{\textrm{LSS}}^{\mu}$ and $F_L^{\mu}$\footnote{It should be noted that even with no $\mu$ present, replacing $\boldsymbol{C}_{\textrm{LSS}}$ and $F_L$ with $\boldsymbol{C}_{\textrm{LSS}}^{\mu}$ and $F_L^{\mu}$ results in a small change in the bias, though the effect turns out to be smaller than the $\mu$ induced bias.}, thereby avoiding the normalization issues highlighted in Fig.~\ref{fig:magbias}. Fig.~\ref{fig:magbias_bias} demonstrates through  $\Delta N^{\mu}$ that magnification bias does induce a small, but non-negligible, contribution to the quadratic estimator biases. The size and sign of these new terms are dependent on $s$ and the experimental configuration. 

Theoretical modelling of the additional $\mu$ bias requires simple modification to the analytic $N^{(1)}_{1\kappa}$, $N^{(2)}_{A1}$, and $N^{(2)}_{C1}$ terms on top of the changes to $\boldsymbol{C}_{\textrm{LSS}}^{\mu}$ and $F_L^{\mu}$. Equation 6.6 in \RL requires the two $\kappa$ cross-spectra factors to be updated in accordance with Eq.~\eqref{eq:mu_cross} if either (or both) involve $g$. The same insertion is applied to the cross-spectrum factor for Equation 6.10 in \RL. These new theory predictions qualitatively agree with $\Delta N^{\mu}$ in Fig.~\ref{fig:magbias}. The size and sign of the bias is well modelled at large scales and becomes negligible on small scales. The agreement with theory is linked to $s$ and tentatively worsens with increasing $\mu$, such as for $s=1$. These differences may arise from un-modelled higher order biases (similar to the discrepancies observed earlier in Fig.~\ref{fig:phi_bias}), non-Gaussian $\mu$ biases, or misnormalization leakage. Nonetheless, the $\mu$ biases are small with respect to either $C^{\omega\omega}$ or $N^{\textrm{tem}}$, and in practice can be evaluated and subtracted using simulations.

\section{Curl cosmology}\label{sec:fisher}
Measurement of the CMB curl through a field-rotation cross-spectrum will principally act as a strong consistency check of gravitational lensing techniques, systematics, and the underlying cosmology. Post-Born corrections, which include non-linear contributions from higher-order lensing effects, are neglected in current weak lensing analyses. However, near-future CMB lensing measurements will require modelling of post-born effects at the level of the convergence bispectrum to remain unbiased \cite{Beck:2018wud, Fabbian:2019tik}. The CMB curl cross-spectrum, which could be the first direct detection of a post-Born signal\footnote{One could also argue that observing small $N^{(3/2)}$ lensing bias could be turned into an indirect post-Born detection, because larger $N^{(3/2)}$ is expected if the only contribution is from the LSS bispectrum.}, thereby provides a strong consistency check of the higher order lensing signals important for next-generation CMB lensing observations.

Importantly, $\omega$ captures cosmological information from the underlying matter distribution similar to $\kappa$, as they are both generated by the gravitational potential of matter along the line of sight. However, their radial sensitivities differ: while $\kappa$'s sensitivity to the Weyl potential, $\Psi$, peaks exactly halfway between the source and the observer, $\omega$ does not. Due to the lens-lens coupling, $\omega$ has peak sensitivities at radial distances of one-third and two-thirds between the source and observer. As a result, lens-lens signals are more sensitive to high-$z$ physics than leading-order Born observables, assuming both can be detected with similar significance. Moreover, as a second-order observable, $\omega \propto P_{\Psi\Psi}^2$, whereas Born $\kappa$ only probes $\propto P_{\Psi\Psi}$. Although $\omega$ is approximately $10^{-4}$ times smaller than $\kappa$, its increased sensitivity to $P_{\Psi\Psi}$, especially at high redshift, could make it a valuable complementary cosmological probe. In this context, we explore whether additional cosmological information could be obtained from a near-future curl detection using Fisher forecasts \cite{Tegmark:1997rp}.

\subsection{Fisher parameter forecasts}
We forecast the constraining power of $C^{\hat{\omega}\hat{\omega}^{\textrm{tem}}}$ for the usual $\Lambda$CDM parameters $\alpha,\beta\in\{\theta_*,\Omega_bh^2, \Omega_ch^2, n_s, \sigma_8\}$ assuming flat geometry and Planck 2018 cosmology \cite{PCP2018} -- fiducial values summarized in Table.~\ref{tab:fish}. Note that the reionization optical depth, $\tau$, is not included as it is not constrained by CMB lensing. Instead, we adopt a Gaussian prior from Planck of $\tau=0.054\pm0.007$ and marginalize over it. This prior is appropriate even for general LSS and CMB primary anisotropy forecasts, as future surveys will be insensitive to the largest modes required to constrain $\tau$. Galaxy bias is also included as a parameter for forecasts involving galaxy density so that it is also marginalized. We only model the linear contribution to the galaxy bias.

\begin{table*}[tp]
\resizebox{\linewidth}{!}{
\begin{tabular}{|l|c||c|c|c||c|c|}
    \hline
Parameter & Fiducial & $\hat{\omega}\hat{\omega}$ [1-$\sigma$]& $\hat{\kappa}\hat{\kappa}$ ($+\hat{\omega}\hat{\omega}$) [1-$\sigma$]& $\hat{\kappa}\hat{\kappa}+\textrm{CMB}$ ($+\hat{\omega}\hat{\omega}$) [1-$\sigma$]&$\hat{\omega}\hat{i}\hat{j}$ [1-$\sigma$]&$\textrm{LSS}$ ($+\hat{\omega}\hat{i}\hat{j}$) [1-$\sigma$]\\ \hline \hline
$100\theta_*$ & 1.041084 & 21 & 0.26 (0.24) & 0.000092 (0.000092) & 0.11 & 0.00050 (0.00050) \\
$\Omega_bh^2$ & 0.022383 & 4.4& 0.053 (0.049) & 0.000036 (0.000036) & 0.027 & 0.00012 (0.0012) \\
$\Omega_ch^2$ & 0.12011 & 12&  0.14 (0.13)  & 0.00039 (0.00039) & 0.027 & 0.00013 (0.00013) \\
$n_s$ & 0.9661 & 14& 0.15 (0.15) & 0.0019 (0.0019) & 0.24 & 0.0010 (0.0010) \\
$\sigma_8$ & 0.8124 & 19&  0.24 (0.23)  & 0.0024 (0.0024) & 0.20 & 0.0010 (0.0010) \\\hline
\end{tabular}}
\caption{\small{{Fisher forecasts for the 1-$\sigma$ errors on 5 standard $\Lambda$CDM parameters in a Planck 2018 cosmology \cite{PCP2018} for CMB-S4. The reionization optical depth has Gaussian prior from Planck at $\tau=0.054\pm0.007$ and is marginalized out. The $\hat{\omega}\hat{\omega}$}} results represent the $\omega$ auto-spectrum Fisher assuming reconstruction variance of the cross-spectrum (Eq.~\ref{eq:fish_omega_auto}). The $\hat{\omega}\hat{i}\hat{j}$ results are for the optimal rotation-tracer-tracer bispectrum Fisher (Eq.~\ref{eq:fisher_bi}). For the LSS and $\hat{\omega}\hat{i}\hat{j}$ forecasts we also marginalize over the linear galaxy bias. }\label{tab:fish}
\end{table*}

\begin{figure*}[t]
 	\includegraphics[width=\linewidth]{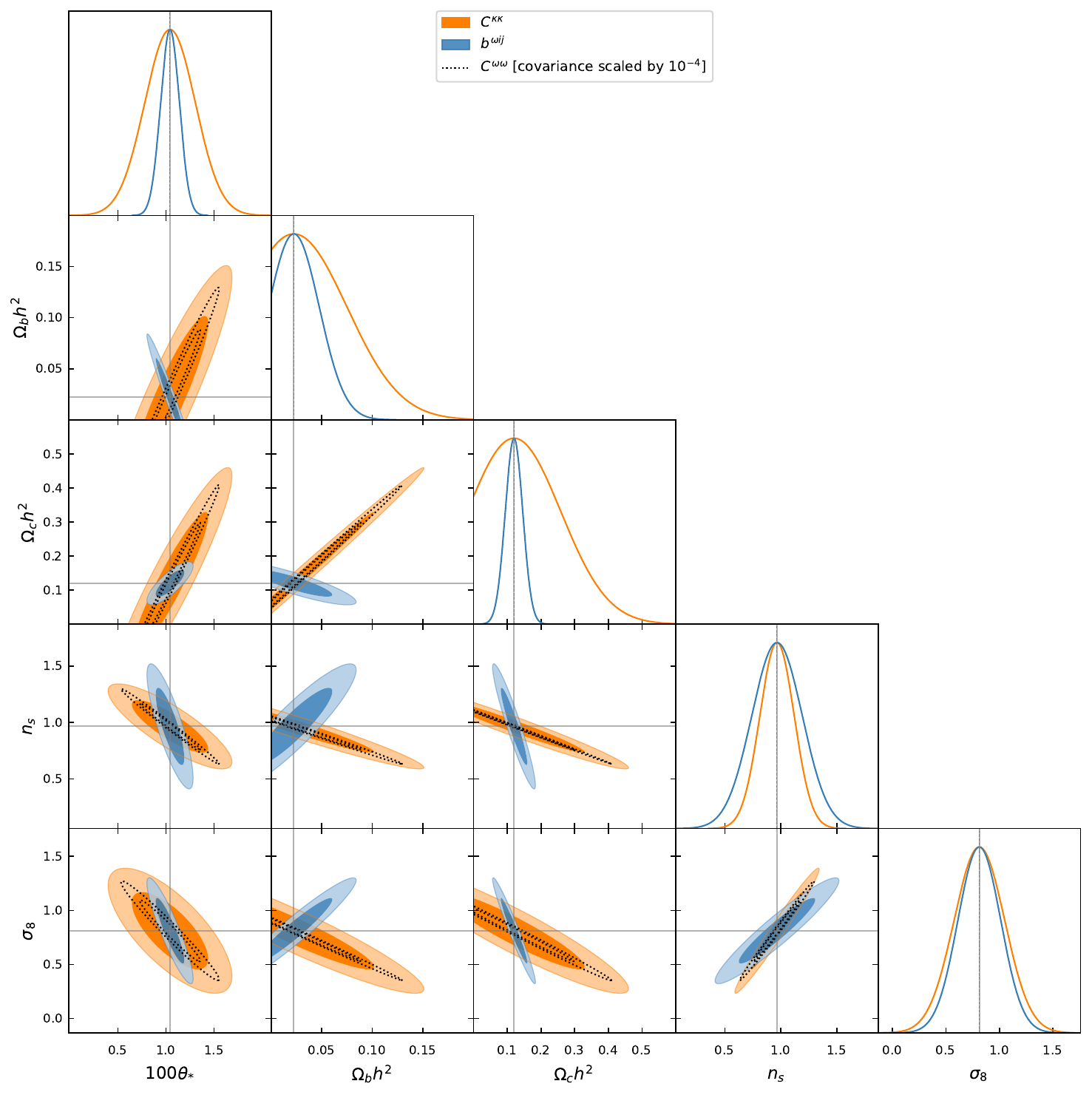}
	\caption{\small{Fisher forecast $68\%$ and $95\%$ integrated probability constraints for CMB-S4 lensing measurements on the standard $\Lambda$CDM parameters. The forecasts for field rotation, $\omega$ (dotted black), assume the auto-spectrum is directly measurable at the variance of the cross-spectrum, $C^{\hat{\omega}\hat{\omega}^{\textrm{tem}}}$. The degeneracy directions are similar to the CMB lensing convergence, $\kappa$ (orange), however the $\omega$ forecasts are $\sim100$ times larger than $\kappa$ (we have scaled the $C^{\omega\omega}$ constraints by factor of 100 to fit onto within the 2D contour panels, and to avoid confusion we do not project them onto the 1D panels). Although, the optimal rotation-tracer-tracer bispectrum constraints (blue) naively appear to break degeneracies with $\kappa$, all of the constraining power comes from the quadratic tracer part of the bispectrum and not from $\omega$. The fiducial values are represented by the grey lines.}}
	\label{fig:triangle}
\end{figure*}

We consider two conceptually different Fisher information matrices in an attempt to give richer understanding to the cosmological gain from measuring the curl cross-spectrum. First, we assume the field-rotation auto-spectrum, $C^{\hat{\omega}\hat{\omega}}$, is directly measurable from a CMB lensing reconstruction but with the lower variance of the template cross-spectrum, $\langle C^{\hat{\omega}\hat{\omega}}\rangle=\langle C^{\hat{\omega}\hat{\omega}^{\textrm{tem}}}\rangle$. The corresponding Fisher
\begin{equation}\label{eq:fish_omega_auto}
F^{\hat{\omega}\hat{\omega}}_{\alpha\beta}=2f_{\textrm{sky}}\int LdL\frac{C^{\omega\omega,\alpha}_LC^{\omega\omega,\beta}_L}{C^{\omega\omega}_LN^{\omega}_0(L)F_L^{-1}},
\end{equation}
probes the cosmological information gained if it were possible to observe $\omega$ directly at the equivalent sensitivity of the cross-spectrum. Note that the commas in Eq.~\eqref{eq:fish_omega_auto} represent partial derivatives. For context, the results are compared to the CMB $\kappa$ Fisher
\begin{equation}
F^{\hat{\kappa}\hat{\kappa}}_{\alpha\beta}=f_{\textrm{sky}}\int LdL\frac{C^{\kappa\kappa,\alpha}_LC^{\kappa\kappa,\beta}_L}{\left[C^{\kappa\kappa}_L+N^{\kappa}_0(L)\right]^2}.
\end{equation}
Fig.~\ref{fig:triangle} shows the Fisher contours for the 5 $\Lambda$CDM parameters at the noise sensitivity of CMB-S4. The degeneracy directions appear very similar for $\kappa$ and $\omega$, but the size of the errors on $\omega$ are $\sim100$ times greater than those for $\kappa$ at the same noise level. The full joint constraints combining $\kappa$ and $\omega$ were achieved through adding both matrices, and the results are summarized in Table \ref{tab:fish}. Joint $\omega$ and $\kappa$ does improve several standard cosmological parameters constraints at $\sim5\%$ compared to $\kappa$ alone for CMB-S4. However, when adding in CMB information from primary temperature and polarization anisotropies
\cite{Hannestad:2006as}
\begin{multline}\label{eq:fisher_cmb}
F^{\textrm{CMB}}_{\alpha\beta}=f_{\textrm{sky}}\int LdL \\
\times\textrm{Tr}\left[\boldsymbol{C}_{\textrm{CMB}}^{-1}(L)\frac{\partial \boldsymbol{C}^{\textrm{CMB}}_L}{\partial \alpha}\boldsymbol{C}_{\textrm{CMB}}^{-1}(L)\frac{\partial \boldsymbol{C}^{\textrm{CMB}}_L}{\partial \beta}\right],
\end{multline}
any improvements vanish. Hence, the curl would fail to contribute any cosmological information for CMB experiments --- in the standard picture --- even if it were directly observable at the reduced variance levels of the template cross-spectrum. 

Measurement of the $\omega$ auto-spectrum remains challenging for the near-future, so our focus returns to the detection of $C^{\hat{\omega}\hat{\omega}^{\textrm{tem}}}$. This cross-spectrum is really an optimal combination of cross-bisectra, $b^{\hat{\omega}\hat{i}\hat{j}}$ with different density tracers, $\hat{i},\hat{j}\in\{\hat{\kappa},\hat{g},\hat{I}\}$. Thus, instead of pondering the cosmological gain of measuring $\omega$ at the noise levels of the cross-spectrum, an alternative approach asks how much information is contained in the optimal rotation-tracer-tracer bispectrum. This is the second forecasting method we consider, and the optimal bispectrum Fisher is defined by \cite{Lewis:2011fk} 
\begin{multline}\label{eq:fisher_bi}
F^{\hat{\omega}\hat{i}\hat{j}}_{\alpha\beta}=\frac{f_{\textrm{sky}}}{2\pi}\int d^2\boldsymbol{L}\left[C^{\omega\omega}_L+N^{\omega}_0(L)\right]^{-1}\\
\times\int \frac{d^2\boldsymbol{l}_1}{(2\pi)^2} 
(\boldsymbol{C}^{-1}_{\textrm{LSS}})^{ip}_{l_1}
(\boldsymbol{C}^{-1}_{\textrm{LSS}})^{jq}_{l_2}b^{\omega ij,\alpha}_{(-\boldsymbol{L})\boldsymbol{l}_1\boldsymbol{l}_2}b^{\omega pq,\beta}_{(-\boldsymbol{L})\boldsymbol{l}_1\boldsymbol{l}_2}.
\end{multline}
Both Fig.~\ref{fig:triangle} and Table \ref{tab:fish} show that the optimal rotation bispectrum does constrain information at similar level to $C^{\kappa\kappa}$ for CMB-S4 noise levels, which is impressive considering $\kappa$ will be detectable at $100$s of $\sigma$. However, considering $\boldsymbol{F}^{\hat{\omega}\hat{i}\hat{j}}$ alone is misleading. We are interested in quantifying the {\it new} information $b^{\hat{\omega}\hat{i}\hat{j}}$ adds to constraints on standard cosmology parameters, so it is not enough to only state the results from Eq.~\eqref{eq:fisher_bi}. To illustrate this point, consider that the rotation-tracer-tracer bispectrum is sensitive to the galaxy bias, $b_g$. While, the amplitude of $b^{\hat{\omega}\hat{i}\hat{j}}$ is indeed dependent on galaxy bias, $\omega$ is not. So the information on $b_g$ from the bispectrum is coming from the tracer legs within the three point contraction and not from $\omega$. It would therefore be misleading to claim either $b^{\hat{\omega}\hat{i}\hat{j}}$ or $C^{\hat{\omega}\hat{\omega}^{\textrm{tem}}}$ provides new information on $b_g$ despite being sensitive to it. Just as with $b_g$, care must be maintained when evaluating the sensitivity of the rotation bispectrum to other cosmological parameters. To properly ascertain the additional information gained from measuring $b^{\hat{\omega}\hat{i}\hat{j}}$, a baseline understanding of the constraining power of the various auto and cross LSS spectra is required. The Fisher matrix for the LSS density tracers is
\begin{multline}\label{eq:fisher_lss}
F^{\textrm{LSS}}_{\alpha\beta}=f_{\textrm{sky}}\int LdL \\
\times\textrm{Tr}\left[\boldsymbol{C}_{\textrm{LSS}}^{-1}(L)\frac{\partial \boldsymbol{C}^{\textrm{LSS}}_L}{\partial \alpha}\boldsymbol{C}_{\textrm{LSS}}^{-1}(L)\frac{\partial \boldsymbol{C}^{\textrm{LSS}}_L}{\partial \beta}\right].
\end{multline}
Note that it would be equally important to also include information from the $\omega$ auto-spectrum if it were actually detectable. Hence $\boldsymbol{F}^{\hat{\omega}\hat{i}\hat{j}}$ was simply added to $\boldsymbol{F}^{\textrm{LSS}}$, and Table.~\ref{tab:fish} shows that the rotation bispectrum does not improve constraints on any parameter when prior information from the tracer $C_{\ell}$s is included. Therefore, the $b^{\hat{\omega}\hat{i}\hat{j}}$ statistic does not provide new information, and $\omega$ specifically will not improve standard $\Lambda$ cosmology constraints. This is consistent with the weak $\boldsymbol{F}^{\hat{\omega}\hat{\omega}}$ forecasts. Even though $\omega$ adds little direct information, the parity odd nature of the rotation-tracer-tracer bispectra would have very different systematics to a full LSS analysis, thus $C^{\hat{\omega}\hat{\omega}^{\textrm{tem}}}$ could still be a useful consistency statistic, and it also has different redshift sensitivity to extended models.

\providecommand{\aj}{Astron. J. }\providecommand{\apj}{ApJ
  }\providecommand{\apjl}{ApJ
  }\providecommand{\mnras}{MNRAS}\providecommand{\prl}{PRL}\providecommand{\prd}{PRD}\providecommand{\jcap}{JCAP}\providecommand{\aap}{A\&A}

\end{document}

%% file: texbase/macros.tex
\newcommand{\la}{\langle}
\newcommand{\ra}{\rangle}

\newcommand{\mksym}[1]{\ifmmode {\rm #1}\else #1\fi}

\providecommand{\gea}{\ga}

\providecommand{\agt}{\gea}
\providecommand{\text}[1]{\rm{#1}}

\newcommand{\grad}{\nabla}

\newcommand{\begm}{\begin{pmatrix}}
\newcommand{\enm}{\end{pmatrix}}

\newcommand\ba{\begin{eqnarray}}
\newcommand\ea{\end{eqnarray}}
\newcommand\bea{\begin{eqnarray}}
\newcommand\eea{\end{eqnarray}}

\newcommand\be{\begin{equation}}
\newcommand\ee{\end{equation}}

%% file: curl2.bbl
\begin{thebibliography}{82}%
\makeatletter
\providecommand \@ifxundefined [1]{%
 \@ifx{#1\undefined}
}%
\providecommand \@ifnum [1]{%
 \ifnum #1\expandafter \@firstoftwo
 \else \expandafter \@secondoftwo
 \fi
}%
\providecommand \@ifx [1]{%
 \ifx #1\expandafter \@firstoftwo
 \else \expandafter \@secondoftwo
 \fi
}%
\providecommand \natexlab [1]{#1}%
\providecommand \enquote  [1]{``#1''}%
\providecommand \bibnamefont  [1]{#1}%
\providecommand \bibfnamefont [1]{#1}%
\providecommand \citenamefont [1]{#1}%
\providecommand \href@noop [0]{\@secondoftwo}%
\providecommand \href [0]{\begingroup \@sanitize@url \@href}%
\providecommand \@href[1]{\@@startlink{#1}\@@href}%
\providecommand \@@href[1]{\endgroup#1\@@endlink}%
\providecommand \@sanitize@url [0]{\catcode `\\12\catcode `\$12\catcode
  `\&12\catcode `\#12\catcode `\^12\catcode `\_12\catcode `\%12\relax}%
\providecommand \@@startlink[1]{}%
\providecommand \@@endlink[0]{}%
\providecommand \url  [0]{\begingroup\@sanitize@url \@url }%
\providecommand \@url [1]{\endgroup\@href {#1}{\urlprefix }}%
\providecommand \urlprefix  [0]{URL }%
\providecommand \Eprint [0]{\href }%
\providecommand \doibase [0]{https://doi.org/}%
\providecommand \selectlanguage [0]{\@gobble}%
\providecommand \bibinfo  [0]{\@secondoftwo}%
\providecommand \bibfield  [0]{\@secondoftwo}%
\providecommand \translation [1]{[#1]}%
\providecommand \BibitemOpen [0]{}%
\providecommand \bibitemStop [0]{}%
\providecommand \bibitemNoStop [0]{.\EOS\space}%
\providecommand \EOS [0]{\spacefactor3000\relax}%
\providecommand \BibitemShut  [1]{\csname bibitem#1\endcsname}%
\let\auto@bib@innerbib\@empty
%</preamble>
\bibitem [{\citenamefont {Lewis}\ and\ \citenamefont
  {Challinor}(2006)}]{Lewis:2006fu}%
  \BibitemOpen
  \bibfield  {author} {\bibinfo {author} {\bibfnamefont {A.}~\bibnamefont
  {Lewis}}\ and\ \bibinfo {author} {\bibfnamefont {A.}~\bibnamefont
  {Challinor}},\ }\bibfield  {title} {\bibinfo {title} {{Weak gravitational
  lensing of the CMB}},\ }\href {https://doi.org/10.1016/j.physrep.2006.03.002}
  {\bibfield  {journal} {\bibinfo  {journal} {Phys. Rept.}\ }\textbf {\bibinfo
  {volume} {429}},\ \bibinfo {pages} {1} (\bibinfo {year} {2006})},\ \Eprint
  {https://arxiv.org/abs/astro-ph/0601594} {arXiv:astro-ph/0601594 [astro-ph]}
  \BibitemShut {NoStop}%
%%CITATION = ASTRO-PH/0601594;%%
\bibitem [{\citenamefont {Ade}\ \emph {et~al.}(2016{\natexlab{a}})\citenamefont
  {Ade} \emph {et~al.}}]{BICEP2:2016rpt}%
  \BibitemOpen
  \bibfield  {author} {\bibinfo {author} {\bibfnamefont {P.~A.~R.}\
  \bibnamefont {Ade}} \emph {et~al.} (\bibinfo {collaboration} {BICEP2, Keck
  Array}),\ }\bibfield  {title} {\bibinfo {title} {{BICEP2 / Keck Array VIII:
  Measurement of gravitational lensing from large-scale B-mode polarization}},\
  }\href {https://doi.org/10.3847/1538-4357/833/2/228} {\bibfield  {journal}
  {\bibinfo  {journal} {Astrophys. J.}\ }\textbf {\bibinfo {volume} {833}},\
  \bibinfo {pages} {228} (\bibinfo {year} {2016}{\natexlab{a}})},\ \Eprint
  {https://arxiv.org/abs/1606.01968} {arXiv:1606.01968 [astro-ph.CO]}
  \BibitemShut {NoStop}%
\bibitem [{\citenamefont {Adachi}\ \emph {et~al.}(2020)\citenamefont {Adachi}
  \emph {et~al.}}]{POLARBEAR:2019snn}%
  \BibitemOpen
  \bibfield  {author} {\bibinfo {author} {\bibfnamefont {S.}~\bibnamefont
  {Adachi}} \emph {et~al.} (\bibinfo {collaboration} {POLARBEAR}),\ }\bibfield
  {title} {\bibinfo {title} {{Internal delensing of Cosmic Microwave Background
  polarization $B$-modes with the POLARBEAR experiment}},\ }\href
  {https://doi.org/10.1103/PhysRevLett.124.131301} {\bibfield  {journal}
  {\bibinfo  {journal} {Phys. Rev. Lett.}\ }\textbf {\bibinfo {volume} {124}},\
  \bibinfo {pages} {131301} (\bibinfo {year} {2020})},\ \Eprint
  {https://arxiv.org/abs/1909.13832} {arXiv:1909.13832 [astro-ph.CO]}
  \BibitemShut {NoStop}%
\bibitem [{\citenamefont {Carron}\ \emph {et~al.}(2022)\citenamefont {Carron},
  \citenamefont {Mirmelstein},\ and\ \citenamefont {Lewis}}]{Carron:2022eyg}%
  \BibitemOpen
  \bibfield  {author} {\bibinfo {author} {\bibfnamefont {J.}~\bibnamefont
  {Carron}}, \bibinfo {author} {\bibfnamefont {M.}~\bibnamefont
  {Mirmelstein}},\ and\ \bibinfo {author} {\bibfnamefont {A.}~\bibnamefont
  {Lewis}},\ }\bibfield  {title} {\bibinfo {title} {{CMB lensing from Planck
  PR4~maps}},\ }\href {https://doi.org/10.1088/1475-7516/2022/09/039}
  {\bibfield  {journal} {\bibinfo  {journal} {JCAP}\ }\textbf {\bibinfo
  {volume} {09}},\ \bibinfo {pages} {039}},\ \Eprint
  {https://arxiv.org/abs/2206.07773} {arXiv:2206.07773 [astro-ph.CO]}
  \BibitemShut {NoStop}%
\bibitem [{\citenamefont {Qu}\ \emph {et~al.}(2024)\citenamefont {Qu} \emph
  {et~al.}}]{ACT:2023dou}%
  \BibitemOpen
  \bibfield  {author} {\bibinfo {author} {\bibfnamefont {F.~J.}\ \bibnamefont
  {Qu}} \emph {et~al.} (\bibinfo {collaboration} {ACT}),\ }\bibfield  {title}
  {\bibinfo {title} {{The Atacama Cosmology Telescope: A Measurement of the DR6
  CMB Lensing Power Spectrum and Its Implications for Structure Growth}},\
  }\href {https://doi.org/10.3847/1538-4357/acfe06} {\bibfield  {journal}
  {\bibinfo  {journal} {Astrophys. J.}\ }\textbf {\bibinfo {volume} {962}},\
  \bibinfo {pages} {112} (\bibinfo {year} {2024})},\ \Eprint
  {https://arxiv.org/abs/2304.05202} {arXiv:2304.05202 [astro-ph.CO]}
  \BibitemShut {NoStop}%
\bibitem [{\citenamefont {Pan}\ \emph {et~al.}(2023)\citenamefont {Pan} \emph
  {et~al.}}]{SPT:2023jql}%
  \BibitemOpen
  \bibfield  {author} {\bibinfo {author} {\bibfnamefont {Z.}~\bibnamefont
  {Pan}} \emph {et~al.} (\bibinfo {collaboration} {SPT}),\ }\bibfield  {title}
  {\bibinfo {title} {{Measurement of gravitational lensing of the cosmic
  microwave background using SPT-3G 2018 data}},\ }\href
  {https://doi.org/10.1103/PhysRevD.108.122005} {\bibfield  {journal} {\bibinfo
   {journal} {Phys. Rev. D}\ }\textbf {\bibinfo {volume} {108}},\ \bibinfo
  {pages} {122005} (\bibinfo {year} {2023})},\ \Eprint
  {https://arxiv.org/abs/2308.11608} {arXiv:2308.11608 [astro-ph.CO]}
  \BibitemShut {NoStop}%
\bibitem [{\citenamefont {Aghanim}\ \emph
  {et~al.}(2020{\natexlab{a}})\citenamefont {Aghanim} \emph {et~al.}}]{PL2018}%
  \BibitemOpen
  \bibfield  {author} {\bibinfo {author} {\bibfnamefont {N.}~\bibnamefont
  {Aghanim}} \emph {et~al.} (\bibinfo {collaboration} {Planck}),\ }\bibfield
  {title} {\bibinfo {title} {{Planck 2018 results. VIII. Gravitational
  lensing}},\ }\href {https://doi.org/10.1051/0004-6361/201833886} {\bibfield
  {journal} {\bibinfo  {journal} {\aap}\ }\textbf {\bibinfo {volume} {641}},\
  \bibinfo {pages} {A8} (\bibinfo {year} {2020}{\natexlab{a}})},\ \Eprint
  {https://arxiv.org/abs/1807.06210} {arXiv:1807.06210 [astro-ph.CO]}
  \BibitemShut {NoStop}%
\bibitem [{\citenamefont {Book}\ \emph {et~al.}(2012)\citenamefont {Book},
  \citenamefont {Kamionkowski},\ and\ \citenamefont {Schmidt}}]{Book:2011dz}%
  \BibitemOpen
  \bibfield  {author} {\bibinfo {author} {\bibfnamefont {L.}~\bibnamefont
  {Book}}, \bibinfo {author} {\bibfnamefont {M.}~\bibnamefont {Kamionkowski}},\
  and\ \bibinfo {author} {\bibfnamefont {F.}~\bibnamefont {Schmidt}},\
  }\bibfield  {title} {\bibinfo {title} {{Lensing of 21-cm Fluctuations by
  Primordial Gravitational Waves}},\ }\href
  {https://doi.org/10.1103/PhysRevLett.108.211301} {\bibfield  {journal}
  {\bibinfo  {journal} {Phys. Rev. Lett.}\ }\textbf {\bibinfo {volume} {108}},\
  \bibinfo {pages} {211301} (\bibinfo {year} {2012})},\ \Eprint
  {https://arxiv.org/abs/1112.0567} {arXiv:1112.0567 [astro-ph.CO]}
  \BibitemShut {NoStop}%
\bibitem [{\citenamefont {Yamauchi}\ \emph {et~al.}(2012)\citenamefont
  {Yamauchi}, \citenamefont {Namikawa},\ and\ \citenamefont
  {Taruya}}]{Yamauchi:2012bc}%
  \BibitemOpen
  \bibfield  {author} {\bibinfo {author} {\bibfnamefont {D.}~\bibnamefont
  {Yamauchi}}, \bibinfo {author} {\bibfnamefont {T.}~\bibnamefont {Namikawa}},\
  and\ \bibinfo {author} {\bibfnamefont {A.}~\bibnamefont {Taruya}},\
  }\bibfield  {title} {\bibinfo {title} {{Weak lensing generated by vector
  perturbations and detectability of cosmic strings}},\ }\href
  {https://doi.org/10.1088/1475-7516/2012/10/030} {\bibfield  {journal}
  {\bibinfo  {journal} {JCAP}\ }\textbf {\bibinfo {volume} {10}},\ \bibinfo
  {pages} {030}},\ \Eprint {https://arxiv.org/abs/1205.2139} {arXiv:1205.2139
  [astro-ph.CO]} \BibitemShut {NoStop}%
\bibitem [{\citenamefont {Rotti}\ and\ \citenamefont
  {Souradeep}(2012)}]{Rotti:2011aa}%
  \BibitemOpen
  \bibfield  {author} {\bibinfo {author} {\bibfnamefont {A.}~\bibnamefont
  {Rotti}}\ and\ \bibinfo {author} {\bibfnamefont {T.}~\bibnamefont
  {Souradeep}},\ }\bibfield  {title} {\bibinfo {title} {{A New Window into
  Stochastic Gravitational Wave Background}},\ }\href
  {https://doi.org/10.1103/PhysRevLett.109.221301} {\bibfield  {journal}
  {\bibinfo  {journal} {Phys. Rev. Lett.}\ }\textbf {\bibinfo {volume} {109}},\
  \bibinfo {pages} {221301} (\bibinfo {year} {2012})},\ \Eprint
  {https://arxiv.org/abs/1112.1689} {arXiv:1112.1689 [astro-ph.CO]}
  \BibitemShut {NoStop}%
\bibitem [{\citenamefont {Saga}\ \emph {et~al.}(2015)\citenamefont {Saga},
  \citenamefont {Yamauchi},\ and\ \citenamefont {Ichiki}}]{Saga:2015apa}%
  \BibitemOpen
  \bibfield  {author} {\bibinfo {author} {\bibfnamefont {S.}~\bibnamefont
  {Saga}}, \bibinfo {author} {\bibfnamefont {D.}~\bibnamefont {Yamauchi}},\
  and\ \bibinfo {author} {\bibfnamefont {K.}~\bibnamefont {Ichiki}},\
  }\bibfield  {title} {\bibinfo {title} {{Weak lensing induced by second-order
  vector mode}},\ }\href {https://doi.org/10.1103/PhysRevD.92.063533}
  {\bibfield  {journal} {\bibinfo  {journal} {Phys. Rev. D}\ }\textbf {\bibinfo
  {volume} {92}},\ \bibinfo {pages} {063533} (\bibinfo {year} {2015})},\
  \Eprint {https://arxiv.org/abs/1505.02774} {arXiv:1505.02774 [astro-ph.CO]}
  \BibitemShut {NoStop}%
\bibitem [{\citenamefont {Cooray}\ and\ \citenamefont
  {Hu}(2002)}]{Cooray:2002mj}%
  \BibitemOpen
  \bibfield  {author} {\bibinfo {author} {\bibfnamefont {A.}~\bibnamefont
  {Cooray}}\ and\ \bibinfo {author} {\bibfnamefont {W.}~\bibnamefont {Hu}},\
  }\bibfield  {title} {\bibinfo {title} {{Second order corrections to weak
  lensing by large scale structure}},\ }\href {https://doi.org/10.1086/340892}
  {\bibfield  {journal} {\bibinfo  {journal} {\apj}\ }\textbf {\bibinfo
  {volume} {574}},\ \bibinfo {pages} {19} (\bibinfo {year} {2002})},\ \Eprint
  {https://arxiv.org/abs/astro-ph/0202411} {arXiv:astro-ph/0202411 [astro-ph]}
  \BibitemShut {NoStop}%
%%CITATION = ASTRO-PH/0202411;%%
\bibitem [{\citenamefont {Krause}\ and\ \citenamefont
  {Hirata}(2010)}]{Krause:2009yr}%
  \BibitemOpen
  \bibfield  {author} {\bibinfo {author} {\bibfnamefont {E.}~\bibnamefont
  {Krause}}\ and\ \bibinfo {author} {\bibfnamefont {C.~M.}\ \bibnamefont
  {Hirata}},\ }\bibfield  {title} {\bibinfo {title} {{Weak lensing power
  spectra for precision cosmology: Multiple-deflection, reduced shear and
  lensing bias corrections}},\ }\href
  {https://doi.org/10.1051/0004-6361/200913524} {\bibfield  {journal} {\bibinfo
   {journal} {\aap}\ }\textbf {\bibinfo {volume} {523}},\ \bibinfo {pages}
  {A28} (\bibinfo {year} {2010})},\ \Eprint {https://arxiv.org/abs/0910.3786}
  {arXiv:0910.3786 [astro-ph.CO]} \BibitemShut {NoStop}%
%%CITATION = ARXIV:0910.3786;%%
\bibitem [{\citenamefont {Pratten}\ and\ \citenamefont
  {Lewis}(2016)}]{Pratten:2016dsm}%
  \BibitemOpen
  \bibfield  {author} {\bibinfo {author} {\bibfnamefont {G.}~\bibnamefont
  {Pratten}}\ and\ \bibinfo {author} {\bibfnamefont {A.}~\bibnamefont
  {Lewis}},\ }\bibfield  {title} {\bibinfo {title} {{Impact of post-Born
  lensing on the CMB}},\ }\href {https://doi.org/10.1088/1475-7516/2016/08/047}
  {\bibfield  {journal} {\bibinfo  {journal} {\jcap}\ }\textbf {\bibinfo
  {volume} {1608}},\ \bibinfo {pages} {047} (\bibinfo {year} {2016})},\ \Eprint
  {https://arxiv.org/abs/1605.05662} {arXiv:1605.05662 [astro-ph.CO]}
  \BibitemShut {NoStop}%
%%CITATION = ARXIV:1605.05662;%%
\bibitem [{\citenamefont {Robertson}\ and\ \citenamefont
  {Lewis}(2023)}]{Robertson:2023xkg}%
  \BibitemOpen
  \bibfield  {author} {\bibinfo {author} {\bibfnamefont {M.}~\bibnamefont
  {Robertson}}\ and\ \bibinfo {author} {\bibfnamefont {A.}~\bibnamefont
  {Lewis}},\ }\bibfield  {title} {\bibinfo {title} {{How to detect lensing
  rotation}},\ }\href {https://doi.org/10.1088/1475-7516/2023/08/048}
  {\bibfield  {journal} {\bibinfo  {journal} {JCAP}\ }\textbf {\bibinfo
  {volume} {08}},\ \bibinfo {pages} {048}},\ \Eprint
  {https://arxiv.org/abs/2303.13313} {arXiv:2303.13313 [astro-ph.CO]}
  \BibitemShut {NoStop}%
\bibitem [{\citenamefont {Aguirre}\ \emph {et~al.}(2019)\citenamefont {Aguirre}
  \emph {et~al.}}]{Ade:2018sbj}%
  \BibitemOpen
  \bibfield  {author} {\bibinfo {author} {\bibfnamefont {J.}~\bibnamefont
  {Aguirre}} \emph {et~al.} (\bibinfo {collaboration} {Simons Observatory}),\
  }\bibfield  {title} {\bibinfo {title} {{The Simons Observatory: Science goals
  and forecasts}},\ }\href {https://doi.org/10.1088/1475-7516/2019/02/056}
  {\bibfield  {journal} {\bibinfo  {journal} {JCAP}\ }\textbf {\bibinfo
  {volume} {1902}},\ \bibinfo {pages} {056}},\ \Eprint
  {https://arxiv.org/abs/1808.07445} {arXiv:1808.07445 [astro-ph.CO]}
  \BibitemShut {NoStop}%
%%CITATION = ARXIV:1808.07445;%%
\bibitem [{\citenamefont {Abazajian}\ \emph {et~al.}(2019)\citenamefont
  {Abazajian} \emph {et~al.}}]{Abazajian:2019eic}%
  \BibitemOpen
  \bibfield  {author} {\bibinfo {author} {\bibfnamefont {K.}~\bibnamefont
  {Abazajian}} \emph {et~al.},\ }\bibfield  {title} {\bibinfo {title} {{CMB-S4
  Science Case, Reference Design, and Project Plan}},\ }\href@noop {} {\
  (\bibinfo {year} {2019})},\ \Eprint {https://arxiv.org/abs/1907.04473}
  {arXiv:1907.04473 [astro-ph.IM]} \BibitemShut {NoStop}%
\bibitem [{\citenamefont {Carbone}\ \emph {et~al.}(2016)\citenamefont
  {Carbone}, \citenamefont {Petkova},\ and\ \citenamefont
  {Dolag}}]{Carbone:2016nzj}%
  \BibitemOpen
  \bibfield  {author} {\bibinfo {author} {\bibfnamefont {C.}~\bibnamefont
  {Carbone}}, \bibinfo {author} {\bibfnamefont {M.}~\bibnamefont {Petkova}},\
  and\ \bibinfo {author} {\bibfnamefont {K.}~\bibnamefont {Dolag}},\ }\bibfield
   {title} {\bibinfo {title} {{DEMNUni: ISW, Rees-Sciama, and weak-lensing in
  the presence of massive neutrinos}},\ }\href
  {https://doi.org/10.1088/1475-7516/2016/07/034} {\bibfield  {journal}
  {\bibinfo  {journal} {JCAP}\ }\textbf {\bibinfo {volume} {07}},\ \bibinfo
  {pages} {034}},\ \Eprint {https://arxiv.org/abs/1605.02024} {arXiv:1605.02024
  [astro-ph.CO]} \BibitemShut {NoStop}%
\bibitem [{\citenamefont {Castorina}\ \emph {et~al.}(2015)\citenamefont
  {Castorina}, \citenamefont {Carbone}, \citenamefont {Bel}, \citenamefont
  {Sefusatti},\ and\ \citenamefont {Dolag}}]{Castorina:2015bma}%
  \BibitemOpen
  \bibfield  {author} {\bibinfo {author} {\bibfnamefont {E.}~\bibnamefont
  {Castorina}}, \bibinfo {author} {\bibfnamefont {C.}~\bibnamefont {Carbone}},
  \bibinfo {author} {\bibfnamefont {J.}~\bibnamefont {Bel}}, \bibinfo {author}
  {\bibfnamefont {E.}~\bibnamefont {Sefusatti}},\ and\ \bibinfo {author}
  {\bibfnamefont {K.}~\bibnamefont {Dolag}},\ }\bibfield  {title} {\bibinfo
  {title} {{DEMNUni: The clustering of large-scale structures in the presence
  of massive neutrinos}},\ }\href
  {https://doi.org/10.1088/1475-7516/2015/07/043} {\bibfield  {journal}
  {\bibinfo  {journal} {JCAP}\ }\textbf {\bibinfo {volume} {07}},\ \bibinfo
  {pages} {043}},\ \Eprint {https://arxiv.org/abs/1505.07148} {arXiv:1505.07148
  [astro-ph.CO]} \BibitemShut {NoStop}%
\bibitem [{\citenamefont {Omori}(2024)}]{Omori:2022uox}%
  \BibitemOpen
  \bibfield  {author} {\bibinfo {author} {\bibfnamefont {Y.}~\bibnamefont
  {Omori}},\ }\bibfield  {title} {\bibinfo {title} {{Agora: Multi-Component
  Simulation for Cross-Survey Science}},\ }\href
  {https://doi.org/10.1093/mnras/stae1031} {\bibfield  {journal} {\bibinfo
  {journal} {\mnras}\ }\textbf {\bibinfo {volume} {530}},\ \bibinfo {pages}
  {5030} (\bibinfo {year} {2024})},\ \Eprint {https://arxiv.org/abs/2212.07420}
  {arXiv:2212.07420 [astro-ph.CO]} \BibitemShut {NoStop}%
\bibitem [{\citenamefont {Namikawa}\ \emph {et~al.}(2013)\citenamefont
  {Namikawa}, \citenamefont {Yamauchi},\ and\ \citenamefont
  {Taruya}}]{Namikawa:2013wda}%
  \BibitemOpen
  \bibfield  {author} {\bibinfo {author} {\bibfnamefont {T.}~\bibnamefont
  {Namikawa}}, \bibinfo {author} {\bibfnamefont {D.}~\bibnamefont {Yamauchi}},\
  and\ \bibinfo {author} {\bibfnamefont {A.}~\bibnamefont {Taruya}},\
  }\bibfield  {title} {\bibinfo {title} {{Constraining cosmic string parameters
  with curl mode of CMB lensing}},\ }\href
  {https://doi.org/10.1103/PhysRevD.88.083525} {\bibfield  {journal} {\bibinfo
  {journal} {Phys. Rev. D}\ }\textbf {\bibinfo {volume} {88}},\ \bibinfo
  {pages} {083525} (\bibinfo {year} {2013})},\ \Eprint
  {https://arxiv.org/abs/1308.6068} {arXiv:1308.6068 [astro-ph.CO]}
  \BibitemShut {NoStop}%
\bibitem [{\citenamefont {Kaiser}(1992)}]{Kaiser92}%
  \BibitemOpen
  \bibfield  {author} {\bibinfo {author} {\bibfnamefont {N.}~\bibnamefont
  {Kaiser}},\ }\bibfield  {title} {\bibinfo {title} {Weak gravitational lensing
  of distant galaxies},\ }\href {https://doi.org/10.1086/171151} {\bibfield
  {journal} {\bibinfo  {journal} {\apj}\ }\textbf {\bibinfo {volume} {388}},\
  \bibinfo {pages} {272} (\bibinfo {year} {1992})}\BibitemShut {NoStop}%
%%CITATION = ASJOA,388,272;%%
\bibitem [{\citenamefont {Fabbian}\ \emph {et~al.}(2018)\citenamefont
  {Fabbian}, \citenamefont {Calabrese},\ and\ \citenamefont
  {Carbone}}]{Fabbian:2017wfp}%
  \BibitemOpen
  \bibfield  {author} {\bibinfo {author} {\bibfnamefont {G.}~\bibnamefont
  {Fabbian}}, \bibinfo {author} {\bibfnamefont {M.}~\bibnamefont {Calabrese}},\
  and\ \bibinfo {author} {\bibfnamefont {C.}~\bibnamefont {Carbone}},\
  }\bibfield  {title} {\bibinfo {title} {{CMB weak-lensing beyond the Born
  approximation: a numerical approach}},\ }\href
  {https://doi.org/10.1088/1475-7516/2018/02/050} {\bibfield  {journal}
  {\bibinfo  {journal} {JCAP}\ }\textbf {\bibinfo {volume} {02}},\ \bibinfo
  {pages} {050}},\ \Eprint {https://arxiv.org/abs/1702.03317} {arXiv:1702.03317
  [astro-ph.CO]} \BibitemShut {NoStop}%
\bibitem [{\citenamefont {Fabbian}\ \emph {et~al.}(2019)\citenamefont
  {Fabbian}, \citenamefont {Lewis},\ and\ \citenamefont
  {Beck}}]{Fabbian:2019tik}%
  \BibitemOpen
  \bibfield  {author} {\bibinfo {author} {\bibfnamefont {G.}~\bibnamefont
  {Fabbian}}, \bibinfo {author} {\bibfnamefont {A.}~\bibnamefont {Lewis}},\
  and\ \bibinfo {author} {\bibfnamefont {D.}~\bibnamefont {Beck}},\ }\bibfield
  {title} {\bibinfo {title} {{CMB lensing reconstruction biases in
  cross-correlation with large-scale structure probes}},\ }\href
  {https://doi.org/10.1088/1475-7516/2019/10/057} {\bibfield  {journal}
  {\bibinfo  {journal} {JCAP}\ }\textbf {\bibinfo {volume} {1910}}\bibfield
  {number} {\bibinfo  {number} { (10)},\ \bibinfo {pages} {057}},\ }\Eprint
  {https://arxiv.org/abs/1906.08760} {arXiv:1906.08760 [astro-ph.CO]}
  \BibitemShut {NoStop}%
%%CITATION = ARXIV:1906.08760;%%
\bibitem [{\citenamefont {Calabrese}\ \emph {et~al.}(2015)\citenamefont
  {Calabrese}, \citenamefont {Carbone}, \citenamefont {Fabbian}, \citenamefont
  {Baldi},\ and\ \citenamefont {Baccigalupi}}]{Calabrese:2014gla}%
  \BibitemOpen
  \bibfield  {author} {\bibinfo {author} {\bibfnamefont {M.}~\bibnamefont
  {Calabrese}}, \bibinfo {author} {\bibfnamefont {C.}~\bibnamefont {Carbone}},
  \bibinfo {author} {\bibfnamefont {G.}~\bibnamefont {Fabbian}}, \bibinfo
  {author} {\bibfnamefont {M.}~\bibnamefont {Baldi}},\ and\ \bibinfo {author}
  {\bibfnamefont {C.}~\bibnamefont {Baccigalupi}},\ }\bibfield  {title}
  {\bibinfo {title} {{Multiple lensing of the cosmic microwave background
  anisotropies}},\ }\href {https://doi.org/10.1088/1475-7516/2015/03/049}
  {\bibfield  {journal} {\bibinfo  {journal} {JCAP}\ }\textbf {\bibinfo
  {volume} {03}},\ \bibinfo {pages} {049}},\ \Eprint
  {https://arxiv.org/abs/1409.7680} {arXiv:1409.7680 [astro-ph.CO]}
  \BibitemShut {NoStop}%
\bibitem [{\citenamefont {Hilbert}\ \emph {et~al.}(2009)\citenamefont
  {Hilbert}, \citenamefont {Hartlap}, \citenamefont {White},\ and\
  \citenamefont {Schneider}}]{Hilbert:2008kb}%
  \BibitemOpen
  \bibfield  {author} {\bibinfo {author} {\bibfnamefont {S.}~\bibnamefont
  {Hilbert}}, \bibinfo {author} {\bibfnamefont {J.}~\bibnamefont {Hartlap}},
  \bibinfo {author} {\bibfnamefont {S.~D.~M.}\ \bibnamefont {White}},\ and\
  \bibinfo {author} {\bibfnamefont {P.}~\bibnamefont {Schneider}},\ }\bibfield
  {title} {\bibinfo {title} {{Ray-tracing through the Millennium Simulation:
  Born corrections and lens-lens coupling in cosmic shear and galaxy-galaxy
  lensing}},\ }\href {https://doi.org/10.1051/0004-6361/200811054} {\bibfield
  {journal} {\bibinfo  {journal} {Astron. Astrophys.}\ }\textbf {\bibinfo
  {volume} {499}},\ \bibinfo {pages} {31} (\bibinfo {year} {2009})},\ \Eprint
  {https://arxiv.org/abs/0809.5035} {arXiv:0809.5035 [astro-ph]} \BibitemShut
  {NoStop}%
\bibitem [{\citenamefont {Das}\ and\ \citenamefont {Bode}(2008)}]{Das:2007eu}%
  \BibitemOpen
  \bibfield  {author} {\bibinfo {author} {\bibfnamefont {S.}~\bibnamefont
  {Das}}\ and\ \bibinfo {author} {\bibfnamefont {P.}~\bibnamefont {Bode}},\
  }\bibfield  {title} {\bibinfo {title} {{A Large Sky Simulation of the
  Gravitational Lensing of the Cosmic Microwave Background}},\ }\href
  {https://doi.org/10.1086/589638} {\bibfield  {journal} {\bibinfo  {journal}
  {Astrophys. J.}\ }\textbf {\bibinfo {volume} {682}},\ \bibinfo {pages} {1}
  (\bibinfo {year} {2008})},\ \Eprint {https://arxiv.org/abs/0711.3793}
  {arXiv:0711.3793 [astro-ph]} \BibitemShut {NoStop}%
\bibitem [{\citenamefont {Becker}(2013)}]{Becker:2012qe}%
  \BibitemOpen
  \bibfield  {author} {\bibinfo {author} {\bibfnamefont {M.~R.}\ \bibnamefont
  {Becker}},\ }\bibfield  {title} {\bibinfo {title} {{CALCLENS: Weak Lensing
  Simulations for Large-area Sky Surveys and Second-order Effects in Cosmic
  Shear Power Spectra}},\ }\href {https://doi.org/10.1093/mnras/stt1352}
  {\bibfield  {journal} {\bibinfo  {journal} {Mon. Not. Roy. Astron. Soc.}\
  }\textbf {\bibinfo {volume} {435}},\ \bibinfo {pages} {115} (\bibinfo {year}
  {2013})},\ \Eprint {https://arxiv.org/abs/1210.3069} {arXiv:1210.3069
  [astro-ph.CO]} \BibitemShut {NoStop}%
\bibitem [{\citenamefont {Fabbian}\ and\ \citenamefont
  {Stompor}(2013)}]{Fabbian:2013owa}%
  \BibitemOpen
  \bibfield  {author} {\bibinfo {author} {\bibfnamefont {G.}~\bibnamefont
  {Fabbian}}\ and\ \bibinfo {author} {\bibfnamefont {R.}~\bibnamefont
  {Stompor}},\ }\bibfield  {title} {\bibinfo {title} {{High precision
  simulations of weak lensing effect on Cosmic Microwave Background
  polarization}},\ }\href {https://doi.org/10.1051/0004-6361/201321575}
  {\bibfield  {journal} {\bibinfo  {journal} {Astron. Astrophys.}\ }\textbf
  {\bibinfo {volume} {556}},\ \bibinfo {pages} {A109} (\bibinfo {year}
  {2013})},\ \Eprint {https://arxiv.org/abs/1303.6550} {arXiv:1303.6550
  [astro-ph.CO]} \BibitemShut {NoStop}%
\bibitem [{\citenamefont {Muciaccia}\ \emph {et~al.}(1997)\citenamefont
  {Muciaccia}, \citenamefont {Natoli},\ and\ \citenamefont
  {Vittorio}}]{Muciaccia:1997pi}%
  \BibitemOpen
  \bibfield  {author} {\bibinfo {author} {\bibfnamefont {P.~F.}\ \bibnamefont
  {Muciaccia}}, \bibinfo {author} {\bibfnamefont {P.}~\bibnamefont {Natoli}},\
  and\ \bibinfo {author} {\bibfnamefont {N.}~\bibnamefont {Vittorio}},\
  }\bibfield  {title} {\bibinfo {title} {{Fast spherical harmonic analysis: a
  quick algorithm for generating and/or inverting full sky, high resolution CMB
  anisotropy maps}},\ }\href {https://doi.org/10.1086/310921} {\bibfield
  {journal} {\bibinfo  {journal} {Astrophys. J. Lett.}\ }\textbf {\bibinfo
  {volume} {488}},\ \bibinfo {pages} {L63} (\bibinfo {year} {1997})},\ \Eprint
  {https://arxiv.org/abs/astro-ph/9703084} {arXiv:astro-ph/9703084}
  \BibitemShut {NoStop}%
\bibitem [{\citenamefont {Mead}\ \emph {et~al.}(2016)\citenamefont {Mead},
  \citenamefont {Heymans}, \citenamefont {Lombriser}, \citenamefont {Peacock},
  \citenamefont {Steele},\ and\ \citenamefont {Winther}}]{Mead:2016zqy}%
  \BibitemOpen
  \bibfield  {author} {\bibinfo {author} {\bibfnamefont {A.}~\bibnamefont
  {Mead}}, \bibinfo {author} {\bibfnamefont {C.}~\bibnamefont {Heymans}},
  \bibinfo {author} {\bibfnamefont {L.}~\bibnamefont {Lombriser}}, \bibinfo
  {author} {\bibfnamefont {J.}~\bibnamefont {Peacock}}, \bibinfo {author}
  {\bibfnamefont {O.}~\bibnamefont {Steele}},\ and\ \bibinfo {author}
  {\bibfnamefont {H.}~\bibnamefont {Winther}},\ }\bibfield  {title} {\bibinfo
  {title} {{Accurate halo-model matter power spectra with dark energy, massive
  neutrinos and modified gravitational forces}},\ }\href
  {https://doi.org/10.1093/mnras/stw681} {\bibfield  {journal} {\bibinfo
  {journal} {Mon. Not. Roy. Astron. Soc.}\ }\textbf {\bibinfo {volume} {459}},\
  \bibinfo {pages} {1468} (\bibinfo {year} {2016})},\ \Eprint
  {https://arxiv.org/abs/1602.02154} {arXiv:1602.02154 [astro-ph.CO]}
  \BibitemShut {NoStop}%
\bibitem [{\citenamefont {Takahashi}\ \emph {et~al.}(2012)\citenamefont
  {Takahashi}, \citenamefont {Sato}, \citenamefont {Nishimichi}, \citenamefont
  {Taruya},\ and\ \citenamefont {Oguri}}]{Takahashi:2012em}%
  \BibitemOpen
  \bibfield  {author} {\bibinfo {author} {\bibfnamefont {R.}~\bibnamefont
  {Takahashi}}, \bibinfo {author} {\bibfnamefont {M.}~\bibnamefont {Sato}},
  \bibinfo {author} {\bibfnamefont {T.}~\bibnamefont {Nishimichi}}, \bibinfo
  {author} {\bibfnamefont {A.}~\bibnamefont {Taruya}},\ and\ \bibinfo {author}
  {\bibfnamefont {M.}~\bibnamefont {Oguri}},\ }\bibfield  {title} {\bibinfo
  {title} {{Revising the Halofit Model for the Nonlinear Matter Power
  Spectrum}},\ }\href {https://doi.org/10.1088/0004-637X/761/2/152} {\bibfield
  {journal} {\bibinfo  {journal} {Astrophys. J.}\ }\textbf {\bibinfo {volume}
  {761}},\ \bibinfo {pages} {152} (\bibinfo {year} {2012})},\ \Eprint
  {https://arxiv.org/abs/1208.2701} {arXiv:1208.2701 [astro-ph.CO]}
  \BibitemShut {NoStop}%
\bibitem [{\citenamefont {Reinecke}\ \emph {et~al.}(2023)\citenamefont
  {Reinecke}, \citenamefont {Belkner},\ and\ \citenamefont
  {Carron}}]{Reinecke:2023gtp}%
  \BibitemOpen
  \bibfield  {author} {\bibinfo {author} {\bibfnamefont {M.}~\bibnamefont
  {Reinecke}}, \bibinfo {author} {\bibfnamefont {S.}~\bibnamefont {Belkner}},\
  and\ \bibinfo {author} {\bibfnamefont {J.}~\bibnamefont {Carron}},\
  }\bibfield  {title} {\bibinfo {title} {{Improved cosmic microwave background
  (de-)lensing using general spherical harmonic transforms}},\ }\href
  {https://doi.org/10.1051/0004-6361/202346717} {\bibfield  {journal} {\bibinfo
   {journal} {Astron. Astrophys.}\ }\textbf {\bibinfo {volume} {678}},\
  \bibinfo {pages} {A165} (\bibinfo {year} {2023})},\ \Eprint
  {https://arxiv.org/abs/2304.10431} {arXiv:2304.10431 [astro-ph.CO]}
  \BibitemShut {NoStop}%
\bibitem [{\citenamefont {Maniyar}\ \emph {et~al.}(2021)\citenamefont
  {Maniyar}, \citenamefont {Ali-Ha\"\i{}moud}, \citenamefont {Carron},
  \citenamefont {Lewis},\ and\ \citenamefont
  {Madhavacheril}}]{Maniyar:2021msb}%
  \BibitemOpen
  \bibfield  {author} {\bibinfo {author} {\bibfnamefont {A.~S.}\ \bibnamefont
  {Maniyar}}, \bibinfo {author} {\bibfnamefont {Y.}~\bibnamefont
  {Ali-Ha\"\i{}moud}}, \bibinfo {author} {\bibfnamefont {J.}~\bibnamefont
  {Carron}}, \bibinfo {author} {\bibfnamefont {A.}~\bibnamefont {Lewis}},\ and\
  \bibinfo {author} {\bibfnamefont {M.~S.}\ \bibnamefont {Madhavacheril}},\
  }\bibfield  {title} {\bibinfo {title} {{Quadratic estimators for CMB weak
  lensing}},\ }\href {https://doi.org/10.1103/PhysRevD.103.083524} {\bibfield
  {journal} {\bibinfo  {journal} {Phys. Rev. D}\ }\textbf {\bibinfo {volume}
  {103}},\ \bibinfo {pages} {083524} (\bibinfo {year} {2021})},\ \Eprint
  {https://arxiv.org/abs/2101.12193} {arXiv:2101.12193 [astro-ph.CO]}
  \BibitemShut {NoStop}%
\bibitem [{\citenamefont {Hirata}\ and\ \citenamefont
  {Seljak}(2003)}]{Hirata:2003ka}%
  \BibitemOpen
  \bibfield  {author} {\bibinfo {author} {\bibfnamefont {C.~M.}\ \bibnamefont
  {Hirata}}\ and\ \bibinfo {author} {\bibfnamefont {U.}~\bibnamefont
  {Seljak}},\ }\bibfield  {title} {\bibinfo {title} {{Reconstruction of lensing
  from the cosmic microwave background polarization}},\ }\href
  {https://doi.org/10.1103/PhysRevD.68.083002} {\bibfield  {journal} {\bibinfo
  {journal} {\prd}\ }\textbf {\bibinfo {volume} {68}},\ \bibinfo {pages}
  {083002} (\bibinfo {year} {2003})},\ \Eprint
  {https://arxiv.org/abs/astro-ph/0306354} {arXiv:astro-ph/0306354 [astro-ph]}
  \BibitemShut {NoStop}%
%%CITATION = ASTRO-PH/0306354;%%
\bibitem [{\citenamefont {Carron}\ and\ \citenamefont
  {Lewis}(2017)}]{Carron:2017mqf}%
  \BibitemOpen
  \bibfield  {author} {\bibinfo {author} {\bibfnamefont {J.}~\bibnamefont
  {Carron}}\ and\ \bibinfo {author} {\bibfnamefont {A.}~\bibnamefont {Lewis}},\
  }\bibfield  {title} {\bibinfo {title} {{Maximum a posteriori CMB lensing
  reconstruction}},\ }\href {https://doi.org/10.1103/PhysRevD.96.063510}
  {\bibfield  {journal} {\bibinfo  {journal} {\prd}\ }\textbf {\bibinfo
  {volume} {96}},\ \bibinfo {pages} {063510} (\bibinfo {year} {2017})},\
  \Eprint {https://arxiv.org/abs/1704.08230} {arXiv:1704.08230 [astro-ph.CO]}
  \BibitemShut {NoStop}%
%%CITATION = ARXIV:1704.08230;%%
\bibitem [{\citenamefont {Belkner}\ \emph {et~al.}(2023)\citenamefont
  {Belkner}, \citenamefont {Carron}, \citenamefont {Legrand}, \citenamefont
  {Umilt\`a}, \citenamefont {Pryke},\ and\ \citenamefont
  {Bischoff}}]{Belkner:2023duz}%
  \BibitemOpen
  \bibfield  {author} {\bibinfo {author} {\bibfnamefont {S.}~\bibnamefont
  {Belkner}}, \bibinfo {author} {\bibfnamefont {J.}~\bibnamefont {Carron}},
  \bibinfo {author} {\bibfnamefont {L.}~\bibnamefont {Legrand}}, \bibinfo
  {author} {\bibfnamefont {C.}~\bibnamefont {Umilt\`a}}, \bibinfo {author}
  {\bibfnamefont {C.}~\bibnamefont {Pryke}},\ and\ \bibinfo {author}
  {\bibfnamefont {C.}~\bibnamefont {Bischoff}} (\bibinfo {collaboration}
  {CMB-S4}),\ }\bibfield  {title} {\bibinfo {title} {{CMB-S4: Iterative
  internal delensing and $r$ constraints}},\ }\href@noop {} {\  (\bibinfo
  {year} {2023})},\ \Eprint {https://arxiv.org/abs/2310.06729}
  {arXiv:2310.06729 [astro-ph.CO]} \BibitemShut {NoStop}%
\bibitem [{\citenamefont {Legrand}\ and\ \citenamefont
  {Carron}(2023)}]{Legrand:2023jne}%
  \BibitemOpen
  \bibfield  {author} {\bibinfo {author} {\bibfnamefont {L.}~\bibnamefont
  {Legrand}}\ and\ \bibinfo {author} {\bibfnamefont {J.}~\bibnamefont
  {Carron}},\ }\bibfield  {title} {\bibinfo {title} {{Robust and efficient CMB
  lensing power spectrum from polarization surveys}},\ }\href
  {https://doi.org/10.1103/PhysRevD.108.103516} {\bibfield  {journal} {\bibinfo
   {journal} {Phys. Rev. D}\ }\textbf {\bibinfo {volume} {108}},\ \bibinfo
  {pages} {103516} (\bibinfo {year} {2023})},\ \Eprint
  {https://arxiv.org/abs/2304.02584} {arXiv:2304.02584 [astro-ph.CO]}
  \BibitemShut {NoStop}%
\bibitem [{\citenamefont {Legrand}\ and\ \citenamefont
  {Carron}(2022)}]{Legrand:2021qdu}%
  \BibitemOpen
  \bibfield  {author} {\bibinfo {author} {\bibfnamefont {L.}~\bibnamefont
  {Legrand}}\ and\ \bibinfo {author} {\bibfnamefont {J.}~\bibnamefont
  {Carron}},\ }\bibfield  {title} {\bibinfo {title} {{Lensing power spectrum of
  the cosmic microwave background with deep polarization experiments}},\ }\href
  {https://doi.org/10.1103/PhysRevD.105.123519} {\bibfield  {journal} {\bibinfo
   {journal} {Phys. Rev. D}\ }\textbf {\bibinfo {volume} {105}},\ \bibinfo
  {pages} {123519} (\bibinfo {year} {2022})},\ \Eprint
  {https://arxiv.org/abs/2112.05764} {arXiv:2112.05764 [astro-ph.CO]}
  \BibitemShut {NoStop}%
\bibitem [{\citenamefont {Abell}\ \emph {et~al.}(2009)\citenamefont {Abell}
  \emph {et~al.}}]{LSSTScience:2009jmu}%
  \BibitemOpen
  \bibfield  {author} {\bibinfo {author} {\bibfnamefont {P.~A.}\ \bibnamefont
  {Abell}} \emph {et~al.} (\bibinfo {collaboration} {LSST Science, LSST
  Project}),\ }\bibfield  {title} {\bibinfo {title} {{LSST Science Book,
  Version 2.0}},\ }\href@noop {} {\  (\bibinfo {year} {2009})},\ \Eprint
  {https://arxiv.org/abs/0912.0201} {arXiv:0912.0201 [astro-ph.IM]}
  \BibitemShut {NoStop}%
\bibitem [{\citenamefont {Aghanim}\ \emph {et~al.}(2016)\citenamefont {Aghanim}
  \emph {et~al.}}]{Aghanim:2016pcc}%
  \BibitemOpen
  \bibfield  {author} {\bibinfo {author} {\bibfnamefont {N.}~\bibnamefont
  {Aghanim}} \emph {et~al.} (\bibinfo {collaboration} {Planck}),\ }\bibfield
  {title} {\bibinfo {title} {{Planck intermediate results. XLVIII.
  Disentangling Galactic dust emission and cosmic infrared background
  anisotropies}},\ }\href {https://doi.org/10.1051/0004-6361/201629022}
  {\bibfield  {journal} {\bibinfo  {journal} {\aap}\ }\textbf {\bibinfo
  {volume} {596}},\ \bibinfo {pages} {A109} (\bibinfo {year} {2016})},\ \Eprint
  {https://arxiv.org/abs/1605.09387} {arXiv:1605.09387 [astro-ph.CO]}
  \BibitemShut {NoStop}%
%%CITATION = ARXIV:1605.09387;%%
\bibitem [{\citenamefont {B\"ohm}\ \emph {et~al.}(2020)\citenamefont {B\"ohm},
  \citenamefont {Modi},\ and\ \citenamefont {Castorina}}]{Bohm:2019bek}%
  \BibitemOpen
  \bibfield  {author} {\bibinfo {author} {\bibfnamefont {V.}~\bibnamefont
  {B\"ohm}}, \bibinfo {author} {\bibfnamefont {C.}~\bibnamefont {Modi}},\ and\
  \bibinfo {author} {\bibfnamefont {E.}~\bibnamefont {Castorina}},\ }\bibfield
  {title} {\bibinfo {title} {{Lensing corrections on galaxy-lensing cross
  correlations and galaxy-galaxy auto correlations}},\ }\href
  {https://doi.org/10.1088/1475-7516/2020/03/045} {\bibfield  {journal}
  {\bibinfo  {journal} {JCAP}\ }\textbf {\bibinfo {volume} {03}},\ \bibinfo
  {pages} {045}},\ \Eprint {https://arxiv.org/abs/1910.06722} {arXiv:1910.06722
  [astro-ph.CO]} \BibitemShut {NoStop}%
\bibitem [{\citenamefont {LoVerde}\ and\ \citenamefont
  {Afshordi}(2008)}]{LoVerde:2008re}%
  \BibitemOpen
  \bibfield  {author} {\bibinfo {author} {\bibfnamefont {M.}~\bibnamefont
  {LoVerde}}\ and\ \bibinfo {author} {\bibfnamefont {N.}~\bibnamefont
  {Afshordi}},\ }\bibfield  {title} {\bibinfo {title} {{Extended Limber
  Approximation}},\ }\href {https://doi.org/10.1103/PhysRevD.78.123506}
  {\bibfield  {journal} {\bibinfo  {journal} {\prd}\ }\textbf {\bibinfo
  {volume} {78}},\ \bibinfo {pages} {123506} (\bibinfo {year} {2008})},\
  \Eprint {https://arxiv.org/abs/0809.5112} {arXiv:0809.5112 [astro-ph]}
  \BibitemShut {NoStop}%
%%CITATION = ARXIV:0809.5112;%%
\bibitem [{\citenamefont {Lewis}\ \emph {et~al.}(2011)\citenamefont {Lewis},
  \citenamefont {Challinor},\ and\ \citenamefont {Hanson}}]{Lewis:2011fk}%
  \BibitemOpen
  \bibfield  {author} {\bibinfo {author} {\bibfnamefont {A.}~\bibnamefont
  {Lewis}}, \bibinfo {author} {\bibfnamefont {A.}~\bibnamefont {Challinor}},\
  and\ \bibinfo {author} {\bibfnamefont {D.}~\bibnamefont {Hanson}},\
  }\bibfield  {title} {\bibinfo {title} {{The shape of the CMB lensing
  bispectrum}},\ }\href {https://doi.org/10.1088/1475-7516/2011/03/018}
  {\bibfield  {journal} {\bibinfo  {journal} {\jcap}\ }\textbf {\bibinfo
  {volume} {1103}},\ \bibinfo {pages} {018} (\bibinfo {year} {2011})},\ \Eprint
  {https://arxiv.org/abs/1101.2234} {arXiv:1101.2234 [astro-ph.CO]}
  \BibitemShut {NoStop}%
%%CITATION = 1101.2234;%%
\bibitem [{\citenamefont {Eriksen}\ \emph {et~al.}(2004)\citenamefont
  {Eriksen}, \citenamefont {Banday}, \citenamefont {Gorski},\ and\
  \citenamefont {Lilje}}]{Eriksen:2004jg}%
  \BibitemOpen
  \bibfield  {author} {\bibinfo {author} {\bibfnamefont {H.~K.}\ \bibnamefont
  {Eriksen}}, \bibinfo {author} {\bibfnamefont {A.~J.}\ \bibnamefont {Banday}},
  \bibinfo {author} {\bibfnamefont {K.~M.}\ \bibnamefont {Gorski}},\ and\
  \bibinfo {author} {\bibfnamefont {P.~B.}\ \bibnamefont {Lilje}},\ }\bibfield
  {title} {\bibinfo {title} {{Foreground removal by an internal linear
  combination method: Limitations and implications}},\ }\href
  {https://doi.org/10.1086/422807} {\bibfield  {journal} {\bibinfo  {journal}
  {Astrophys. J.}\ }\textbf {\bibinfo {volume} {612}},\ \bibinfo {pages} {633}
  (\bibinfo {year} {2004})},\ \Eprint {https://arxiv.org/abs/astro-ph/0403098}
  {arXiv:astro-ph/0403098} \BibitemShut {NoStop}%
\bibitem [{\citenamefont {Remazeilles}\ \emph {et~al.}(2011)\citenamefont
  {Remazeilles}, \citenamefont {Delabrouille},\ and\ \citenamefont
  {Cardoso}}]{Remazeilles:2010hq}%
  \BibitemOpen
  \bibfield  {author} {\bibinfo {author} {\bibfnamefont {M.}~\bibnamefont
  {Remazeilles}}, \bibinfo {author} {\bibfnamefont {J.}~\bibnamefont
  {Delabrouille}},\ and\ \bibinfo {author} {\bibfnamefont {J.-F.}\ \bibnamefont
  {Cardoso}},\ }\bibfield  {title} {\bibinfo {title} {{CMB and SZ effect
  separation with Constrained Internal Linear Combinations}},\ }\href
  {https://doi.org/10.1111/j.1365-2966.2010.17624.x} {\bibfield  {journal}
  {\bibinfo  {journal} {Mon. Not. Roy. Astron. Soc.}\ }\textbf {\bibinfo
  {volume} {410}},\ \bibinfo {pages} {2481} (\bibinfo {year} {2011})},\ \Eprint
  {https://arxiv.org/abs/1006.5599} {arXiv:1006.5599 [astro-ph.CO]}
  \BibitemShut {NoStop}%
\bibitem [{\citenamefont {Sailer}\ \emph {et~al.}(2020)\citenamefont {Sailer},
  \citenamefont {Schaan},\ and\ \citenamefont {Ferraro}}]{Sailer:2020lal}%
  \BibitemOpen
  \bibfield  {author} {\bibinfo {author} {\bibfnamefont {N.}~\bibnamefont
  {Sailer}}, \bibinfo {author} {\bibfnamefont {E.}~\bibnamefont {Schaan}},\
  and\ \bibinfo {author} {\bibfnamefont {S.}~\bibnamefont {Ferraro}},\
  }\bibfield  {title} {\bibinfo {title} {{Lower bias, lower noise CMB lensing
  with foreground-hardened estimators}},\ }\href
  {https://doi.org/10.1103/PhysRevD.102.063517} {\bibfield  {journal} {\bibinfo
   {journal} {Phys. Rev. D}\ }\textbf {\bibinfo {volume} {102}},\ \bibinfo
  {pages} {063517} (\bibinfo {year} {2020})},\ \Eprint
  {https://arxiv.org/abs/2007.04325} {arXiv:2007.04325 [astro-ph.CO]}
  \BibitemShut {NoStop}%
\bibitem [{\citenamefont {Schaan}\ and\ \citenamefont
  {Ferraro}(2019)}]{Schaan:2018tup}%
  \BibitemOpen
  \bibfield  {author} {\bibinfo {author} {\bibfnamefont {E.}~\bibnamefont
  {Schaan}}\ and\ \bibinfo {author} {\bibfnamefont {S.}~\bibnamefont
  {Ferraro}},\ }\bibfield  {title} {\bibinfo {title} {{Foreground-immune CMB
  lensing with shear-only reconstruction}},\ }\href
  {https://doi.org/10.1103/PhysRevLett.122.181301} {\bibfield  {journal}
  {\bibinfo  {journal} {\prl}\ }\textbf {\bibinfo {volume} {122}},\ \bibinfo
  {pages} {181301} (\bibinfo {year} {2019})},\ \Eprint
  {https://arxiv.org/abs/1804.06403} {arXiv:1804.06403 [astro-ph.CO]}
  \BibitemShut {NoStop}%
%%CITATION = ARXIV:1804.06403;%%
\bibitem [{\citenamefont {Adam}\ \emph
  {et~al.}(2016{\natexlab{a}})\citenamefont {Adam} \emph
  {et~al.}}]{Planck:2014dmk}%
  \BibitemOpen
  \bibfield  {author} {\bibinfo {author} {\bibfnamefont {R.}~\bibnamefont
  {Adam}} \emph {et~al.} (\bibinfo {collaboration} {Planck}),\ }\bibfield
  {title} {\bibinfo {title} {{Planck intermediate results. XXX. The angular
  power spectrum of polarized dust emission at intermediate and high Galactic
  latitudes}},\ }\href {https://doi.org/10.1051/0004-6361/201425034} {\bibfield
   {journal} {\bibinfo  {journal} {Astron. Astrophys.}\ }\textbf {\bibinfo
  {volume} {586}},\ \bibinfo {pages} {A133} (\bibinfo {year}
  {2016}{\natexlab{a}})},\ \Eprint {https://arxiv.org/abs/1409.5738}
  {arXiv:1409.5738 [astro-ph.CO]} \BibitemShut {NoStop}%
\bibitem [{\citenamefont {Adam}\ \emph
  {et~al.}(2016{\natexlab{b}})\citenamefont {Adam} \emph
  {et~al.}}]{Planck:2015mvg}%
  \BibitemOpen
  \bibfield  {author} {\bibinfo {author} {\bibfnamefont {R.}~\bibnamefont
  {Adam}} \emph {et~al.} (\bibinfo {collaboration} {Planck}),\ }\bibfield
  {title} {\bibinfo {title} {{Planck 2015 results. X. Diffuse component
  separation: Foreground maps}},\ }\href
  {https://doi.org/10.1051/0004-6361/201525967} {\bibfield  {journal} {\bibinfo
   {journal} {Astron. Astrophys.}\ }\textbf {\bibinfo {volume} {594}},\
  \bibinfo {pages} {A10} (\bibinfo {year} {2016}{\natexlab{b}})},\ \Eprint
  {https://arxiv.org/abs/1502.01588} {arXiv:1502.01588 [astro-ph.CO]}
  \BibitemShut {NoStop}%
\bibitem [{\citenamefont {Kandel}\ \emph {et~al.}(2018)\citenamefont {Kandel},
  \citenamefont {Lazarian},\ and\ \citenamefont {Pogosyan}}]{Kandel:2017xjx}%
  \BibitemOpen
  \bibfield  {author} {\bibinfo {author} {\bibfnamefont {D.}~\bibnamefont
  {Kandel}}, \bibinfo {author} {\bibfnamefont {A.}~\bibnamefont {Lazarian}},\
  and\ \bibinfo {author} {\bibfnamefont {D.}~\bibnamefont {Pogosyan}},\
  }\bibfield  {title} {\bibinfo {title} {{Statistical properties of galactic
  CMB foregrounds: dust and synchrotron}},\ }\href
  {https://doi.org/10.1093/mnras/sty1115} {\bibfield  {journal} {\bibinfo
  {journal} {Mon. Not. Roy. Astron. Soc.}\ }\textbf {\bibinfo {volume} {478}},\
  \bibinfo {pages} {530} (\bibinfo {year} {2018})},\ \Eprint
  {https://arxiv.org/abs/1711.03161} {arXiv:1711.03161 [astro-ph.GA]}
  \BibitemShut {NoStop}%
\bibitem [{\citenamefont {Abril-Cabezas}\ \emph {et~al.}(2024)\citenamefont
  {Abril-Cabezas}, \citenamefont {Herv\'\i{}as-Caimapo}, \citenamefont {von
  Hausegger}, \citenamefont {Sherwin},\ and\ \citenamefont
  {Alonso}}]{Abril-Cabezas:2023ftf}%
  \BibitemOpen
  \bibfield  {author} {\bibinfo {author} {\bibfnamefont {I.}~\bibnamefont
  {Abril-Cabezas}}, \bibinfo {author} {\bibfnamefont {C.}~\bibnamefont
  {Herv\'\i{}as-Caimapo}}, \bibinfo {author} {\bibfnamefont {S.}~\bibnamefont
  {von Hausegger}}, \bibinfo {author} {\bibfnamefont {B.~D.}\ \bibnamefont
  {Sherwin}},\ and\ \bibinfo {author} {\bibfnamefont {D.}~\bibnamefont
  {Alonso}},\ }\bibfield  {title} {\bibinfo {title} {{Impact of Galactic dust
  non-Gaussianity on searches for B-modes from inflation}},\ }\href
  {https://doi.org/10.1093/mnras/stad3529} {\bibfield  {journal} {\bibinfo
  {journal} {Mon. Not. Roy. Astron. Soc.}\ }\textbf {\bibinfo {volume} {527}},\
  \bibinfo {pages} {5751} (\bibinfo {year} {2024})},\ \Eprint
  {https://arxiv.org/abs/2309.09978} {arXiv:2309.09978 [astro-ph.CO]}
  \BibitemShut {NoStop}%
\bibitem [{\citenamefont {Farren}\ \emph {et~al.}(2024)\citenamefont {Farren}
  \emph {et~al.}}]{ACT:2023oei}%
  \BibitemOpen
  \bibfield  {author} {\bibinfo {author} {\bibfnamefont {G.~S.}\ \bibnamefont
  {Farren}} \emph {et~al.} (\bibinfo {collaboration} {ACT}),\ }\bibfield
  {title} {\bibinfo {title} {{The Atacama Cosmology Telescope: Cosmology from
  Cross-correlations of unWISE Galaxies and ACT DR6 CMB Lensing}},\ }\href
  {https://doi.org/10.3847/1538-4357/ad31a5} {\bibfield  {journal} {\bibinfo
  {journal} {Astrophys. J.}\ }\textbf {\bibinfo {volume} {966}},\ \bibinfo
  {pages} {157} (\bibinfo {year} {2024})},\ \Eprint
  {https://arxiv.org/abs/2309.05659} {arXiv:2309.05659 [astro-ph.CO]}
  \BibitemShut {NoStop}%
\bibitem [{\citenamefont {Osborne}\ \emph {et~al.}(2014)\citenamefont
  {Osborne}, \citenamefont {Hanson},\ and\ \citenamefont
  {Doré}}]{Osborne:2013nna}%
  \BibitemOpen
  \bibfield  {author} {\bibinfo {author} {\bibfnamefont {S.~J.}\ \bibnamefont
  {Osborne}}, \bibinfo {author} {\bibfnamefont {D.}~\bibnamefont {Hanson}},\
  and\ \bibinfo {author} {\bibfnamefont {O.}~\bibnamefont {Doré}},\ }\bibfield
   {title} {\bibinfo {title} {{Extragalactic Foreground Contamination in
  Temperature-based CMB Lens Reconstruction}},\ }\href
  {https://doi.org/10.1088/1475-7516/2014/03/024} {\bibfield  {journal}
  {\bibinfo  {journal} {\jcap}\ }\textbf {\bibinfo {volume} {1403}},\ \bibinfo
  {pages} {024} (\bibinfo {year} {2014})},\ \Eprint
  {https://arxiv.org/abs/1310.7547} {arXiv:1310.7547 [astro-ph.CO]}
  \BibitemShut {NoStop}%
%%CITATION = ARXIV:1310.7547;%%
\bibitem [{\citenamefont {van Engelen}\ \emph {et~al.}(2014)\citenamefont {van
  Engelen}, \citenamefont {Bhattacharya}, \citenamefont {Sehgal}, \citenamefont
  {Holder}, \citenamefont {Zahn},\ and\ \citenamefont
  {Nagai}}]{vanEngelen:2013rla}%
  \BibitemOpen
  \bibfield  {author} {\bibinfo {author} {\bibfnamefont {A.}~\bibnamefont {van
  Engelen}}, \bibinfo {author} {\bibfnamefont {S.}~\bibnamefont
  {Bhattacharya}}, \bibinfo {author} {\bibfnamefont {N.}~\bibnamefont
  {Sehgal}}, \bibinfo {author} {\bibfnamefont {G.~P.}\ \bibnamefont {Holder}},
  \bibinfo {author} {\bibfnamefont {O.}~\bibnamefont {Zahn}},\ and\ \bibinfo
  {author} {\bibfnamefont {D.}~\bibnamefont {Nagai}},\ }\bibfield  {title}
  {\bibinfo {title} {{CMB Lensing Power Spectrum Biases from Galaxies and
  Clusters using High-angular Resolution Temperature Maps}},\ }\href
  {https://doi.org/10.1088/0004-637X/786/1/13} {\bibfield  {journal} {\bibinfo
  {journal} {\apj}\ }\textbf {\bibinfo {volume} {786}},\ \bibinfo {pages} {13}
  (\bibinfo {year} {2014})},\ \Eprint {https://arxiv.org/abs/1310.7023}
  {arXiv:1310.7023 [astro-ph.CO]} \BibitemShut {NoStop}%
%%CITATION = ARXIV:1310.7023;%%
\bibitem [{\citenamefont {Ade}\ \emph {et~al.}(2014)\citenamefont {Ade} \emph
  {et~al.}}]{Planck:2013wmz}%
  \BibitemOpen
  \bibfield  {author} {\bibinfo {author} {\bibfnamefont {P.~A.~R.}\
  \bibnamefont {Ade}} \emph {et~al.} (\bibinfo {collaboration} {Planck}),\
  }\bibfield  {title} {\bibinfo {title} {{Planck 2013 results. IX. HFI spectral
  response}},\ }\href {https://doi.org/10.1051/0004-6361/201321531} {\bibfield
  {journal} {\bibinfo  {journal} {Astron. Astrophys.}\ }\textbf {\bibinfo
  {volume} {571}},\ \bibinfo {pages} {A9} (\bibinfo {year} {2014})},\ \Eprint
  {https://arxiv.org/abs/1303.5070} {arXiv:1303.5070 [astro-ph.IM]}
  \BibitemShut {NoStop}%
\bibitem [{\citenamefont {Fixsen}(2009)}]{Fixsen:2009ug}%
  \BibitemOpen
  \bibfield  {author} {\bibinfo {author} {\bibfnamefont {D.}~\bibnamefont
  {Fixsen}},\ }\bibfield  {title} {\bibinfo {title} {{The Temperature of the
  Cosmic Microwave Background}},\ }\href
  {https://doi.org/10.1088/0004-637X/707/2/916} {\bibfield  {journal} {\bibinfo
   {journal} {\apj}\ }\textbf {\bibinfo {volume} {707}},\ \bibinfo {pages}
  {916} (\bibinfo {year} {2009})},\ \Eprint {https://arxiv.org/abs/0911.1955}
  {arXiv:0911.1955 [astro-ph.CO]} \BibitemShut {NoStop}%
%%CITATION = ARXIV:0911.1955;%%
\bibitem [{\citenamefont {Reichardt}\ \emph {et~al.}(2021)\citenamefont
  {Reichardt} \emph {et~al.}}]{SPT:2020psp}%
  \BibitemOpen
  \bibfield  {author} {\bibinfo {author} {\bibfnamefont {C.~L.}\ \bibnamefont
  {Reichardt}} \emph {et~al.} (\bibinfo {collaboration} {SPT}),\ }\bibfield
  {title} {\bibinfo {title} {{An Improved Measurement of the Secondary Cosmic
  Microwave Background Anisotropies from the SPT-SZ + SPTpol Surveys}},\ }\href
  {https://doi.org/10.3847/1538-4357/abd407} {\bibfield  {journal} {\bibinfo
  {journal} {Astrophys. J.}\ }\textbf {\bibinfo {volume} {908}},\ \bibinfo
  {pages} {199} (\bibinfo {year} {2021})},\ \Eprint
  {https://arxiv.org/abs/2002.06197} {arXiv:2002.06197 [astro-ph.CO]}
  \BibitemShut {NoStop}%
\bibitem [{\citenamefont {Lagache}\ \emph {et~al.}(2020)\citenamefont
  {Lagache}, \citenamefont {B\'ethermin}, \citenamefont {Montier},
  \citenamefont {Serra},\ and\ \citenamefont {Tucci}}]{Lagache:2019xto}%
  \BibitemOpen
  \bibfield  {author} {\bibinfo {author} {\bibfnamefont {G.}~\bibnamefont
  {Lagache}}, \bibinfo {author} {\bibfnamefont {M.}~\bibnamefont
  {B\'ethermin}}, \bibinfo {author} {\bibfnamefont {L.}~\bibnamefont
  {Montier}}, \bibinfo {author} {\bibfnamefont {P.}~\bibnamefont {Serra}},\
  and\ \bibinfo {author} {\bibfnamefont {M.}~\bibnamefont {Tucci}},\ }\bibfield
   {title} {\bibinfo {title} {{Impact of polarised extragalactic sources on the
  measurement of CMB B-mode anisotropies}},\ }\href
  {https://doi.org/10.1051/0004-6361/201937147} {\bibfield  {journal} {\bibinfo
   {journal} {\aap}\ }\textbf {\bibinfo {volume} {642}},\ \bibinfo {pages}
  {A232} (\bibinfo {year} {2020})},\ \Eprint {https://arxiv.org/abs/1911.09466}
  {arXiv:1911.09466 [astro-ph.CO]} \BibitemShut {NoStop}%
\bibitem [{\citenamefont {Bennett}\ \emph {et~al.}(2003)\citenamefont {Bennett}
  \emph {et~al.}}]{WMAP:2003ivt}%
  \BibitemOpen
  \bibfield  {author} {\bibinfo {author} {\bibfnamefont {C.~L.}\ \bibnamefont
  {Bennett}} \emph {et~al.} (\bibinfo {collaboration} {WMAP}),\ }\bibfield
  {title} {\bibinfo {title} {{First year Wilkinson Microwave Anisotropy Probe
  (WMAP) observations: Preliminary maps and basic results}},\ }\href
  {https://doi.org/10.1086/377253} {\bibfield  {journal} {\bibinfo  {journal}
  {Astrophys. J. Suppl.}\ }\textbf {\bibinfo {volume} {148}},\ \bibinfo {pages}
  {1} (\bibinfo {year} {2003})},\ \Eprint
  {https://arxiv.org/abs/astro-ph/0302207} {arXiv:astro-ph/0302207}
  \BibitemShut {NoStop}%
\bibitem [{\citenamefont {Ade}\ \emph {et~al.}(2016{\natexlab{b}})\citenamefont
  {Ade} \emph {et~al.}}]{Planck:2015emq}%
  \BibitemOpen
  \bibfield  {author} {\bibinfo {author} {\bibfnamefont {P.~A.~R.}\
  \bibnamefont {Ade}} \emph {et~al.} (\bibinfo {collaboration} {Planck}),\
  }\bibfield  {title} {\bibinfo {title} {{Planck 2015 results. XXIII. The
  thermal Sunyaev-Zeldovich effect--cosmic infrared background correlation}},\
  }\href {https://doi.org/10.1051/0004-6361/201527418} {\bibfield  {journal}
  {\bibinfo  {journal} {Astron. Astrophys.}\ }\textbf {\bibinfo {volume}
  {594}},\ \bibinfo {pages} {A23} (\bibinfo {year} {2016}{\natexlab{b}})},\
  \Eprint {https://arxiv.org/abs/1509.06555} {arXiv:1509.06555 [astro-ph.CO]}
  \BibitemShut {NoStop}%
\bibitem [{\citenamefont {Herranz}\ \emph {et~al.}(2002)\citenamefont
  {Herranz}, \citenamefont {Sanz}, \citenamefont {Hobson}, \citenamefont
  {Barreiro}, \citenamefont {Diego}, \citenamefont {Martinez-Gonzalez},\ and\
  \citenamefont {Lasenby}}]{Herranz:2002kg}%
  \BibitemOpen
  \bibfield  {author} {\bibinfo {author} {\bibfnamefont {D.}~\bibnamefont
  {Herranz}}, \bibinfo {author} {\bibfnamefont {J.~L.}\ \bibnamefont {Sanz}},
  \bibinfo {author} {\bibfnamefont {M.~P.}\ \bibnamefont {Hobson}}, \bibinfo
  {author} {\bibfnamefont {R.~B.}\ \bibnamefont {Barreiro}}, \bibinfo {author}
  {\bibfnamefont {J.~M.}\ \bibnamefont {Diego}}, \bibinfo {author}
  {\bibfnamefont {E.}~\bibnamefont {Martinez-Gonzalez}},\ and\ \bibinfo
  {author} {\bibfnamefont {A.~N.}\ \bibnamefont {Lasenby}},\ }\bibfield
  {title} {\bibinfo {title} {{Filtering techniques for the detection of
  Sunyaev-Zel'dovich clusters in multifrequency CMB maps}},\ }\href
  {https://doi.org/10.1046/j.1365-8711.2002.05704.x} {\bibfield  {journal}
  {\bibinfo  {journal} {Mon. Not. Roy. Astron. Soc.}\ }\textbf {\bibinfo
  {volume} {336}},\ \bibinfo {pages} {1057} (\bibinfo {year} {2002})},\ \Eprint
  {https://arxiv.org/abs/astro-ph/0203486} {arXiv:astro-ph/0203486}
  \BibitemShut {NoStop}%
\bibitem [{\citenamefont {Melin}\ \emph {et~al.}(2006)\citenamefont {Melin},
  \citenamefont {Bartlett},\ and\ \citenamefont {Delabrouille}}]{Melin:2006qq}%
  \BibitemOpen
  \bibfield  {author} {\bibinfo {author} {\bibfnamefont {J.-B.}\ \bibnamefont
  {Melin}}, \bibinfo {author} {\bibfnamefont {J.~G.}\ \bibnamefont
  {Bartlett}},\ and\ \bibinfo {author} {\bibfnamefont {J.}~\bibnamefont
  {Delabrouille}},\ }\bibfield  {title} {\bibinfo {title} {{Catalog extraction
  in SZ cluster surveys: a matched filter approach}},\ }\href
  {https://doi.org/10.1051/0004-6361:20065034} {\bibfield  {journal} {\bibinfo
  {journal} {Astron. Astrophys.}\ }\textbf {\bibinfo {volume} {459}},\ \bibinfo
  {pages} {341} (\bibinfo {year} {2006})},\ \Eprint
  {https://arxiv.org/abs/astro-ph/0602424} {arXiv:astro-ph/0602424}
  \BibitemShut {NoStop}%
\bibitem [{\citenamefont {Bucher}\ and\ \citenamefont
  {Louis}(2012)}]{Bucher:2011nf}%
  \BibitemOpen
  \bibfield  {author} {\bibinfo {author} {\bibfnamefont {M.}~\bibnamefont
  {Bucher}}\ and\ \bibinfo {author} {\bibfnamefont {T.}~\bibnamefont {Louis}}
  (\bibinfo {collaboration} {Astrophysics Group, Oxford University}),\
  }\bibfield  {title} {\bibinfo {title} {{Filling in CMB map missing data using
  constrained Gaussian realizations}},\ }\href
  {https://doi.org/10.1111/j.1365-2966.2012.21138.x} {\bibfield  {journal}
  {\bibinfo  {journal} {Mon. Not. Roy. Astron. Soc.}\ }\textbf {\bibinfo
  {volume} {424}},\ \bibinfo {pages} {1694} (\bibinfo {year} {2012})},\ \Eprint
  {https://arxiv.org/abs/1109.0286} {arXiv:1109.0286 [astro-ph.CO]}
  \BibitemShut {NoStop}%
\bibitem [{\citenamefont {MacCrann}\ \emph {et~al.}(2024)\citenamefont
  {MacCrann} \emph {et~al.}}]{ACT:2023ubw}%
  \BibitemOpen
  \bibfield  {author} {\bibinfo {author} {\bibfnamefont {N.}~\bibnamefont
  {MacCrann}} \emph {et~al.} (\bibinfo {collaboration} {ACT}),\ }\bibfield
  {title} {\bibinfo {title} {{The Atacama Cosmology Telescope: Mitigating the
  Impact of Extragalactic Foregrounds for the DR6 Cosmic Microwave Background
  Lensing Analysis}},\ }\href {https://doi.org/10.3847/1538-4357/ad2610}
  {\bibfield  {journal} {\bibinfo  {journal} {Astrophys. J.}\ }\textbf
  {\bibinfo {volume} {966}},\ \bibinfo {pages} {138} (\bibinfo {year}
  {2024})},\ \Eprint {https://arxiv.org/abs/2304.05196} {arXiv:2304.05196
  [astro-ph.CO]} \BibitemShut {NoStop}%
\bibitem [{\citenamefont {Stein}\ \emph {et~al.}(2020)\citenamefont {Stein},
  \citenamefont {Alvarez}, \citenamefont {Bond}, \citenamefont {van Engelen},\
  and\ \citenamefont {Battaglia}}]{Stein:2020its}%
  \BibitemOpen
  \bibfield  {author} {\bibinfo {author} {\bibfnamefont {G.}~\bibnamefont
  {Stein}}, \bibinfo {author} {\bibfnamefont {M.~A.}\ \bibnamefont {Alvarez}},
  \bibinfo {author} {\bibfnamefont {J.~R.}\ \bibnamefont {Bond}}, \bibinfo
  {author} {\bibfnamefont {A.}~\bibnamefont {van Engelen}},\ and\ \bibinfo
  {author} {\bibfnamefont {N.}~\bibnamefont {Battaglia}},\ }\bibfield  {title}
  {\bibinfo {title} {{The Websky Extragalactic CMB Simulations}},\ }\href
  {https://doi.org/10.1088/1475-7516/2020/10/012} {\bibfield  {journal}
  {\bibinfo  {journal} {JCAP}\ }\textbf {\bibinfo {volume} {10}},\ \bibinfo
  {pages} {012}},\ \Eprint {https://arxiv.org/abs/2001.08787} {arXiv:2001.08787
  [astro-ph.CO]} \BibitemShut {NoStop}%
\bibitem [{\citenamefont {Sehgal}\ \emph {et~al.}(2010)\citenamefont {Sehgal},
  \citenamefont {Bode}, \citenamefont {Das}, \citenamefont
  {Hernandez-Monteagudo}, \citenamefont {Huffenberger}, \citenamefont {Lin},
  \citenamefont {Ostriker},\ and\ \citenamefont {Trac}}]{Sehgal:2009xv}%
  \BibitemOpen
  \bibfield  {author} {\bibinfo {author} {\bibfnamefont {N.}~\bibnamefont
  {Sehgal}}, \bibinfo {author} {\bibfnamefont {P.}~\bibnamefont {Bode}},
  \bibinfo {author} {\bibfnamefont {S.}~\bibnamefont {Das}}, \bibinfo {author}
  {\bibfnamefont {C.}~\bibnamefont {Hernandez-Monteagudo}}, \bibinfo {author}
  {\bibfnamefont {K.}~\bibnamefont {Huffenberger}}, \bibinfo {author}
  {\bibfnamefont {Y.-T.}\ \bibnamefont {Lin}}, \bibinfo {author} {\bibfnamefont
  {J.~P.}\ \bibnamefont {Ostriker}},\ and\ \bibinfo {author} {\bibfnamefont
  {H.}~\bibnamefont {Trac}},\ }\bibfield  {title} {\bibinfo {title}
  {{Simulations of the Microwave Sky}},\ }\href
  {https://doi.org/10.1088/0004-637X/709/2/920} {\bibfield  {journal} {\bibinfo
   {journal} {Astrophys. J.}\ }\textbf {\bibinfo {volume} {709}},\ \bibinfo
  {pages} {920} (\bibinfo {year} {2010})},\ \Eprint
  {https://arxiv.org/abs/0908.0540} {arXiv:0908.0540 [astro-ph.CO]}
  \BibitemShut {NoStop}%
\bibitem [{\citenamefont {Hamana}\ \emph {et~al.}(2015)\citenamefont {Hamana},
  \citenamefont {Sakurai}, \citenamefont {Koike},\ and\ \citenamefont
  {Miller}}]{Hamana:2015bwa}%
  \BibitemOpen
  \bibfield  {author} {\bibinfo {author} {\bibfnamefont {T.}~\bibnamefont
  {Hamana}}, \bibinfo {author} {\bibfnamefont {J.}~\bibnamefont {Sakurai}},
  \bibinfo {author} {\bibfnamefont {M.}~\bibnamefont {Koike}},\ and\ \bibinfo
  {author} {\bibfnamefont {L.}~\bibnamefont {Miller}},\ }\bibfield  {title}
  {\bibinfo {title} {{Cosmological constraints from Subaru weak lensing cluster
  counts}},\ }\href {https://doi.org/10.1093/pasj/psv034} {\bibfield  {journal}
  {\bibinfo  {journal} {Publ. Astron. Soc. Jap.}\ }\textbf {\bibinfo {volume}
  {67}},\ \bibinfo {pages} {34} (\bibinfo {year} {2015})},\ \Eprint
  {https://arxiv.org/abs/1503.01851} {arXiv:1503.01851 [astro-ph.CO]}
  \BibitemShut {NoStop}%
\bibitem [{\citenamefont {Shirasaki}\ \emph {et~al.}(2015)\citenamefont
  {Shirasaki}, \citenamefont {Hamana},\ and\ \citenamefont
  {Yoshida}}]{Shirasaki:2015dga}%
  \BibitemOpen
  \bibfield  {author} {\bibinfo {author} {\bibfnamefont {M.}~\bibnamefont
  {Shirasaki}}, \bibinfo {author} {\bibfnamefont {T.}~\bibnamefont {Hamana}},\
  and\ \bibinfo {author} {\bibfnamefont {N.}~\bibnamefont {Yoshida}},\
  }\bibfield  {title} {\bibinfo {title} {{Probing cosmology with weak lensing
  selected clusters \textendash{} I. Halo approach and all-sky simulations}},\
  }\href {https://doi.org/10.1093/mnras/stv1854} {\bibfield  {journal}
  {\bibinfo  {journal} {Mon. Not. Roy. Astron. Soc.}\ }\textbf {\bibinfo
  {volume} {453}},\ \bibinfo {pages} {3043} (\bibinfo {year} {2015})},\ \Eprint
  {https://arxiv.org/abs/1504.05672} {arXiv:1504.05672 [astro-ph.CO]}
  \BibitemShut {NoStop}%
\bibitem [{\citenamefont {Madhavacheril}\ \emph {et~al.}(2020)\citenamefont
  {Madhavacheril}, \citenamefont {Smith}, \citenamefont {Sherwin},\ and\
  \citenamefont {Naess}}]{Madhavacheril:2020ido}%
  \BibitemOpen
  \bibfield  {author} {\bibinfo {author} {\bibfnamefont {M.~S.}\ \bibnamefont
  {Madhavacheril}}, \bibinfo {author} {\bibfnamefont {K.~M.}\ \bibnamefont
  {Smith}}, \bibinfo {author} {\bibfnamefont {B.~D.}\ \bibnamefont {Sherwin}},\
  and\ \bibinfo {author} {\bibfnamefont {S.}~\bibnamefont {Naess}},\ }\bibfield
   {title} {\bibinfo {title} {{CMB lensing power spectrum estimation without
  instrument noise bias}}\ }\href
  {https://doi.org/10.1088/1475-7516/2021/05/028}
  {10.1088/1475-7516/2021/05/028} (\bibinfo {year} {2020}),\ \Eprint
  {https://arxiv.org/abs/2011.02475} {arXiv:2011.02475 [astro-ph.CO]}
  \BibitemShut {NoStop}%
\bibitem [{\citenamefont {Lembo}\ \emph {et~al.}(2022)\citenamefont {Lembo},
  \citenamefont {Fabbian}, \citenamefont {Carron},\ and\ \citenamefont
  {Lewis}}]{Lembo:2021kxc}%
  \BibitemOpen
  \bibfield  {author} {\bibinfo {author} {\bibfnamefont {M.}~\bibnamefont
  {Lembo}}, \bibinfo {author} {\bibfnamefont {G.}~\bibnamefont {Fabbian}},
  \bibinfo {author} {\bibfnamefont {J.}~\bibnamefont {Carron}},\ and\ \bibinfo
  {author} {\bibfnamefont {A.}~\bibnamefont {Lewis}},\ }\bibfield  {title}
  {\bibinfo {title} {{CMB lensing reconstruction biases from masking
  extragalactic sources}},\ }\href
  {https://doi.org/10.1103/PhysRevD.106.023525} {\bibfield  {journal} {\bibinfo
   {journal} {Phys. Rev. D}\ }\textbf {\bibinfo {volume} {106}},\ \bibinfo
  {pages} {023525} (\bibinfo {year} {2022})},\ \Eprint
  {https://arxiv.org/abs/2109.13911} {arXiv:2109.13911 [astro-ph.CO]}
  \BibitemShut {NoStop}%
\bibitem [{\citenamefont {Bartelmann}\ and\ \citenamefont
  {Schneider}(2001)}]{Bartelmann:1999yn}%
  \BibitemOpen
  \bibfield  {author} {\bibinfo {author} {\bibfnamefont {M.}~\bibnamefont
  {Bartelmann}}\ and\ \bibinfo {author} {\bibfnamefont {P.}~\bibnamefont
  {Schneider}},\ }\bibfield  {title} {\bibinfo {title} {Weak gravitational
  lensing},\ }\href@noop {} {\bibfield  {journal} {\bibinfo  {journal} {Phys.
  Rept.}\ }\textbf {\bibinfo {volume} {340}},\ \bibinfo {pages} {291} (\bibinfo
  {year} {2001})},\ \Eprint {https://arxiv.org/abs/astro-ph/9912508}
  {astro-ph/9912508} \BibitemShut {NoStop}%
%%CITATION = ASTRO-PH 9912508;%%
\bibitem [{\citenamefont {Bartelmann}\ and\ \citenamefont
  {Maturi}(2016)}]{Bartelmann:2016dvf}%
  \BibitemOpen
  \bibfield  {author} {\bibinfo {author} {\bibfnamefont {M.}~\bibnamefont
  {Bartelmann}}\ and\ \bibinfo {author} {\bibfnamefont {M.}~\bibnamefont
  {Maturi}},\ }\bibfield  {title} {\bibinfo {title} {{Weak gravitational
  lensing}}\ }(\bibinfo {year} {2016})\ \Eprint
  {https://arxiv.org/abs/1612.06535} {arXiv:1612.06535 [astro-ph.CO]}
  \BibitemShut {NoStop}%
\bibitem [{\citenamefont {Villumsen}(1995)}]{Villumsen:1995ar}%
  \BibitemOpen
  \bibfield  {author} {\bibinfo {author} {\bibfnamefont {J.~V.}\ \bibnamefont
  {Villumsen}},\ }\bibfield  {title} {\bibinfo {title} {{Clustering of faint
  galaxies: omega (Theta), induced by weak gravitational lensing}},\
  }\href@noop {} {\  (\bibinfo {year} {1995})},\ \Eprint
  {https://arxiv.org/abs/astro-ph/9512001} {arXiv:astro-ph/9512001}
  \BibitemShut {NoStop}%
\bibitem [{\citenamefont {Moessner}\ \emph {et~al.}(1998)\citenamefont
  {Moessner}, \citenamefont {Jain},\ and\ \citenamefont
  {Villumsen}}]{Moessner:1997qs}%
  \BibitemOpen
  \bibfield  {author} {\bibinfo {author} {\bibfnamefont {R.}~\bibnamefont
  {Moessner}}, \bibinfo {author} {\bibfnamefont {B.}~\bibnamefont {Jain}},\
  and\ \bibinfo {author} {\bibfnamefont {J.~V.}\ \bibnamefont {Villumsen}},\
  }\bibfield  {title} {\bibinfo {title} {{The effect of weak lensing on the
  angular correlation function of faint galaxies}},\ }\href
  {https://doi.org/10.1046/j.1365-8711.1998.01225.x} {\bibfield  {journal}
  {\bibinfo  {journal} {Mon. Not. Roy. Astron. Soc.}\ }\textbf {\bibinfo
  {volume} {294}},\ \bibinfo {pages} {291} (\bibinfo {year} {1998})},\ \Eprint
  {https://arxiv.org/abs/astro-ph/9708271} {arXiv:astro-ph/9708271}
  \BibitemShut {NoStop}%
\bibitem [{\citenamefont {Kaiser}(1987)}]{Kaiser:1987qv}%
  \BibitemOpen
  \bibfield  {author} {\bibinfo {author} {\bibfnamefont {N.}~\bibnamefont
  {Kaiser}},\ }\bibfield  {title} {\bibinfo {title} {{Clustering in real space
  and in redshift space}},\ }\href {https://doi.org/10.1093/mnras/227.1.1}
  {\bibfield  {journal} {\bibinfo  {journal} {Mon. Not. Roy. Astron. Soc.}\
  }\textbf {\bibinfo {volume} {227}},\ \bibinfo {pages} {1} (\bibinfo {year}
  {1987})}\BibitemShut {NoStop}%
\bibitem [{\citenamefont {Tanidis}\ and\ \citenamefont
  {Camera}(2019)}]{Tanidis:2019teo}%
  \BibitemOpen
  \bibfield  {author} {\bibinfo {author} {\bibfnamefont {K.}~\bibnamefont
  {Tanidis}}\ and\ \bibinfo {author} {\bibfnamefont {S.}~\bibnamefont
  {Camera}},\ }\bibfield  {title} {\bibinfo {title} {{Developing a unified
  pipeline for large-scale structure data analysis with angular power spectra
  \textendash{} I. The importance of redshift-space distortions for galaxy
  number counts}},\ }\href {https://doi.org/10.1093/mnras/stz2366} {\bibfield
  {journal} {\bibinfo  {journal} {Mon. Not. Roy. Astron. Soc.}\ }\textbf
  {\bibinfo {volume} {489}},\ \bibinfo {pages} {3385} (\bibinfo {year}
  {2019})},\ \Eprint {https://arxiv.org/abs/1902.07226} {arXiv:1902.07226
  [astro-ph.CO]} \BibitemShut {NoStop}%
\bibitem [{\citenamefont {Tanidis}\ \emph {et~al.}(2024)\citenamefont {Tanidis}
  \emph {et~al.}}]{Euclid:2023pyq}%
  \BibitemOpen
  \bibfield  {author} {\bibinfo {author} {\bibfnamefont {K.}~\bibnamefont
  {Tanidis}} \emph {et~al.} (\bibinfo {collaboration} {Euclid}),\ }\bibfield
  {title} {\bibinfo {title} {{Euclid preparation - XXXIV. The effect of linear
  redshift-space distortions in photometric galaxy clustering and its
  cross-correlation with cosmic shear}},\ }\href
  {https://doi.org/10.1051/0004-6361/202347870} {\bibfield  {journal} {\bibinfo
   {journal} {Astron. Astrophys.}\ }\textbf {\bibinfo {volume} {683}},\
  \bibinfo {pages} {A17} (\bibinfo {year} {2024})},\ \Eprint
  {https://arxiv.org/abs/2309.00052} {arXiv:2309.00052 [astro-ph.CO]}
  \BibitemShut {NoStop}%
\bibitem [{\citenamefont {Aghanim}\ \emph
  {et~al.}(2020{\natexlab{b}})\citenamefont {Aghanim} \emph
  {et~al.}}]{PCP2018}%
  \BibitemOpen
  \bibfield  {author} {\bibinfo {author} {\bibfnamefont {N.}~\bibnamefont
  {Aghanim}} \emph {et~al.} (\bibinfo {collaboration} {Planck}),\ }\bibfield
  {title} {\bibinfo {title} {{Planck 2018 results. VI. Cosmological
  parameters}},\ }\href {https://doi.org/10.1051/0004-6361/201833910}
  {\bibfield  {journal} {\bibinfo  {journal} {\aap}\ }\textbf {\bibinfo
  {volume} {641}},\ \bibinfo {pages} {A6} (\bibinfo {year}
  {2020}{\natexlab{b}})},\ \Eprint {https://arxiv.org/abs/1807.06209}
  {arXiv:1807.06209 [astro-ph.CO]} \BibitemShut {NoStop}%
\bibitem [{\citenamefont {Beck}\ \emph {et~al.}(2018)\citenamefont {Beck},
  \citenamefont {Fabbian},\ and\ \citenamefont {Errard}}]{Beck:2018wud}%
  \BibitemOpen
  \bibfield  {author} {\bibinfo {author} {\bibfnamefont {D.}~\bibnamefont
  {Beck}}, \bibinfo {author} {\bibfnamefont {G.}~\bibnamefont {Fabbian}},\ and\
  \bibinfo {author} {\bibfnamefont {J.}~\bibnamefont {Errard}},\ }\bibfield
  {title} {\bibinfo {title} {{Lensing Reconstruction in Post-Born Cosmic
  Microwave Background Weak Lensing}},\ }\href
  {https://doi.org/10.1103/PhysRevD.98.043512} {\bibfield  {journal} {\bibinfo
  {journal} {Phys. Rev.}\ }\textbf {\bibinfo {volume} {D98}},\ \bibinfo {pages}
  {043512} (\bibinfo {year} {2018})},\ \Eprint
  {https://arxiv.org/abs/1806.01216} {arXiv:1806.01216 [astro-ph.CO]}
  \BibitemShut {NoStop}%
%%CITATION = ARXIV:1806.01216;%%
\bibitem [{\citenamefont {Tegmark}(1997)}]{Tegmark:1997rp}%
  \BibitemOpen
  \bibfield  {author} {\bibinfo {author} {\bibfnamefont {M.}~\bibnamefont
  {Tegmark}},\ }\bibfield  {title} {\bibinfo {title} {{Measuring cosmological
  parameters with galaxy surveys}},\ }\href
  {https://doi.org/10.1103/PhysRevLett.79.3806} {\bibfield  {journal} {\bibinfo
   {journal} {Phys. Rev. Lett.}\ }\textbf {\bibinfo {volume} {79}},\ \bibinfo
  {pages} {3806} (\bibinfo {year} {1997})},\ \Eprint
  {https://arxiv.org/abs/astro-ph/9706198} {arXiv:astro-ph/9706198}
  \BibitemShut {NoStop}%
\bibitem [{\citenamefont {Hannestad}\ \emph {et~al.}(2006)\citenamefont
  {Hannestad}, \citenamefont {Tu},\ and\ \citenamefont
  {Wong}}]{Hannestad:2006as}%
  \BibitemOpen
  \bibfield  {author} {\bibinfo {author} {\bibfnamefont {S.}~\bibnamefont
  {Hannestad}}, \bibinfo {author} {\bibfnamefont {H.}~\bibnamefont {Tu}},\ and\
  \bibinfo {author} {\bibfnamefont {Y.~Y.~Y.}\ \bibnamefont {Wong}},\
  }\bibfield  {title} {\bibinfo {title} {{Measuring neutrino masses and dark
  energy with weak lensing tomography}},\ }\href
  {https://doi.org/10.1088/1475-7516/2006/06/025} {\bibfield  {journal}
  {\bibinfo  {journal} {JCAP}\ }\textbf {\bibinfo {volume} {06}},\ \bibinfo
  {pages} {025}},\ \Eprint {https://arxiv.org/abs/astro-ph/0603019}
  {arXiv:astro-ph/0603019} \BibitemShut {NoStop}%
\end{thebibliography}
